\documentclass[12pt,a4paper]{JHEP3}
 
\usepackage{amscd,amsmath,amssymb,amsfonts,xspace, mathtools}
  \usepackage{xcolor}
  \usepackage{tikz-cd}

\usepackage{latexsym}  
 \usepackage{graphicx}

\newcommand{\be}{\begin{equation}}  
\newcommand{\ee}{\end{equation}} 
\newcommand{\bes}{\begin{equation*}} 
\newcommand{\ees}{\end{equation*}}

\newcommand{\CH}{\mathcal{H}}

\newcommand{\CK}{\mathcal{K}}
 
\newcommand{\CM}{\mathcal{M}}  
\newcommand{\CN}{\mathcal{N}}

\newcommand{\CV}{\mathcal{V}}

\newcommand{\CZ}{\mathcal{Z}}

\newcommand{\bfl}{{\boldsymbol l}} 
\newcommand{\bfLambda}{{\boldsymbol \Lambda}}
\newcommand{\bfK}{{\boldsymbol K}}

\newcommand{\bfnu}{{\boldsymbol \nu}}

\newcommand{\bfk}{{\boldsymbol k}}

\newcommand{\bfQ}{{\boldsymbol Q}}
\newcommand{\bfu}{{\boldsymbol u}}

\newcommand{\bfmu}{{\boldsymbol \mu}}

\newcommand{\I}{{\rm i}}

\newcommand{\refb}[1]{(\ref{#1})}

\newcommand{\sgn}{\mathop{\mathrm{sgn}}}

\newcommand{\nn}{\ensuremath{\nonumber{}}}

\title{Scaling Black Holes and Modularity}

\author{Aradhita Chattopadhyaya, Jan Manschot, Swapnamay Mondal
\\
$^1$ {\it School of Mathematics, Trinity College, Dublin 2, Ireland}
\\
$^2$ {\it Hamilton Mathematical Institute, Trinity College, Dublin 2, Ireland}
\vspace*{2mm}\\
\vspace*{-3mm}
}  
 
\abstract{
Scaling black holes are solutions of supergravity with multiple
  black hole singularities, which can be adiabatically connected to a
  single center black hole solution. We develop techniques to determine
partition functions for such scaling black holes, if each constituent carries a
  non-vanishing magnetic charge corresponding to a D4-brane in string
  theory, or equivalently M5-brane in M-theory. For three
  constituents, we demonstrate that the partition function is a mock
  modular form of  depth two, and we determine the appropriate non-holomorphic
  completion using generalized error functions. From the
  four-dimensional perspective, the modular parameter is the
  axion-dilaton, and our results show that $S$-duality leaves this
  subset of the spectrum invariant. From the five-dimensional perspective,
the modular parameter is the complex structure of a torus
$T^2$, and the scaling black holes are dual to states in the
dimensional reduction of the M5-brane worldvolume theory to $T^2$.
As a case study, we specialize the compactification manifold to a K3
  fibration, and explicitly evaluate holomorphic parts of
  partition functions. 
\vspace{.3cm} \\
\today\\ 
}

\begin{document}
 
\maketitle
 
\section{Introduction} 

Solutions of supergravity with multiple black hole singularities provide
interesting insights on the spectrum of quantum gravity and the
dependence on the compactification moduli \cite{Maldacena:1998uz, Maloney:1999dv, Denef:2000nb,   
  Denef:2001xn, Bena:2006kb, Denef:2007vg, Cheng:2007ch,
  deBoer:2008zn, Manschot:2010qz, Dabholkar:2012nd}. Such multi-center solutions are
particularly intriguing if their solution space contains a region,
where the distances between the centers becomes arbitrarily small.
In other words, these solutions, known
as  ``scaling'' solutions \cite{Denef:2007vg, Bena:2012hf} in four dimensions, are adiabatically connected to
a supergravity solution with a single black hole singularity.
  
We will study such scaling black holes formed from M5-branes in 
M-theory, or equivalently D4-branes in Type IIA string theory
\cite{Maldacena:1997de}. Such black holes are well studied using the
AdS$_3$/CFT$_2$ correspondence \cite{Maldacena:1997de, Gaiotto:2006wm, deBoer:2006vg,
  Kraus:2006nb}, as well as the hypermultiplet
geometry of the IIB string theory \cite{Alexandrov:2012au, Alexandrov:2013yva, Alexandrov:2016tnf}.
The D4/M5-branes wrap a four-cycle (or divisor)
$P\in H_4(X,\mathbb{Z})$ of the Calabi-Yau threefold $X$, which can also be studied within algebraic geometry
\cite{Gholampour:2013hfa, Gholampour:2013ifa, Bouchard:2016lfg, Feyzbakhsh:2020wvm, Feyzbakhsh:2021rcv}.

Multi-center solutions exist with each
center having a positive D4-brane charge but vanishing D6-brane
charge. Their walls of marginal stability extend to
the large volume regime of the K\"ahler moduli space 
\cite{Andriyash:2008it, Manschot:2009ia, Manschot:2010xp, deBoer:2008fk}, which
 corresponds to $\mathrm{Im}(t)=J=\lambda\, \underline{J}$,
where $\underline{J}$ is the normalized, dimensionless K\"ahler
modulus, $\underline J^3=1$, and $\lambda \gg 1$. While attractor points lie typically in the interior
of the K\"ahler moduli space, there is an analogue of the attractor point at large
volume $t^\lambda_\gamma$ defined in Eq. (\ref{LVattractor})
\cite{Alexandrov:2016tnf, Manschot:2009ia, deBoer:2008fk}. Similarly,
single center black holes, whose internal
degrees of freedom are independent of the asymptotic moduli, have an
analogue at large volume, where families of black hole
solutions appear effectively as single center black holes. Much like nucleons can
be considered as elementary particles at sufficiently low energies. 

To deal with these effectively single center solutions, we introduce
the notion of ``$\lambda$-core'', or ``core'' for short. The
defining property of a ``$\lambda$-core'' is that the 
distances between constituents in a $\lambda$-core are bounded by
$C\lambda^{-3}$ for sufficiently large $\lambda$, and for some fixed length $C$. Examples of such bound
black holes states are of course proper single center black holes, as
well as bound states of a D6-brane and anti-D6-brane,
for which the wall of marginal stability lies in the interior of the K\"ahler moduli
space.

The $\lambda$-core solutions are also distinguished in the uplift to
five dimensions, and decoupling limit. Recall that the four
dimensional Newton's constant equals $G_4=\ell_5^3/R$, with $\ell_5$ the five dimensional Planck length and $R$ the radius of the
M-theory circle. Then, $\lambda$ is related to these variables by
$\lambda= R/\ell_5$ \cite{deBoer:2008fk}.
 When uplifted to five dimensions and in the decoupling limit $\ell_5\to 0$, states with D4-D2-D0 charges and within a $\lambda$-core will develop an asymptotically AdS$_3$
throat, whereas bound states with larger separation will decouple from
the spectrum since their energy diverges \cite{deBoer:2008fk}.

In a series of works \cite{Alexandrov:2016tnf, Alexandrov:2017qhn, Alexandrov:2018lgp, Alexandrov:2019rth}, the modular properties of
partition functions enumerating D4-D2-D0 black holes are studied from
a complementary perspective, namely by mapping this D-brane system to
D3-D1-D(-1) instantons. These D-instantons correct the hypermultiplet
geometry \cite{Alexandrov:2013yva, Robles-Llana:2006hby, Alexandrov:2008gh,
  Alexandrov:2009zh}, which is constrained by IIB $SL(2, \mathbb{Z})$ S-duality
group. This duality group acts on the axion-dilation field $\chi:= C^{(0)} + i e^{-\phi}$ by
linear fractional transformations. Here $C^{(0)}$ is the
RR scalar and $\phi$ is the dilaton. Alternatively, the duality can be
identified with the modular symmetry of the worldvolume reduction of M5-branes. These have proven to be fruitful connections to determine the
modular and analytic properties of the partition functions for
D4-D2-D0 black holes. In this way, non-holomorphic contributions to the 
partition function are determined, which imply that the partition
functions involve mock modular forms \cite{ZwegersThesis, MR2605321}, and mock
modular forms of higher depth \cite{Alexandrov:2016enp, Nazaroglu:2016lmr, kudla2016theta, funke_kudla_2019}.
This implies potentially interesting arithmetic of the BPS indices,
while the non-holomorphic contribution is also interesting
independently, and a generalization of similar non-holomorphic terms in
partition functions of $\CN=4$ Yang-Mills on four-manifolds
\cite{Vafa:1994tf, Minahan:1998vr}. The origins and explicit
expressions of these terms have been understood better recently \cite{ Alexandrov:2019rth, Manschot:2011dj, Manschot:2017xcr,
 Dabholkar:2020fde, Alexandrov:2020bwg,
  Alexandrov:2020dyy,  Bonelli:2020xps, Manschot:2021qqe}.     
    
We will study in this paper the partition function of
 scaling black holes with three constituents. Each constituent is a $\lambda$-core carrying a positive D4-brane charge,
but vanishing D6-brane charge. Intriguingly, each core gives rise to an associated
AdS$_3$/CFT$_2$ after uplifting to M-theory, while the near coincident
region gives rise to an asymptotic AdS$_3$ solution for the total
magnetic charge, and should thus be captured by the
AdS$_3$/CFT$_2$ correspondence for the total charge. In other words,
they are examples of AdS fragmentation, and
attributed to the Coulomb branch of the CFT \cite{Maldacena:1998uz}.
The existence of scaling black holes with these charges puts
constraints on the topology of the Calabi-Yau threefold. In
particular, these only exist if the second Betti
number of the Calabi-Yau manifold $b_2\geq 2$.   

Our explicit analysis of scaling solutions gives rise to a decomposition for the partition
function of attractor black holes $\CZ^\lambda_P$ with fixed magnetic
charge $P$ reads (See also Eq. (\ref{AttrPDecomp}).) in terms of
partition functions of single core and $n$-core scaling black holes
$\CZ^{nT}_P$. This reads explicitly,
\be
\label{CZlambdaDec}
\CZ^\lambda_P= \CZ^T_P+\CZ^{3T}_P+\dots ,
\ee
where $ \CZ^T_P$ is the partition function of single core black
holes, $\CZ^{3T}_P$ the partition function of scaling black
holes consistenting of three cores, and
the dots stand for scaling black holes with $n>3$ cores.  We determine the holomorphic part of the partition function using
 formula's for the degeneracies of black hole bound states such as
 studied in  \cite{Denef:2007vg, Bena:2012hf, Lee:2012sc, Anninos:2013nra,
   Manschot:2013sya, Mirfendereski:2020rrk, Beaujard:2021fsk}, which
 gives rise to a holomorphic indefinite theta series $\Psi_\bfmu$ of signature
 $(2,2b_2-2)$. Since its coefficients grow polynomially, this demonstrates that the entropy arising from these solutions is
 exponentially smaller than the entropy of a single center black
 hole. This is expected since we have not included pure Higgs
 degeneracies \cite{Bena:2012hf, Lee:2012sc}  or single center
 \cite{Manschot:2013sya}. 
We have worked out two explicit case studies, where we specialize the CY three-fold to a K3 fibration and determine
explicit $q$-series for the partition function.
 We do observe that the exponent of the
 leading term in the $q$-expansions is rather large.

 We have moreover demonstrated that the partition function
 $\CZ^{3T}_P$ admits a theta function decomposition as a consequence
 of a spectral flow symmetry familiar from  MSW CFT. This in contrast
 to generic bound states with non-vanishing D4-brane charge for which
 this symmetry is not present \cite{Manschot:2009ia}. On the other
 hand for modular transformations, $\CZ^{3T}_P$ needs to be
 complemented with additional non-holomorphic terms, a phenomena which
 is also familiar for the attractor partition function $
 \CZ^\lambda_P$ as mentioned above. We distinguish the completed
 functions from the original functions by a hat, thus $
 \widehat \CZ^\lambda_P$ for $
 \CZ^\lambda_P$ and $\widehat \CZ^{3T}_P$ for $\CZ^{3T}_P$ and
 similarly for other functions. We then establish that $\widehat \CZ^{\lambda}_P$ and $\widehat
 \CZ^{3T}_P$ transform identically. Alexandrov and Pioline have
derived in Ref. \cite{Alexandrov:2018lgp} the non-holomorphic terms for the rhs, $\widehat\CZ^\lambda_P$. 
 It would be interesting to combine this with the non-holomorphic
 terms derived in the present paper for $\widehat \CZ^{3T}_P$ to deduce the
 non-holomorphic terms of $\widehat \CZ^T_P$.

The identical transformation properties of $\widehat \CZ^{\lambda}_P$ and $\widehat
 \CZ^{nT}_P$ raises the question, which term(s) correspond to the
 partition function of the conformal field theory. Since upon taking the decoupling limit to vanishing 5d
Planck length $\ell_5\to 0$ \cite{deBoer:2008fk}, the supergravity
solutions contributing to  $\widehat\CZ^{3T}_P$ decouple from the
AdS$_3$ geometry. This term is therefore not expected to correspond to
states within the MSW conformal field theory, and it seems therefore
plausible to us to expect that
$\widehat\CZ^T_P$ is to be identified with the MSW CFT partition
function. It would be interesting to understand whether the finite
difference between $\widehat\CZ^T_P$ and $\widehat\CZ^\lambda_P$ can
arise by turning on the irrelevant perturbation in the CFT, which
corresponds to moving away from the near horizon geometry and up the
attractor flow \cite{deBoer:2008ss}.

Determination of the completion $\widehat\CZ^{3T}_P$ amounts to
determining the completion of the indefinite theta series $\Psi_\bfmu$.
We determine furthermore the non-holomorphic completion of a closely related
function $\Phi_\bfmu$ (\ref{DefPhi}), which enumerates the number of ``scaling'' charge
configurations. Although still involved, determination of the
non-holomorphic completion of this function is simpler than for
generating function of the BPS indices. The analysis of $\Phi_\bfmu$
and $\Psi_\bfmu$
demonstrates that these are mock modular forms of depth 2. There are
different representations of the completion:
\begin{enumerate}
\item As a non-holomorphic kernel of the theta series. This involves
  (generalized) error functions \cite{Alexandrov:2016enp}. The modular properties of the
  completion follow from application of results by Vign\'eras
  \cite{Vigneras:1977}. This is applied to the partition functions for
  scaling black holes in Section \ref{ssmodcom}. For the partition function
  $\widehat \Phi_\bfmu$ see Equation (\ref{compterms}), and for $\widehat \Psi_\bfmu$, see
  Equation (\ref{completePsi}). 
\item Another useful representation, well-known for mock modular forms
  \cite{ZwegersThesis, MR2605321}, is as a holomorphic $q$-series
  plus an (iterated) integral of modular forms \cite{Alexandrov:2016enp, Manschot:2017xcr, bringmann2018higher}. This form is determined for three-center scaling black holes
in equation (\ref{defhatPhinew}) and (\ref{Rphifinalnew}). 
The modular transformations of the $q$-series 
follow directly from those of the iterated integral. This 
representation is also relevant physically, where the non-holomorphic
part is contributed to continuum of states \cite{Pioline:2015wza}, or the Coulomb branch
\cite{Dabholkar:2020fde, Manschot:2021qqe}. 
\item The third representation we mention here is as an integral over a domain in a
  symmetric space, studied by Funke, Kudla and Millson
  \cite{kudla2016theta, funke_kudla_2019, Kudla1986}, also known as Narain
  moduli space in the context of conformal field theory. In the case
  of a 3-center scaling solution, this would be a union of triangles in $SO(2,2b_2-2;\mathbb{Z})\backslash
  SO(2,2b_2-2;\mathbb{R})/SO(2;\mathbb{R})\times
  SO(2b_2-2;\mathbb{R})$. While we have not explored this representation in detail, it
  is interesting to mention in light of recent discussion on averaging
  over Narian moduli space and the AdS$_3$/CFT$_2$ correspondence
  \cite{Afkhami-Jeddi:2020ezh, Maloney:2020nni}. 
 \end{enumerate}

Though technically involved, we believe that our results can be
extended to scaling solutions with $n>3$ cores, and will give rise
to mock modular forms of depth $n-1$. Any Calabi-Yau manifold
with $b_2\geq 2$ gives rise to such mock modular forms, which thus
provides a large resource of holomorphic higher depth mock modular
forms. Further, it will be interesting to include single center degeneracies. 
Moreover, we hope that our results could be used for the study of AdS$_3$ fragmention, and the interpretation of these solutions in the dual
CFT. In this way, it may be possible to derive the non-holomorphic terms within gravity or the worldvolume theory of
intersecting D-branes.

The outline of this paper is as follows. Section \ref{BHsols} reviews
aspects of multi-center black holes, and in particular the index of
scaling solutions. Section \ref{s2} reviews partition functions of
D4-brane black holes. Section \ref{s3gen} discusses charge lattices
for D4-brane bound states, and defines the partition
functions of scaling black holes. We define here also the partition
functions $\Phi_\bfmu$ and $\Psi_\bfmu$, which enumerate scaling
configurations and their BPS indices, and determine its modular
completion. Section \ref{sec:Mtheory} discusses the relation to
M-theory and the decoupling limit AdS$_3$. 
Section \ref{sexample} considers two case studies for a
specific Calabi-Yau 3-fold and charges, and presents the holomorphic
$q$-series which are mock modular forms of depth 2.

\section{Black hole solutions in $\CN=2$ Supergravity}  
\label{BHsols}
We briefly review in this section supersymmetric black holes in $\CN=2$
supergravity and partition functions.

\subsection{Black hole bound states}
Let $X$ be a simply connected Calabi-Yau threefold, with triple
intersection product $d_{abc}$, $a,b,c=1,\dots,b_2$. The intersection product $d_{abc}$ is symmetric in its indices. 
The classical central charge of a BPS state is given by
\be
\label{defCZ}
Z(\gamma,t)=-\int_X e^{-t}\wedge \gamma,
\ee
where $\gamma$ on the rhs is the Poincar\'e dual differential form of the homology class of the cycle which supports
the D-branes, and is in 1-to-1 correspondence with the electric-magnetic
charge of the BPS state. Moreover, $t$ is the K\"ahler modulus of the Calabi-Yau three-fold. 

The scalar fields $X^I$, $I=0,\dots, b_2$ of the vector
multiplets are related to the Calabi-Yau moduli, $t^a=B^a+iJ^a$, as $t^a=X^a/X^0$ for $a=1,\dots,b_2$. 
Near the horizon, their values are determined by the attractor
equations at the horizon in terms of the
electric-magnetic charges of the black hole \cite{Ferrara:1995ih, Ferrara:1996dd,
  Shmakova:1996nz}. On the other hand, their asymptotic values for
$|\vec r|\to \infty$ are boundary conditions for the equations of motion.

Besides the single center black hole, the equations of motion of
$\CN=2$ supergravity give rise to intricate multi-center black hole solutions
\cite{Denef:2000nb, Denef:2007vg, Denef:2002ru, Bates:2003vx}. Upon
varying the asymptotic values of the scalar fields, multi-center solutions can cease to exist as proper
solutions to the supergravity equations of motions, or reversely new
solutions can appear. If the asymptotic values are chosen equal to the
attractor values only few multi-center solutions exist. That is to
say, only multi-center solutions which can be continuously connected to a
single center black hole exist. These are the scaling 
 solutions mentioned above, and are the main focus of this paper. 

To understand this more explicitly, recall that an
$n$-center solution, is required to satisfy the following $n-1$ Denef equations,
\be
\label{Deqs}
\sum_{i\neq j} \frac{\left< \gamma_{i},\gamma_j\right>}{r_{ij}}=
2\,\left.\mathrm{Im}\left( e^{-\I\alpha} Z(\gamma_i,t)\right)\right\vert_{r=\infty},
\ee
where $\left<, \right>$ is the symplectic innerproduct between the
charges, 
\be
\left< \gamma_1, \gamma_2 \right>=-P_1^0Q_{0,2}+P_1\cdot Q_2-P_2\cdot Q_1+P_2^0Q_{0,1}.
\ee
Moreover, $r_{ij}=|\vec r_i-\vec r_j|$ is the distance between the centers $i$ and $j$, and $\alpha$ is the phase of the central charge
$Z(\gamma,t)$ for the total charge $\gamma=\sum_{i=1}^n \gamma_i$. We
set  
\be   
\label{ciZ} 
c_{j}=2\,\left.\mathrm{Im}\left( e^{-\I\alpha} Z(\gamma_j,t)\right)\right\vert_{r=\infty}.
\ee
We fix $\vec r_1$ at the origin $\mathbb{R}^3$ and let $\CM_n$ be the
solution space for $\vec r_{j}\in \mathbb{R}^3$, $j=2,\dots,n$ to
(\ref{Deqs}). Then $\CM_n$ has dimension $2n-2$. The low energy
degrees of freedom of the supersymmetric multi-center black hole give
rise to $\CN=4$ quiver quantum
mechanics \cite{Denef:2002ru}. The quiver for an $n$-center bound
state with charges $\{\gamma_j\}$ consists of $n$ nodes, and $\gamma_{ij}>0$ arrows from node
$i$ to node $j$.

Note that the equations \eqref{Deqs} are necessary but not
sufficient for the multi-center solution to exist, to this end one
needs to verify that the full supergravity solution is regular away
from the black hole singularities, and without time-like curves \cite{Denef:2000nb}.
Since we restrict to the large volume limit of
the Calabi-Yau moduli space, we assume that this is the case in the
following, and that we can determine the existence of bound states
from (\ref{Deqs}).

The gravity perspective has led to the following form for the index of
an $n$-center bound state \cite{Denef:2007vg, Manschot:2010qz}. To
express this, we first introduce the rational index $\bar
\Omega(\gamma)$ associated to the integer index $\Omega(\gamma)$,
\be
\label{ratIndex} 
\bar \Omega(\gamma)=\sum_{m\vert \gamma} \frac{\Omega(\gamma/m)}{m^2}.
\ee
The single center invariants $\Omega_S(\gamma)$ are the internal degeneracies
of a supersymmetric particle or black hole with charge $\gamma$. It is
expected to be a positive integer for a black hole. 
To analyze the spectrum of bound states it is convenient to introduce
a fugacity $y$ for angular momentum. The rational variant of the
refined index is defined as
\be 
\label{ratIndexy} 
\bar \Omega(\gamma,y)=\sum_{m\vert \gamma} \frac{1}{m}\frac{y-y^{-1}}{y^m-y^{-m}}\,\Omega(\gamma/m,y^m).
\ee
This reproduces \eqref{ratIndex} in the limit $y\to 1$. 
A few variants of BPS indices will be important for us. We mention,
\begin{itemize}
\item The BPS invariant $\Omega(\gamma;t)$, which enumerates BPS
  states for a given value $t$ of the asymptotic moduli. This include
  single-center BPS states as well as bound states.
\item The single-center invariant $\Omega_S(\gamma)$, which is the
  internal degeneracy of a BPS particle or black hole center. This
  invariant is independent of the moduli $t$.
\item The total invariant $\Omega_T(\gamma)$, which is a composite of
  $\Omega_S(\gamma)$ and independent of the moduli $t$. We give the
  expression below in (\ref{defOmTy}).
\end{itemize}

The rational refined BPS index $\bar \Omega(\gamma,y;t)$ can be
expressed as a sum over partitions of $\gamma$, each representing a
BPS bound state. It takes the form \cite{Manschot:2010qz}
\be
\label{gCformula}
\bar \Omega(\gamma,y;t)=\sum_{\gamma=\sum_{i=1}^n\gamma_i} \frac{ g_C(\{\gamma_j\},\{c_j\},y)}{|{\rm Aut}(\{\gamma_j\})|}\,\prod_{j=1}^n \bar \Omega_T(\gamma_j,y),
\ee
with, 
\begin{itemize}
\item $|{\rm Aut}(\{\gamma_j\})|$ is the order of the subgroup of
the permutation group, which preserves the ordered set
$\{\gamma_1,\dots,\gamma_n\}$. 
\item The index $g_C$ can be determined using localization of the black hole
solution with respect to rotation around a fixed  axis generated by $J_3$, say the
$z$-axis \cite{Manschot:2011xc}. A fixed point $p\in \CM_n$ corresponds to a collinear
solution with all centers placed on the $z$-axis. If the associated
bound state quiver has no oriented loop, $g_C$ is the refined index of the $\CN=4$ quiver quantum 
mechanics describing the bound state \cite{Manschot:2010qz},
\be
g_C(\{\gamma_j\},\{c_j\},y)= \mathrm{Tr}'_{\CH_{\rm qm}} (-y)^{2J_3},
\ee
where the trace is over the BPS Hilbert space $\CH_{\rm qm}$ of the quiver
quantum mechanics, and $J_3$ is one of the generators of $SU(2)$. 
\item  $\Omega_T$ the {\it total invariant} defined by
\be
\label{defOmTy}
\Omega_T(\gamma,y)=\Omega_S(\gamma,y)+\sum_{\sum_{j=1}^n
  m_j\gamma_j=\gamma} H(\{\gamma_i,m_i\},y)\,\prod_{i=1}^n \Omega_S(\gamma_i,y^{m_i}),
\ee 
where $m_j\in \mathbb{N}$ are multiplicities of the charges in the
partition of $\gamma$. 
For bound states whose associated quiver has no closed loops, the
$H(\{\gamma_i,m_i\},y)$ vanish. Otherwise they are determined by the
``minimal modification hypothesis''. This has the effect that if we
express (\ref{gCformula}) as
\be
\label{bgCformula}
\bar \Omega(\gamma,y;t)=\sum_{\gamma=\sum_{i=1}^n\gamma_i} \frac{\bar g_C(\{\gamma_j\},\{c_j\},y)}{|{\rm Aut}(\{\gamma_j\})|}\,\prod_{j=1}^n \bar \Omega_S(\gamma_j,y),
\ee
then the $\bar g_C$ are $SU(2)$ characters.
\end{itemize}

To determine $g_C$ using localization, it is convenient to introduce
the refined index 
$$g_C(\{\gamma_j\},\{c_j\},y)=\mathrm{Tr}'_{\CH_{\rm qm}} (-y)^{2J_3}.$$
Let $z_j$ be the position of the center with charge
$\gamma_j$. The localization technique then gives the following sum
over collinear fixed points with respect to rotation around this axis,
\be
\label{gCnoloops}
g_C(\{\gamma_j\},\{c_j\};y)=(-1)^{n-1} (y - y^{-1})^{-n+1} \sum_{p\in \CM_n} s(p)\, (-y)^{\sum_{i<j} \gamma_{ij}\sgn(z_j-z_i)},
\ee 
where $s(p)\in \pm 1$ is a sign depending on the details of the fixed
point, and $z_j$ is the $z$-coordinate of center $j$.
If the associated quiver does not contain loops, this is the
complete index and the $y\to 1$ limit is well-defined. However, if the
quiver contains loops, the distances between the black hole centers may 
be arbitrarily small \cite{Denef:2007vg}. Such solutions are known as scaling
solutions, and additional fixed points need to be
included in (\ref{gCnoloops}). An ($n-$center) {\it scaling black hole} is a multi-center solution of $n$
black holes, whose phase space $\CM_n$ contains a region where the
centers can approach each other arbitrarily close. Thus, while the
centers are spatially separated for generic points of $\CM_n$, they
are adiabatically connected to the black hole solution with a single center.

While many BPS bound states decay if we tune the moduli to their
attractor values, scaling solutions remain part of the BPS
spectrum. Since the index is evaluated at the attractor point $c_j^*$, each term
on the rhs of (\ref{gCformula}) with $n\geq 3$ corresponds to a scaling solution.

Various quantities may diverge in the limit $y\to 1$, such as $\Omega_T$ and $g_C$. In order to
arrive at numerical counterparts for these quantities, we propose to regularize
a rational function of the form
\be
\frac{f(y)}{(y-y^{-1})^\ell},\qquad \text {with}\qquad \lim_{y\to 1} f(y)\neq 0,
\ee
as follows 
\be
\label{regindex}
\frac{f(y)}{(y-y^{-1})^\ell} \longrightarrow \frac{1}{2^\ell\,\ell!}\,\left.
\left(y \frac{d}{dy}\right)^\ell f(y) \right\vert_{y=1}.
\ee

\subsection{Bound state indices}\label{bsi}
Let us consider the equations \eqref{Deqs} for small values of $n$. For $n=2$,
there is a single equation,
\be
\label{2center}
\frac{\gamma_{12}}{r_{12}}=c_{1}.
\ee  
We deduce that the two-center solution only exists as a physical
solution if $\gamma_{12}\,c_{1}>0$. This depends on the moduli $t$. If
$t$ approaches a value where $c_1$ vanishes, $r_{12}$ diverges and the
solution disappears as a solution to the supergravity equations of
motion. In particular for the attractor point, the two-center solution
never exists.

The quantum states of the two-center solution 
correspond to the product of the internal degeneracies of the centers times the states of a spin $(|\gamma_{12}|-1)/2$ multiplet, which arises due to the
electric-magnetic fields sourced by the charges of the two-centers \cite{Denef:2002ru}. We express it here as the product
\be
\begin{split}
\Omega_{2}(\gamma_1+\gamma_2;t)&= g_C(\{\gamma_1,\gamma_2\},\{c_1,c_2\})\,
\Omega(\gamma_1)\,\Omega(\gamma_2), 
\end{split}
\ee
with 
\be
\begin{split}
&g_C(\{\gamma_1,\gamma_2\},\{c_1,c_2\})=\tfrac{1}{2} \left(
  \sgn(\gamma_{12}) +\sgn(c_1)\right) \, (-1)^{\gamma_{12}-1}\,\gamma_{12},
\end{split}
\ee
and $\Omega(\gamma_j)$ are degeneracies of the individual centers. For $\sgn$, we use the definition
\be
\sgn(x)=\left\{ \begin{array}{rl} 1,\quad & x>0, \\ 0, \quad & x=0, \\ -1,\quad &x<0. \end{array}\ \right. 
\ee
The function $\frac{1}{2}\left(
  \sgn(\gamma_{12}) +\sgn(c_1)\right)$ equals 1 if the solution to
(\ref{2center}) is physical, i.e. $r_{12}>0$, and it vanishes if the sign of $r_{12}$ is unphysical,
$r_{12}<0$. The factor $(-1)^{\gamma_{12}-1}\gamma_{12}$ is the number
of states of the bound state, assuming that it exists. The case that
$c_1=0$ is a special case, we aim to avoid. At the attractor point for the total charge, the $c_j$ (\ref{ciZ}) are equal to $c^*_j$, given by \cite{Denef:2000nb, Denef:2001xn}
\be 
\label{c*j}
c^*_j=|Z(\gamma,t^*_\gamma)|\,\left<\gamma,\gamma_j\right>.
\ee
As a result, two-center solutions do not exist at the attractor point,
because substituting $c^*_1$ in (\ref{2center}) gives a negative
value for $r_{12}$ which is unphysical.

For $n=3$ distinct charges, \eqref{Deqs} gives two independent equations,
\be
\label{n=3Deqs}
\begin{split}
\frac{\gamma_{12}}{r_{12}}+\frac{\gamma_{13}}{r_{13}}=c_1,\\
\frac{\gamma_{21}}{r_{12}}+\frac{\gamma_{23}}{r_{23}}=c_2.\\
\end{split}
\ee  
An intriguing aspect of these equations is that for appropriate values of $\gamma_{ij}$, they can be satisfied with
positive $r_{ij}$ for all pairs $i\neq j$, for $t$ at the attractor
point. Then, there is a
one-parameter family of solutions \cite{Denef:2007vg}, with
\be   
\label{rijeps}
r_{ij}=\varepsilon\,\gamma_{ij}+O(\varepsilon^2),\qquad \varepsilon\to 0,
\ee 
for $ij$ equal to $12$, $23$ and $31$.  

We can in fact do better, and give an all order solution in
$\varepsilon$. The parameter $\varepsilon$ together with three angular
variables form the 4-dimensional solution space to
(\ref{n=3Deqs}). We set $\gamma_{12}=a$, $\gamma_{23}=b$ and
$\gamma_{31}=c$ in the following, and assume $a,b,c>0$.\footnote{We apologize for the multiple use of $a$, $b$ and $c$.} 
, we can then verify that the following 1-parameter family of
distances $r_{ij}$ satisfy Denef's equations \refb{Deqs}, 
\begin{align}
\frac{1}{r_{12}} &= \frac{1}{a \varepsilon} - \rho_a \, , \quad\,
\frac{1}{r_{23}} = \frac{1}{b \varepsilon} - \rho_b \, , \quad \, \frac{1}{r_{31}} = \frac{1}{c \varepsilon} - \rho_c \, ,
\end{align}
where $\rho_a, \rho_b, \rho_c$ satisfy $ \left( \rho_c c -
  \rho_a a  \right) = c_1 , \, \left( \rho_a a - \rho_b b\right) =
c_2 , \,  \left( \rho_b b - \rho_c c \right) =
c_3$. The range of
$\varepsilon$ is determined by triangle inequalities and positivity of
$r_{12}, r_{23}, r_{31}$. The equations for $\rho_{a,b,c}$ allow for a shift, 
which modifies the range of $\varepsilon$ accordingly. 

We discuss this now in detail for the attractor poin, where we can use
(\ref{c*j}) such that $|Z(\gamma,t^*_\gamma)|=M$ and $\rho_a=\rho_b=\rho_c=M$. We then have,
 \be
\frac{1}{r_{12}} = \frac{1}{a \varepsilon} -M \, , \quad \frac{1}{r_{23}} = \frac{1}{b \varepsilon} -M \, , \quad \frac{1}{r_{31}} = \frac{1}{c \varepsilon} -M \, .   \label{scalesol}
\ee
The free parameter $\varepsilon$ is bounded from below by 0. In
$\varepsilon \ll 1/M$ regime, this solution reduces to Eq. (\ref{rijeps});
thus existence of the scaling solution requires $a,b,c$ to obey
triangle inequalities \cite{Denef:2007vg}. As $\varepsilon$ is
increased, the shape of the triangle changes, and we need to determine
the correct maximum of $\varepsilon$ domain.
First, positivity of $r_{12},
r_{23}, r_{31}$ imposes the upper bound $\varepsilon \leq \frac{1}{M
  \max{} (a,b,c)}$. However, we also need to impose that $r_{ij}$
satisfy the triangle inequalities, which imposes an even stronger
upper bound. E.g. the condition $r_{12} + r_{23} \geq r_{31}$
requires 
\be
\label{poleps}
(a+b-c) - 2 a b \varepsilon M +  abc \varepsilon^2 M^2 \geq
0.
\ee
If $(c-a)(c-b)<0$, this condition is always satisfied since
$a+b-c>0$ and the lhs does not have real roots in this case. Moreover if $(c-a)(c-b)\geq 0$, the condition is saturated for $\varepsilon_c^\pm = \frac{1}{M c}
\left[ 1 \pm \sqrt{\frac{(c-a)(c-b)}{ab}}\right]$ and violated for
$\varepsilon_c^- < \varepsilon < \varepsilon_c^+$. Both roots are
non-negative provided $(a,b,c)$ obeys the triangle inequality $a+b
\geq c$. Noting that $\varepsilon_c^+ > \frac{1}{M c}$, we see $\varepsilon
\geq \varepsilon_c^+$ makes $r_{31}$ negative. Thus we must have
$\varepsilon \leq \varepsilon_c^-$. Using two other triangle
inequalities, we have $\varepsilon \in (0, \min{} (\varepsilon_a^-,
\varepsilon_b^-, \varepsilon_c^-)_{\mathbb{R}}]$ where  
$\varepsilon_a^- = \frac{1}{M a} \left[ 1 - \sqrt{\frac{(a-b)(a-c)}{bc}} \right], \, 
\varepsilon_b^- = \frac{1}{M b} \left[ 1 -
  \sqrt{\frac{(b-a)(b-c)}{ac}} \right]$, and $\min{} (\varepsilon_a^-,
\varepsilon_b^-, \varepsilon_c^-)_{\mathbb{R}}$ means the minimum
among the $\varepsilon^-_{a,b,c}\in \mathbb{R}$. One of
$\varepsilon^-_{a,b,c}$ may be complex. At the maximal value of
$\varepsilon$ the configuration is collinear. With the three additional
angular variables, one can aligned the solution along the $z$-axis,
thus giving a fixed point for rotations around this axis.

For three centers, the sum over fixed points reads
\begin{align} 
\nn
&g_C \left( \{ \gamma_j \}; \{ c_j \}; y\right) 
= \frac{(-1)^{a+b+c}}{(y - y^{-1})^{2}}  
\Bigg[ 
F(123) y^{a+b-c}
+ F(321) y^{-a-b+c} 
+ F(213) y^{-a+b-c} \\
&+ F(312) y^{ a -b+c}
+ F(132) y^{a-b -c}
+ F(231) y^{-a+b +c}
\Bigg],
\end{align}
where
\be 
F(ijk):=F(ijk;\{c_j\})=\left\{ \begin{array}{rl} s(p), & \qquad \exists\, {\rm a\,\, fixed\,\, point\,\,} p\in
    \CM_n\,\,{\rm with\,\,}  z_i<z_j<z_k, \\
    0, & \qquad  \nexists \, {\rm a\,\, fixed\,\, point\,\,} p\in \CM_n{\rm \,\,with\,\,}z_i<z_j<z_k.\end{array}\right.
\ee
Since $a,b,c\in \mathbb{Z}$, the signs $(-1)^{a\pm b \pm c}$ equal
$(-1)^{a+b+c}$, and we can factor this out from the sum over
permutations. The dependence on the rhs on $\{c_j\}$ is through $\CM_n$. 

Since $F(123;\{c_j\})=F(321;\{c_j\})$, we can shorten $g_C$ to
\be
\label{defgCy}
\begin{split}
&g_C \left( \{ \gamma_j\}; \{ c_j \}; y\right) 
= \frac{(-1)^{a+b+c}}{(y - y^{-1})^{2}} 
\Bigg[ 
F(123;\{c_j\}) \left( y^{a+b-c} 
+ y^{-a-b+c} \right) \\
&+ F(213;\{c_j\}) \left( y^{-a+b-c} 
+  y^{a-b+c} \right)
+ F(132;\{c_j\}) \left( y^{a-b -c}
+  y^{-a+b +c} \right) 
\Bigg].
\end{split}
\ee
This expression does not have a smooth $y \rightarrow 1$ limit if $\CM_n$
is non-compact and contains a scaling region. 
In that case, only one of $F(ijk;\{c_j\})$'s is non-vanishing. Turning on a fugacity is
indeed known to be subtle for non-compact phase spaces \cite{Lee:2016dbm}. 

In the present context, the minimal
modification hypothesis is put forward in \cite{Manschot:2011xc} to correct this. It
adds a term with minimal angular momentum corresponding to the 
fixed point with coincident centers. The effect of the minimal
modification hypothesis for three distinct charges is that the refined index $g_C
\left( \{ \gamma_j\}; \{ c_j \}\right)$  is completed to 
\be
\bar g_C \left( \{ \gamma_j\}; \{ c_j \},y\right)=g_C
\left( \{ \gamma_j\}; \{ c_j \},y \right)+H(\{\gamma_j\},y),
\ee
with
\be
\label{3centerH}
H(\{\gamma_j\},y)=\left\{ \begin{array}{ll} -\frac{2}{(y-y^{-1})^2},&\quad  {\rm if}~  a+b+c\in 2\mathbb{Z}, \\  \frac{y+y^{-1}}{(y-y^{-1})^2},&\quad  {\rm if}~ a+b+c\in 2\mathbb{Z}+1, \end{array}\right.
\ee
$\bar g_C$ has a well-defined $y\to 1$ limit, which reads,
\be
\label{gCFs}
\begin{split}
\bar g_C \left( \{ \gamma_j\}; \{ c_j \}\right)&=\lim_{y\to 1} \bar g_C \left( \{
  \gamma_j\}; \{ c_j \}; y\right)\\
&=(-1)^{a+b-c} \left[ F(123)\,s(a,b,c)+F(213)\,s(a,c,b)+F(132)\,s(c,b,a)\right],
\end{split}
\ee  
with 
\be    
s(a,b,c) = \left\{\begin{array}{rl} \frac{(a+b-c)^2}{4}, & \qquad
    \text{if}~a+b+c\in 2\mathbb{Z},\\ \frac{(a+b-c)^2-1}{4}, &  \qquad
    \text{if}~a+b+c\in 2\mathbb{Z}+1. \end{array} \right. \,  
\ee
We note that for degenerate cases, where one or more
among $a,b$ and $c$ vanish, $g_C$ can be a multiple of $1/2$ or $1/4$
rather than in $\{-1,0,1\}$. 

Using the regularization (\ref{regindex}), we obtaine for the numerical
version (\ref{3centerH}) of $H$,
\be
H(\{\gamma_j\})=\left\{ \begin{array}{rl}  0,&\quad  {\rm if}~  a+b+c\in 2\mathbb{Z}, \\  \frac{1}{4},&\quad  {\rm if}~ a+b+c\in 2\mathbb{Z}+1, \end{array}\right.
\ee 
which in turn gives the numerical counterpart for $\Omega_T$,
\be
\label{3centerOmT}
\Omega_T(\gamma)=\Omega_S(\gamma)+\left\{ \begin{array}{rl}  0,&\quad
    {\rm if}~  a+b+c\in 2\mathbb{Z}, \\  \frac{1}{4}\prod_{j=1}^3 \Omega_S(\gamma_j),&\quad  {\rm
      if}~ a+b+c\in 2\mathbb{Z}+1. \end{array}\right. 
\ee
We obtain similarly using (\ref{regindex}) for the numerical version of $g_C$,
\be
\label{gCFnumerical}
\begin{split}
&g_C \left( \{ \gamma_j\}; \{ c_j \}\right)=\frac{(-1)^{a+b+c}}{4}\times \left[ F(123; \{ c_j \})\,(a+b-c)^2\right.\\
&\quad \left.+F(213, \{ c_j
  \})\,(a-b+c)^2+F(132, \{ c_j \})\,(a-b-c)^2\right].
\end{split} 
\ee  
Thus the numerical $g_C$ essentially corresponds to the $y\to 1$ limit
of the equivariant volume of the solution space $\CM_3$ \cite{Manschot:2011xc}.

The term $F(123, \{c_j\})$ is determined in \cite[Equation
  (2.57)]{Manschot:2013sya}. For our purposes we rewrite this in terms
  of $\sgn$ rather than the step function. This reads, 
\be
\label{F*123}
F(123,\{c_j\})=F_1(a,b,c,\{c_j\})+F_2(a,b,c),
\ee
with 
\be
\begin{split}
F_1(a,b,c,\{c_j\})&=\frac{1}{4}(\sgn(a)+\sgn(c_1))\,(\sgn(b)-\sgn(c_3)),\\
F_2(a,b,c)&= \frac{1}{4} \left( \sgn(a) + \sgn(b) \right) \left( \sgn(a + b - c) -\sgn(a + b)  \right).
\label{2f} 
\end{split}   
\ee 
At special charge configurations, where one or more arguments of the
signs vanish, this may differ from \cite[Equation
  (2.57)]{Manschot:2013sya}. In such cases, the $F_j$ can be a fraction rather
than an integer. To deal properly with such cases, we include
additional terms below in Eqs (\ref{defgC}) and (\ref{deff}).

Our interest in this paper is the index at the attractor point, thus
\be
g_C \left( \{ \gamma_j\}; \{ c_j^* \}\right),
\ee 
where $c_j^*$ (\ref{c*j}) is $c_j$ evaluated at the attractor point
$t^*_\gamma$.  Then $F_1$ reads 
\be
\begin{split}
F_1(a,b,c) &= \frac{1}{4} \left( \sgn(a) + \sgn(c - a) \right) \left( \sgn(b) + \sgn(c-b) \right). 
\end{split}
\ee 

Assuming non-vanishing arguments of the sign
functions, we can simplify the sum $F^*(123)=F_1+F_2$ using the
identity \cite[Eq. (A.1)]{Alexandrov:2018iao}\footnote{We thank Sergey
Alexandrov for stressing the simplifying power of this identity.} 
\be
(\sgn(x_1)+\sgn(x_2))\,\sgn(x_1+x_2)-\sgn(x_1)\,\sgn(x_2)=1,\qquad
(x_1,x_2)\neq (0,0), \label{signid}
\ee
to $F^*(123)=F_1(a,b,c)+F_2(a,b,c)$ \cite[Eq. (4.10)]{Alexandrov:2018iao},
\be
\label{F0123}
F^*(123)=\frac{1}{4}(1+\sgn(a-c)\,\sgn(b-c)+\sgn(b-c)\,\sgn(c-a-b)+\sgn(c-a-b)\,\sgn(a-c)\,).
\ee
We stress that this expression may differ from (\ref{F*123}) if the
arguments of some products of $\sgn$'s vanish. For example for
$a=0,b=c=1$, $F(123)=0$, while $F^*(123)$ equals $\frac{1}{4}$.  

For the other permutations, we also define
\be
\begin{split}
F_3(a,b,c) &= \frac{1}{4}\left( \sgn(a) + \sgn(b - a) \right) \left( \sgn(c) + \sgn(b-c) \right), \\
F_4(a,b,c) & =  \frac{1}{4}\left( \sgn(a) + \sgn(c) \right) \left( \sgn(a + c - b) -\sgn(a + c)  \right), \\
F_5(a,b,c) &= \frac{1}{4}\left( \sgn(c) + \sgn(a - c) \right) \left( \sgn(b) + \sgn(a-b) \right), \\
F_6(a,b,c) & = \frac{1}{4} \left( \sgn(c) + \sgn(b) \right) \left( \sgn(c + b - a) -\sgn(c + b)  \right), \label{6f}
\end{split}
\ee
and 
\be
\nn
F^*(213)=\frac{1}{4}(1+\sgn(a-b)\,\sgn(c-b)+\sgn(c-b)\,\sgn(b-a-c)+\sgn(b-a-c)\,\sgn(a-b)\,),
\ee
\be
\nn
F^*(132)=\frac{1}{4}(1+\sgn(c-a)\,\sgn(b-a)+\sgn(b-a)\,\sgn(a-b-c)+\sgn(a-b-c)\,\sgn(c-a)\,).
\ee

Having defined the $F^*(ijk)$, we turn to
$g_C(\{\gamma_j\};\{c_j^*\})$ and take care of the special cases where
both arguments of a products of $\sgn$'s vanish. Let us first
determine for which values of $a$, $b$ and $c$, 
$g_C(\{\gamma_j\};\{c_j^*\})$ is affected. When the last two
products of $\sgn$'s of $F^*(123)$ (\ref{F0123}) vanish, the
angular momentum factor $(a + b - c)^2/4$ also vanishes. Thus replacing
these products of $\sgn$'s with a non-vanishing value will not affect
$g_C$. For the remaining product, $\sgn(a-c)\,\sgn(b-c)$'s, the
arguments vanish in the equilateral case $a=b=c$ for which the angular
momentum factor is $a^2/4$. This is the same as
for the other permutations, $F^*(213)$ and $F^*(132)$, such that we
can take all three into account by adding a single additional term with yet
undetermined coefficient $A$. We obtain thus for $g_C$ at the attractor point,
\be
\label{defgC}
\begin{split}
g_C(\{\gamma_j\};\{c_j^*\})&=\frac{(-1)^{a+b+c}}{4}\left[F^*(123)\,(a+b-c)^2\,+F^*(213)\,(a-b+c)^2\,\right.\\
&\quad \left.+ F^*(132)\,(-a+b+c)^2\,+\frac{1}{4} A\,\delta_{a,c}\,\delta_{b,c}\,a^2 \right].
\end{split}
\ee

At this point, we can make a
``guess'' of the preferred ``physical'' value for $A$. From the gravity
perspective, the equilateral configurations with $a=b=c$ are perfectly
fine multi-center solutions, such that it is most natural that these 
contribute with $1\times (-1)^a\,a^2/4$ to $g_C$. For \eqref{defgC},
we have instead $(3+A)\times (-1)^a\,a^2/16$, thus suggesting
$A=1$. We will demonstrate in Section \ref{modcompPsi} that modularity
of the $q$-series leads to exactly the same value.
 
Besides the bound state index $g_C$, we are also interested to
enumerate the number of charge configurations giving rise to
scaling black holes.
Up to vanishing arguments, the sum $F_{\rm total}(a,b,c)=\sum_{j=1}^6
F_j(a,b,c)$ can be further simplified to
\be
\begin{split}
F_{\rm total}(a,b,c) &= \frac{1}{4} \Big[ 1 + \sgn{} (a+b-c) \sgn{} (a+c-b) \\
&+ \sgn{} (a+c-b) \sgn{} (b+c-a) +  \sgn{} (b+c-a) \sgn{} (a+b-c) 
 \Big] \, . 
\end{split}\label{F0fin}
\ee
This expression has the advantage that for $a+b+c$ odd, the arguments
of the $\sgn$'s never vanish. With inclusion of additional terms to deal with
vanishing arguments, we define number $f_C$,
\be
\label{deff}
\begin{split}
f_C(\{\gamma_j\},\{c_j^*\})&= F_{\rm total}(a,b,c)\,(-1)^{a+b+c}\\
&+ \tfrac{1}{4}A_1\,\delta_{a,0}\,\delta_{b,c} +\tfrac{1}{4}A_2\,\delta_{c,0}\,\delta_{a,b}+\tfrac{1}{4}A_3\,\delta_{b,0}\,\delta_{a,c}, 
\end{split}
\ee 
where the constants $A_j$ are yet to be determined. We will
determine these from the modular completion. To our surprise, these
numbers are typically {\it irrational} for $a+b+c\in 2\mathbb{Z}$,
while they do not contribute for $a+b+c\in 2\mathbb{Z}+1$. We find the
irrationality quite
peculiar. On the other hand, $f_C$ is not a BPS index, such that it is
not really violating any physical principles.

\section{Review of D4-brane black holes}  \label{s2}

We review in this section aspects of partition functions of D4-brane
black holes. We start in subsection \ref{sec:Mtheory} by reviewing
the uplift of D4-brane black holes to M-theory.
In Subsection \ref{SugraPart}, we discuss the ``supergravity'' partition
function, which enumerates D4-brane BPS indices $\Omega(\gamma;t)$ for fixed K{\"a}hler
modulus $t$. In Subsection \ref{AttrPart}, we discuss the ``attractor'' partition
function, which is a generating function of BPS indices $\Omega(\gamma;t_\gamma^*)$ evaluated at
the attractor point $t_\gamma^*$ of the corresponding charge $\gamma$.
 
\subsection{Supergravity partition function}  
\label{SugraPart}  
From the perspective of asymptotically flat $\mathbb{R}^4$, the most natural BPS partition
function is enumerates the BPS indices for a fixed, asymptotic value of the (K\"ahler)
moduli $t$, and in the mixed ensemble where the magnetic charge $P$ is
fixed \cite{Ooguri:2004zv, Gaiotto:2006wm, deBoer:2006vg, Denef:2007vg} and the electric charge $Q$ is varied. 
The electric charge takes value in the lattice $\Lambda \simeq \mathbb{Z}^{b_2}$ with bi-linear form $D_{ab}=d_{abc}P^c$. 
For a positive magnetic charge $P^a$,
$d_{abc}P^c$ provides a non-degenerate quadratic form with signature $(1,b_2-1)$ for the lattice $\Lambda$ of
magnetic charges. 
The electric charges $Q_a$ take values in dual
lattice $\Lambda^*$, with quadratic form
$D^{ab}=(d_{abc}P^c)^{-1}$. We abbreviate the pairing between an element $Q\in
\Lambda^*$ and $k \in \Lambda$ as
\be
\sum_{a=1}^{b_2} Q_aP^a=Q.P
\ee
and extend this by linearity in each argument to $\Lambda^*\otimes \mathbb{R}$ and  $\Lambda\otimes \mathbb{R}$. For later use, we also introduce the notation $\mu\in \Lambda^*/\Lambda$, 
\be
\label{Lmustar}
\Lambda_\mu^*=\left\{ Q\in \Lambda + \mu + P/2\right\}.
\ee 
We stress that the elements of $\Lambda_\mu^*$ do not necessarily lie in $\Lambda^*$ due to the shift by $P/2$. 
Using this notation, the partition function for D4-D2-D0 black holes reads schematically
\begin{align}
\label{ZSG}
\mathcal{Z}_{SG}(\tau, C, t) 
&= \sum_{Q_0, Q_a} \bar \Omega(\gamma,t)\, (-1)^{P.Q}\,e^{-2\pi \tau_2 |Z(\gamma,t)| + 2\pi i \tau_1 \left( Q_0 - Q.B + B^2/2 \right) + 2\pi i C. (Q-B/2)},
\end{align}
where $\bar \Omega(\gamma,t)$ is the rational index (\ref{ratIndex}), $\tau_1\in \mathbb{R}$ is the RR 1-form $C_1$, and $\tau_2=e^{-\phi}\in \mathbb{R}_+$ with $\phi$ being the dilaton, $C$ the RR 3-form and $B$ the $B$-field.

Here $Z(\gamma,t)$ is the central charge of the $\CN=2$ algebra. Ignoring the
non-perturbative terms in the strict large volume limit, $Z$ reads 
\begin{align}
Z(\gamma,t) &= \frac{1}{2} P. (J^2 - B^2) + Q.B - Q_0  + i(Q-BP).J \, .
\end{align}
The BPS mass becomes in this limit,
\begin{align}
|Z(\gamma,t)| &= \frac{1}{2} P. (J^2 - B^2) + Q.B - Q_0 + \frac{\left( (Q-BP).J\right)^2}{P.J^2} \, ,
\end{align} 
up to inverse powers of $J$. By assumption $J^2>0$, hence $J^a$ lies
in the positive cone of $\Lambda$. Thus $\frac{k.J\,J}{J^2}$ is the
component of the vector $k$ along $J$. 

In the large volume limit, supergravity has an $Sp(2b_2 + 2, \mathbb{Z})$ symmetry, generated by matrices 
\begin{align}
\mathbb{K}(k) &=
\begin{pmatrix}
1 & 0 & 0 & 0 \\
k^a & \mathbb{I}_{b_2} & 0 & 0 \\
\frac{1}{2} d_{abc} k^b k^c & d_{abc} k^c & \mathbb{I}_{b_2} & 0 \\
\frac{1}{6} d_{abc} k^a k^b k^c & \frac{1}{2} d_{abc} k^b k^c & k^a & 1
\end{pmatrix}
, \, \,  k \in \mathbb{Z}^{b_2} \, , \label{spectrans}
\end{align}
and acts linearly on $\gamma = (P^0, P^a, Q_a, Q_0)$. For $P^0=0$, these
transformations preserve the magnetic charge, and act on remaining charges and moduli
as follows: 
\begin{align}
\nn
Q_a &\rightarrow Q_a + d_{abc} k^b P^c \, , \\
Q_0 &\rightarrow Q_0 + k^a Q_a + \frac{1}{2} d_{abc} k^a k^b P^c \, , \label{spectransimple} \\
\nn t^a &\rightarrow t^a + k^a \, . 
\end{align}
We introduce the abbreviations:
\be
\begin{split}
\hat{Q}_{\bar{0}} &:= - Q_0 + \frac{1}{2} Q^2 \, ,\\
 \, \hat{Q} &:= Q-B \, ,\\
\, \hat{Q}^2_- &:= \hat{Q}^2 - \hat{Q}^2_+ \, ,\\
\, (Q-B)^2_+ &:= \frac{\left( (Q-BJ).P\right)^2}{PJ^2} \, .
\end{split}
\ee
Note that the combinations $\hat{Q}, \hat{Q}_{\bar{0}}$ and the
conjugacy class $\mu$ of electric charge vector are invariant under
the transformations (\ref{spectransimple}). Invariance of the conjugacy class is seen by noting
that spectral flow shifts $Q$ by integer lattice vectors, when mapped
to magnetic lattice $\Lambda$ and such shifts do not change the conjugacy
class.

The holomorphic and anti-holomorphic dependence on $\tau$ can be made more manifest, if we rewrite $\mathcal{Z}_{SG}$ as
\begin{align}
\label{ZSGdef}
\mathcal{Z}_{SG}(\tau, C, t) 
&= e^{-\pi \tau_2 J^2} \sum_{Q_0, Q} \bar \Omega (P, Q, Q_0;t)\, (-1)^{P.Q} \bar{q}^{\hat{Q}_{\bar{0}} - \hat{Q}^2_-/2} q^{\hat{Q}^2_+/2}  e^{2i \pi C. (Q-B/2)},
\end{align}
where $q=e^{2\pi i \tau}$. The transformations \refb{spectransimple} act on $\mathcal{Z}_{SG}$ as
\begin{align}
\mathcal{Z}_{SG}(\tau, C, t) &\rightarrow (-1)^{P.k} e^{\pi i C.k} \mathcal{Z}_{SG}(\tau, C, t) \, , 
\end{align}
under the action of $\mathbb{K}(k)$.

We can map the system of D4-D2-D0 branes to IIB string theory, 
by a T-duality along time circle. The RR 1-form $C_1$ is mapped to the RR 0-form $C_0$, while the D4, D2 and D0-branes respectively become D3, D1 branes and D(-1)-instantons. Moreover, the action of the IIB strong-weak duality on the instantonic branes carries over to the spectrum of D4-D2-D0 branes in IIA string theory. The duality acts as follows. The modular parameter is $\tau= C_1 + i e^{-\phi}$, with $\phi$ being the dilaton and $C_1$ the Ramond-Ramond 1-form flux along $S^1_t$. For an element $\begin{pmatrix}
a & b \\
c & d
\end{pmatrix} \in SL(2, \mathbb{Z})$.
 This duality acts as
\be
\begin{split}
& \tau \rightarrow \frac{a \tau + b}{c \tau +d} , \\
& \begin{pmatrix}
C \\
B
\end{pmatrix}
\rightarrow
\begin{pmatrix}
a & b \\
c & d
\end{pmatrix} 
\begin{pmatrix} 
C \\
B
\end{pmatrix}
, \\
& 
J \rightarrow |c \tau + d| J \, ,
\end{split} 
\ee 
In a series of papers \cite{Denef:2007vg, Gaiotto:2006wm, deBoer:2006vg, Alexandrov:2016tnf, Manschot:2009ia, Manschot:2010xp, Alexandrov:2018lgp}, the supergravity partition function (\ref{ZSG}) has been analyzed.
We summarize the main properties:
\begin{itemize}
\item  Quasi-periodicity in the two-form fields $B$ and $C$:\\
For $k\in \Lambda$, we have
\be
\begin{split}
&\mathcal{Z}_{SG}(\tau, C,t+k)=(-1)^{P.k}\,e^{\pi i C.k}\,\mathcal{Z}_{SG} (\tau,
C,t),\\
&\mathcal{Z}_{SG}(\tau, C+k,t)=(-1)^{P.k}\,e^{-\pi i B.k}\,\mathcal{Z}_{SG} (\tau,
C,t).
\end{split}
\ee 
These can be seen as large gauge transformations. 
\item $SL(2,\mathbb{Z})$ S-duality:
\be
\label{ZSGtrafos}
\begin{split}
S &:\qquad  \, \mathcal{Z}_{SG}(-1/\tau, -B, C+|\tau| J) = \tau^{1/2} \bar{\tau}^{-3/2} \varepsilon(S) \,\mathcal{Z}_{SG}(\tau, C,t) \, , \\
T &: \qquad \, \mathcal{Z}_{SG}(\tau+1, C+B, t) = \varepsilon(T)\, \mathcal{Z}_{SG}(\tau, C,t) \, ,
\end{split} 
\ee
where $\varepsilon(S)= \varepsilon(T)^{-3}, \varepsilon(T) = e^{- i \pi
  \frac{c_2(X).P}{12}}$ are phases. The factor $\bar{\tau}^{-3/2}$ can be understood from the non-compact bosons in the CFT due to the center of mass in $\mathbb{R}^3$. 
\item The partition function involves an intricate dependence on the
  moduli $t$ through wall-crossing. The proper modular invariant
  partition function differs from (\ref{ZSGdef}) by additional
  non-holomorphic, subleading terms which are non-holomorphic in $\tau$.
\end{itemize}

\subsection{Attractor partition function}  
\label{AttrPart}
The supergravity partition function is a rather complicated
function. It has become clear that an alternative partition function is a more amenable object to study \cite{Alexandrov:2016tnf}. This is the attractor partititon function
$\mathcal{Z}_{P}^\lambda$, which is the generating function of indices $\bar \Omega(\gamma;t^*_\gamma)$, where
each index is evaluated at its (large volume) attractor point
$t^\lambda_\gamma$. The indices $\bar \Omega(\gamma;t^\lambda_\gamma)$ are
also referred to as MSW invariants \cite{Alexandrov:2016tnf}. For
irreducible magnetic charge $P$ and magnetic charges which can be 
written as a sum of at most 2 magnetic charges, these indices are
conjectured to coincide
with the CFT indices. However as discussed in Section \ref{sec:Mtheory}, our findings in this paper
suggest that there is a difference for generic magnetic charges due to
scaling black holes.
 
This partition is obtained by replacing $\Omega(\gamma,t)$ in the
definition of $\mathcal{Z}_{SG}$ by its attractor value
$\Omega(\gamma, t^\lambda_\gamma)$. 
For a D4-D2-D0 black
hole with charge $\gamma=(0,P^a,Q_a,Q_0)$, we have for the ``large volume'' attractor
value $(t_\gamma^*)^a$, 
\be
\label{LVattractor}
(t^\lambda_\gamma)^a= D^{ab}Q_b+i \lambda \,P^a,
\ee
with $\lambda$ sufficiently large, such that subleading terms in a $\lambda$ expansion can be ignored.
Eq. (\ref{LVattractor}) is equivalent to
\be
\label{LVAttractor}
B^a_\gamma=D^{ab} Q_b,\qquad (J^\lambda_\gamma)^a= \lambda\, P^a.
\ee
The precise proportionality constant $\lambda$ between $J^a$ and $P^a$ will not
be important for us, since we will restrict to the large volume limit,
$|J|\gg 1$. On the other hand, we do not take the limit $\lambda\to
\infty$, since physical quantities such as the BPS mass diverge in
this limit. 

The attractor or MSW partition function then reads, 
\begin{align}
\mathcal{Z}^\lambda_{P}(\tau, C, t)  
&= \sum_{Q_0, Q_a} \bar \Omega(\gamma,t_\gamma^\lambda)\, (-1)^{P.Q}\, e^{-2\pi \tau_2 |Z(\gamma,t)| + 2\pi i \tau_1 \left( Q_0 - Q.B + B^2/2 \right) + 2\pi i C. (Q-B/2)}.
\end{align}
Note that although the degeneracy is evaluated at attractor point
$t^\lambda_\gamma$ , the mass in the exponent is the ADM mass evaluated for
moduli at infinity being $t$. The other moduli dependence in the
exponent is also similar to that of $\mathcal{Z}_{SG}(\tau, C,t)$.  

Since black holes states contributing to the attractor index exist everywhere in moduli space,
their quantum degeneracies should be moduli independent. 
Moreover, one can show that $\Omega(\gamma;t^\lambda_\gamma)$ are
invariant under spectral flow transformations
(\ref{spectransimple}). This entails that
$\Omega(\gamma;t^\lambda_\gamma)$ depends on the charges only through the invariant
combination $\hat{Q}_{\bar{0}}$ and the conjugacy class $\mu$. 
These imply the sum
\begin{align}
h_{P, \mu}(\tau) &:= \sum_{Q_0} \overline{\Omega}(\gamma;t^\lambda_\gamma)\, q^{\hat{Q}_{\bar{0}}} \, , 
\end{align}
for fixed $Q$ is also invariant under spectral flow transformations and apart
from the magnetic charge, depends solely on the conjugacy class
$\mu$. It has been understood from wall-crossing and the perspective
of D-instantons, that $h_{P,\mu}$ receives 
additional non-holomorphic contributions $h_{P,\mu}$
\cite{Alexandrov:2016tnf, Alexandrov:2018lgp}. The resultant is the 
non-holomorphic function $\widehat h_{P,\mu}(\tau,\bar \tau)$. 

Hence one is led to the following theta function decomposition of the attractor partition function,
\begin{align}
\mathcal{Z}^\lambda_P(\tau, C, t) &= e^{-\pi \tau_2 J^2} \sum_{\mu \in
  \Lambda^*/\Lambda} \overline{h_{P,\mu} (\tau)}\, \Theta_\mu (\tau,
\bar \tau, C, B) \, , \label{Zdecomp} \\ 
\text{where} \, \,
\Theta_\mu (\tau, \bar \tau, C, B) &= \sum_{Q \in \Lambda^*_\mu} (-1)^{P.Q} q^{\hat{Q}^2_+/2} \bar{q}^{- \hat{Q}^2_-/2} e^{2\pi i C. (Q - B/2)} \, , \label{thetadefn}
\end{align}
with the set $\Lambda^*_\mu$ is given in (\ref{Lmustar}). We define analogously the completion
$\widehat{\mathcal{Z}}^\lambda_P$ by replacing $h_{P,\mu}$ by
$\widehat h_{P,\mu}$ in \eqref{Zdecomp}.

Various factors in the right hand side of \refb{Zdecomp} also have definite modular transformation properties. The prefactor $e^{-\pi \tau_2 J^2}$ is modular invariant.
Moreover, the Siegel-Narain theta function $\Theta_\mu(\tau, C,B)$ transforms as
\be
\begin{split}
S &: \, \Theta_\mu (-1/\tau,-1/\bar \tau, -B, C) = \frac{1}{\sqrt{|\Lambda^*/\Lambda|}}  (-i \tau)^{b_2^+/2} (i \bar{\tau})^{b_2^-/2} e^{- i \pi P^2/2} \\
& \qquad \qquad \qquad \qquad \times\sum_{\nu \in \Lambda^*/\Lambda} e^{-2\pi i \mu.\nu} \Theta_\nu (\tau, \bar \tau,C, B) \, , \\
T &: \, \Theta_\mu (\tau+1,\bar \tau+1, C+B, B) = e^{i \pi (\mu + P/2)^2} \Theta_\mu (\tau,\bar \tau, B, C) \, .
\end{split}
\ee
This implies that the appropriate completion $\widehat h_{P,\mu}$ transforms as a vector valued modular form,
\be
\label{STh}
\begin{split}
S &:  \, \widehat h_{P,\mu} (-1/\tau,-1/\bar \tau) = -
\frac{1}{\sqrt{|\Lambda^*/\Lambda|}} (-i \tau)^{-b_2/2-1}
\varepsilon(S)^* e^{- i \pi P^2/2} \sum_{\delta \in \Lambda^*/\Lambda}
e^{- 2\pi i \delta.\mu} \widehat h_{P, \delta} (\tau,\bar \tau) \, , \\
T &: \, \widehat h_{P,\mu} (\tau +1,\bar \tau+1) = \varepsilon(T)^* e^{i \pi (\mu
  + P/2)^2}\, \widehat h_{P,\mu}(\tau, \bar \tau), 
\end{split} 
\ee 
where $\varepsilon(S)$ and $\varepsilon(T)$ are as below
(\ref{ZSGtrafos}).

\section{Partition functions for scaling black holes} \label{s3gen}
We consider multi-core black holes, where each core carries a positive
D4-brane charge and vanishing D6-brane charge. The notion of ``core''
is as introduced in the Introduction; it is a
bound state with spatial magnitude of the order $\ell_5^3/R^2$, which
decreases as $\lambda^{-3}$ in the large $\lambda$ limit. We will
determine the partition function at the large volume
attractor point (\ref{LVAttractor}) for such black holes. For this value of the moduli and
assuming that $\gamma$ and $\gamma_j$ both carry D4-brane charge, the
$c_j$ (\ref{ciZ}) are to leading order equal to $c^{\lambda}_{j}$,
\be
\label{LVc*j}
c^{\lambda}_{j}=2\lambda\left<\gamma,\gamma_j\right>.
\ee
Since the factor $2\lambda$
on the rhs is positive, the large volume attractor point is similar in
nature to the $c_j^*$ (\ref{c*j}) for the proper attractor point. Restricting to three center solutions with
D4-brane charge, we have 
for $a$, $b$ and $c$,
\be
\label{abcPQ}
\begin{split}
a&=P_1 Q_2-P_2Q_1,\\
b&=P_2 Q_3-P_3Q_2,\\
c&=P_3Q_1-P_1Q_3.
\end{split}
\ee
This gives for the $c_j^\lambda$, $c_1^\lambda=2\lambda\,(c-a),
c_2^\lambda=2\lambda\,(a-b)$. Since the $c^{*}_{j}$ (\ref{c*j}) and
$c^{\lambda}_{j}$ (\ref{LVc*j}) are simply related by the
substitution $|Z(\gamma;t_\gamma^*)|\to 2\lambda$, the discussion on
the solution to Denef's equations below (\ref{scalesol}) is
applicable, and in particular gives the separation of collinear solutions.

At the attractor point, many multi-center configurations do not exist
as physical solutions. However we have seen that some multi-center solutions do
exist. These solutions are distinguished from generic solutions, since
the centers can approach each other arbitrarily close. The scaling
solutions do respect the spectral flow symmetry, since the symplectic
innerproducts are invariant under (\ref{spectransimple}). 

We note that the existence scaling solutions poses constraints on the magnetic charges $P_j$. 
For example since for a scaling solution $a,b,c$ must have the same sign, we deduce that there are no
scaling solutions for $P_1=P_2=P_3=P/3$, since this gives rise to
$\gamma_{12}+\gamma_{23}+\gamma_{31}=a+b+c=0$. Therefore, there must be an asymmetry
in the magnetic charge of the centers for such scaling solutions to
exist. To see that $F$
vanishes in the case of three equal magnetic charges, note that the second factors in $F_2,F_4,F_6$ vanish
immediately as ${\rm sgn}(2x)-{\rm sgn}(x)=0$. Moreover, in $F_1$ we can replace
$c=-a-b$ and when $a>0$ the first factor imposes the constraint
$-2a-b\ge 0$ so $b<-2a$ and $b$ must be negative. Hence the second
factor in $F_1$ vanishes. The argument goes similarly for
$F_3,F_5$. Similarly one may show that such scaling black holes only
exist for $b_2>1$.

\subsection{Bound states and lattices}
\label{BoStaLat}
We discuss in this subsection various aspects of charge lattices of
bound states, and the projection of the charge lattice to a sublattice with fixed
total charge. The discussion in this subsection does not rely on the existence of scaling
black holes for these charges and lattices.
\\

\noindent {\it Lattice decomposition}\\
We apply some techniques of decompositions and gluings of integral
lattices to charge lattices of bound states. See for example \cite{conway, Gannon1989}. We will consider a black hole bound state of $n$ centers, with
non-vanishing, positive magnetic charge $P_j\in \Lambda_j,$ $j=1,2,\dots,n$, where  $\Lambda_j$
is the $b_2$-dimensional lattice associated to the $j$'th center with innerproduct $D_j$,
\be
D_{jab}=d_{abc}P^c_j,\qquad a,b,c=1,\dots,b_2.
\ee
The total magnetic charge is $P=\sum_{j=1}^n P_j$ with quadratic form
$D$ and $b_2$-dimensional lattice $\Lambda$. The signature $\Lambda_j$
and $\Lambda$ is $(1,b_2-1)$.

We use boldface notation for the lattices for boundstates. 
For $n$ centers, we introduce $(n\,b_2)$-dimensional lattices and vectors,
\be
\begin{split}  
\vec k&=(k_1,k_2,\dots ,k_n)\in {\boldsymbol \Lambda}:=
\Lambda_1\oplus \Lambda_2\oplus \dots \oplus \Lambda_n,\\
\vec x&=(x_1,x_2,\dots, x_n)\in{\boldsymbol \Lambda^*}:=
\Lambda_1^*\oplus \Lambda_2^*\oplus \dots \oplus \Lambda_n^*,\\
\vec Q&=(Q_1,Q_2,\dots, Q_n) \in {\boldsymbol
   \Lambda}^*+\vec P/2=\Lambda_1^*\oplus \Lambda_2^*\oplus \dots \oplus \Lambda_n^*+(P_1,P_2,\dots,P_n)/2.
\end{split}
\ee
We denote the quadratic form for $\bfLambda$ by $\vec D={\rm diag}(D_1,D_2,\dots,D_n)$. 

Since we typically sum over bound states with fixed total electric
charge, we aim to decompose the lattice $\bfLambda$ in a $b_2$-dimensional sublattice
$\overline \bfLambda \subset \bfLambda$ representing the total charge, and its
orthogonal complement $\underline \bfLambda$ with dimension $(n-1)b_2$
representing the relative charge distribution over the constituents.
To introduce this properly, let $\overline \bfLambda$ be the sublattice $\overline
\bfLambda\subset \bfLambda$,
\be
\overline \bfLambda=\{ \vec k=(k,k,\dots,k)\in \bfLambda\,\vert\, k\in \mathbb{Z}^{b_2} \}.
\ee
The lattice $\bfLambda$ induces a quadratic form on $\overline \bfLambda$. Namely for $\vec k\in \bfLambda$, $\vec D(\vec
k)=\sum_{j=1}^nD_j(k)=D(k)$. For $k\in \mathbb{Z}^{b_2}$, this is the quadratic form for the total
charge $P$ as desired. As a result, we have the group isomorphism $\Lambda^*/\Lambda=\overline \bfLambda^*/\overline \bfLambda$. 

Let $\overline \pi$ be the orthogonal projection,
\be
\overline \pi: \bfLambda \to \overline \bfLambda\otimes \mathbb{Q}.
\ee
For $\vec k \in \bfLambda$ and $\vec m\in \overline \bfLambda$, we have
\be
\label{vecDpi}
\vec D(\overline \pi(\vec k),\vec m)=\vec D(\vec k,\overline \pi(\vec m))=\vec D(\vec k,
\vec m)\in \mathbb{Z}.
\ee 
Therefore, $\overline \pi$ gives an injection of $\bfLambda$ to the dual lattice $\overline \bfLambda^*$,
\be
\overline \pi: \bfLambda \to \overline \bfLambda^*.
\ee
Moreover, if we extend $\overline \pi$ by linearity to $\bfLambda^*$,
a similar argument to (\ref{vecDpi}) shows that $\overline \pi: \bfLambda^* \to \overline \bfLambda^*$.
The projection $\overline \pi$ is explicitly given for $\vec
k=(k_1,k_2,\dots,k_n)\in \bfLambda$ by, 
\be
\overline \pi(\vec k)=(k,k,\dots,k)\in (\overline \bfLambda)^*\quad {\rm with}\quad k=D^{-1}\sum_{j=1}^n D_jk_j.
\ee
We define furthermore the kernel $\underline \bfLambda={\rm Ker}(\overline \pi)$,
\be
\label{DefunderLambda}
\begin{split}
&\underline \bfLambda \coloneqq \left\{ \vec k\in \bfLambda\, \left\vert\, \sum_{j=1}^n D_jk_j=0 \right.\right\}.
\end{split}
\ee
Elements in $\underline \bfLambda$ have vanishing innerproduct with
$\overline \bfLambda$, such that they are indeed each others
orthogonal complement in $\bfLambda$. 

Since $\overline \pi(\vec k)\in \overline \bfLambda$ for $\vec k\in \bfLambda$,
we have that $\vec k-\overline \pi(\vec k)\in
\underline \bfLambda$. Thus $\overline \bfLambda \oplus \underline
\bfLambda$ is in the kernel of the homomorphism $\overline h:\bfLambda\to \overline
\bfLambda ^*/\overline \bfLambda$. We call
\be
G=\bfLambda/(\overline \bfLambda \oplus \underline \bfLambda),
\ee
the glue group for the decomposition of $\bfLambda$, and the image of $G$ under $\overline h$,
$\overline h(G)\subset \overline \bfLambda^*/\overline \bfLambda$, the glue group for
$\overline \bfLambda$ \cite{conway}. The homomorphism $\overline h$ gives an
injection of $G$ to the subgroup $\overline h(G) \subset\overline \bfLambda^*/\overline \bfLambda$.
Therefore, the number of glue vectors, $N_g=|\bfLambda/(\overline
\bfLambda \oplus \underline \bfLambda)|$ is a factor in $|\overline
\bfLambda^*/\overline \bfLambda|=\det(\overline D)$. In the
special case that $\det(\overline D)$ is prime, or more generally co-prime with $\det(D_j)$ for all
$j$, $N_g=\det(\overline D)$. By the same arguments, there is a
projection $\underline \pi:\bfLambda\to \underline \bfLambda^*$, and a homomorphism $\underline h:G\to \underline \bfLambda^*/\underline \bfLambda$.
Therefore, $N_g$ is also a factor in $\det(\underline D)$. This gives
for the number of glue vectors $N_g$ in general
\be
\label{Ng}
N_g=\sqrt{\frac{\det(\overline D)\,\det(\underline D)}{\prod_{j=1}^n\det(D_j)}}.
\ee
The order of the quotient group $(\underline
\bfLambda^*/\underline \bfLambda)/\underline h(G)$ is
\be
\label{Nq}
N_q=\frac{\det(\underline
D)}{N_g}.
\ee
This will be useful for us in the following way. If we consider the
class of vectors $\vec k\in \bfLambda$ with fixed projection to $\overline
\bfLambda^*$, for example $\overline \pi(\vec k)=0$, this fixes an element of
the glue group $G$, and thus also of the images of $G$ in $\overline
\bfLambda^*/\overline \bfLambda$ and $\underline \bfLambda^*/\underline \bfLambda$. As a
result, the number of possible conjugacy classes of $\underline \pi(\vec k)\in
\underline \bfLambda^*/\underline \bfLambda$ of $\underline \pi(\vec
k)$ is $N_q$ (\ref{Nq}). This is also the case for a vector $\vec x\in
\bfLambda^*$ with fixed projection $\overline \pi(\vec x)\in \overline
\bfLambda^*$.
 
 Similarly to $\Lambda^*_\mu$ (\ref{Lmustar}), we introduce the
notation $\underline {\boldsymbol \Lambda}^*_{\bfmu}$,
\be
\label{defunderLambda} 
\begin{split}
&\underline {\boldsymbol \Lambda}^*_{\bfmu} \coloneqq\left\{
  \vec Q=(Q_1,Q_2,\dots,Q_n)\in  {\boldsymbol
    \Lambda}+(\mu_1,\mu_2,\dots,\mu_n)+\vec P/2\,\left\vert\,
  \sum_{j=1}^n Q_j=\mu+P/2 \right.\right\}, 
\end{split} 
\ee    
where the subscript $\bfmu\in \underline {\boldsymbol \Lambda}^*/\underline {\boldsymbol \Lambda}$, $\mu=\sum_{j=1}^n\mu_j\in \Lambda^*$ and $P=\sum_{j=1}^nP_j\in \Lambda$.
We define the quadratic form $\bfQ^2$ for  $\underline \bfLambda^*_\bfmu$,
\be
\label{bfQ2}
\begin{split}
\bfQ^2&=-Q^2+\sum_{j=1}^3 (Q_j)_j^2\\
&=-(\mu+P/2)^2+\sum_{j=1}^3 (Q_j)_j^2.
\end{split}
\ee
This form of the quadratic form appears naturally, when we determine
the partition function $h^s_\mu$ (\ref{defhs}) of scaling solutions in the
next subsection. On the other hand, it should match with the inverse
of the quadratic form for $\underline \bfLambda$. In the cases
considered below, we find that this is indeed the case.

Note that in components, $P$ in (\ref{defunderLambda}) has a lower index and equals $P_a=d_{abc}
P^bP^c=d_{abc} (\sum_j P_j)^b(\sum_j P_j)^c$. To understand constraint
$\sum_j Q_j=\mu+P/2$ better, we 
change variables from $Q_j\in \Lambda^*_j$ to $k_j\in \Lambda_j$ using 
\be
\label{Qjkj}
\begin{split}
&Q_{1,a}=\mu_{1,a}+D_{1ab}(k^b_1+P_2^b+P_1^b/2), \\ 
&Q_{2,a}=\mu_{2,a}+D_{2ab}(k^b_2+P_3^b+P_2^b/2), \\
&\qquad \vdots \\
&Q_{n-1,a}=\mu_{n-1,a}+D_{(n-1)ab}(k^b_n+P_n^b+P_{n-1}^b/2),\\
&Q_{n,a}=\mu_{n,a}+D_{nab}(k^b_n+P_1^b+P_n^b/2),
\end{split} 
\ee
with $k_j^b\in \mathbb{Z}^{b_2}$ for $j=1,\dots,n$. The shifts of $k_j^b$ by $P_j^b$ are included such that the required identity for the $k_j$ takes a compact form,
\be
\label{keq}
\sum_{j=1}^n D_{jab}\,k_j^b=0,\qquad k_j\in \mathbb{Z}^{b_2},
\ee
which is indeed identical to the defining condition for the lattice $\underline
\bfLambda$ in (\ref{DefunderLambda}). Solving this relation over the
integers, $k_j\in \mathbb{Z}^{b_2}$, is in general a complicated problem depending on the $D_j$. Solving say for $k_n$, we have
\be
\label{k3=k1k2}
k_n=-D_n^{-1}\sum_{j=1}^{n-1}D_jk_j.
\ee 
Thus we find that if $D_n^{-1}D_j$, $j=1,\dots n-1$ are not integral
matrices, not all $k_j\in \mathbb{Z}^{b_2}$ can correspond to bound
state charges for these conjugacy classes. In some cases, this problem
can be avoid by solving for another $k_j$ instead of $k_n$, but not in
general. We restrict to special cases in the following.
\vspace{.4cm}\\
\noindent {\it 2- and 3-center bound states for $b_2=1$}\\
For $b_2=1$, the $D_j$ are simply positive numbers. For $n=2$, the relation (\ref{keq})
becomes
\be
D_1k_1+D_2k_2=0.
\ee
The solutions with $k_{1,2}\in \mathbb{Z}$ are
\be 
\label{n2solutions}
k_1=\frac{D_2}{\gcd(D_1,D_2)}\,m,\qquad
k_2=-\frac{D_1}{\gcd(D_1,D_2)}\,m,\qquad m\in \mathbb{Z},
\ee
where $\gcd$ stands for the greatest common divisor. Substituting
(\ref{n2solutions}) in the quadratic form $\sum_{j=1,2} D_jk_j^2$, gives for the quadratic form $\underline D$ on $\underline \bfLambda$,
\be
\label{uDb21}
\underline D=\frac{D_1D_2(D_1+D_2)}{\gcd(D_1,D_2)^2}.
\ee
Thus the number of glue vectors (\ref{Ng}) and order $N_q$ (\ref{Nq})
are in this case,
\be
n=2:\qquad N_g=\frac{D_1+D_2}{\gcd(D_1,D_2)}, \qquad N_q=\frac{D_1D_2}{\gcd(D_1,D_2)}.
\ee

Moving on to $n=3$, we need to find the integral solutions to
\be
\label{eq3}
D_1k_1+D_2k_2+D_3k_3=0.
\ee
To this end, we first consider $k_3=0$. Then, the solutions are
obviously given by (\ref{n2solutions}). Next for a fixed non-vanishing $k_3$,
there are only solutions if $\gcd(D_1,D_2)$ divides $D_3k_3$ by
B\'ezout's identity. As a result, if we choose
$\gcd(D_1,D_2)/\gcd(D_1,D_2,D_3)$ for $k_3$, B\'ezout's identity
asserts that there is an integral solution $(\ell_1,\ell_2)$ for
$(k_1,k_2)$, 
\be
\label{eqn3}
D_1\ell_1+D_2\ell_2+D_3\frac{\gcd(D_1,D_2)}{\gcd(D_1,D_2,D_3)}=0.
\ee
Since this choice for $k_3$ has the smallest non-vanishing magnitude with
integral solutions, the other choices for $k_3$ follow by
multiplication by an integer $m_1$. Including also the solutions
(\ref{n2solutions}), we find for the general solution,
\be
\label{n3solution}
\begin{split}
k_1&=\ell_1 m_1+\frac{D_2}{\gcd(D_1,D_2)}\,m_2,\\
k_2&=\ell_2 m_1-\frac{D_1}{\gcd(D_1,D_2)}\,m_2,\\
k_3&=\frac{\gcd(D_1,D_2)}{\gcd(D_1,D_2,D_3)}m_1.
\end{split}
\ee
Substitution of (\ref{n3solution}) in $\sum_{j=1}^3 D_jk_j^2$, one
finds for the 2-dimensional quadratic form $\underline
D$ for $(m_1,m_2)$,
\be
\underline D=\left(\begin{array}{cc}
    D_1\ell_1^2+D_2\ell_2^2+D_3\frac{\gcd(D_1,D_2)^2}{\gcd(D_1,D_2,D_3)^2}
    &\quad  \frac{D_1D_2(\ell_1-\ell_2)}{\gcd(D_1,D_2)} \\
    \frac{D_1D_2(\ell_1-\ell_2)}{\gcd(D_1,D_2)} & \frac{D_1D_2(D_1+D_2)}{\gcd(D_1,D_2)^2}\end{array}\right).
\ee
Using (\ref{eqn3}), we find for its determinant,
\be
\label{detb21n3}
\det(\underline D)=\frac{D_1D_2D_3(D_1+D_2+D_3)}{\gcd(D_1,D_2,D_3)^2},
\ee
which is symmetric in the $D_j$. For the number of glue vectors $N_g$
and order $N_q$, we now obtain,
\be
n=3:\qquad N_g=\frac{D_1+D_2+D_3}{\gcd(D_1,D_2,D_3)}, \qquad N_q=\frac{D_1D_2D_3}{\gcd(D_1,D_2,D_3)}.
\ee
Together with the discussion below, this is very suggestive that the
generalization to $n>3$ is 
\be
\det(\underline D)=\frac{(\sum_{j=1}^nD_j)\,\prod_{j=1}^nD_j}{\gcd(D_1,D_2,\dots,D_n)^2}.
\ee
\vspace{.4cm}
\\
\noindent {\it 2- and 3-center bound states for $b_2\geq 1$ with simplifications}\\
We continue to discuss the general case $b_2\geq 1$. Let us first consider
$n=2$. We make the assumption that $D_2^{-1}D_1$ is an integral
matrix. The equation (\ref{keq}) can be solved by setting
$k_2=-D_2^{-1}D_1k_1$. Substituting this in the quadratic form $\vec D$, gives us for the
$b_2$-dimensional quadratic form on $\underline \bfLambda$
\be
\underline D=D_1+D_1D_2^{-1}D_1.
\ee
This agrees with (\ref{uDb21}) upon specialization to $b_2=1$. We find
then for $N_g$ and $N_q$,
\be
n=2:\qquad N_g={\rm det}(D_2^{-1}D),\qquad N_q={\rm det}(D_1).
\ee
 
To deal with our main case, $n=3$, we make two technical simplifications: 
\begin{enumerate}
\item We assume that $D_3^{-1}D_1$ and $D_3^{-1}D_2$ are integral
  matrices, such that $k_3\in \mathbb{Z}^{b_2}$ in (\ref{k3=k1k2}) if
  $k_1$ and $k_2\in \mathbb{Z}^2$.
\item We assume that $d_{abc}\in 2\mathbb{Z}$ for all $a$, $b$, $c$. Then $\Lambda^* +P/2= \Lambda^*$ for any $P$ such that shifts by $P$ (and $P_j$) in (\ref{defunderLambda}), (\ref{bfQ2}) and (\ref{Qjkj}) are unnecessary.
\end{enumerate}
These assumptions are satisfied in the examples in Section \ref{sexample}.

 If we substitute now (\ref{k3=k1k2}) in $\bfQ^2$, we arrive at
\be 
\label{bfQk}
\begin{split}
\bfQ^2&=-\mu^2+(\mu_1)^2_1+(\mu_2)^2_2+ (\mu_3)^2_3 +  2(\mu_1-D_3^{-1}D_1\mu_3).k_1+2(\mu_2-D_3^{-1}D_2\mu_3).k_2\\
&\quad +(k_1,k_2)\,\underline D\,(k_1,k_2)^T,
\end{split}
\ee
where $\underline D$ is the quadratic form of the lattice $\underline\bfLambda$,
\be
\label{underD}
\underline D=\left( \begin{array}{cc} D_1 + D_1D_3^{-1}D_1 & \quad D_1D_3^{-1}D_2 \\ D_2D_3^{-1}D_1 & \quad  D_2+D_2D_3^{-1}D_2  \end{array}\right).
\ee  
The determinant of $\underline D$ is given by\footnote{We use that for an
   invertible $n\times n$ matrix $A$ and $n\times m$ matrices $U$ and $V$, we
   have $\det(A+UV^T)=\det(A)\,\det(1_{m}+V^TA^{-1}U)$, en.wikipedia.org/wiki/Matrix$\_$determinant$\_$lemma}
\be  
\label{uDdet} 
\begin{split}
\det(\underline D)&=\det(D_1)\,\det(D_2)\,\det(D_3^{-1})\,\det(D_1+D_2+D_3)\\
&=\det(D_1)\,\det(D_2)\,\det(D_3^{-1} D).
\end{split} 
\ee
We have for $N_g$ and $N_q$ in this case,
\be
\label{SimpNgNq}
N_g=\det(D_3^{-1} D),\qquad N_q=\det(D_1)\det(D_2).
\ee
These formulas are in agreement with (\ref{detb21n3}) for $b_2=1$, and
$\gcd(D_1,D_2,D_3)=D_3$. Moreover for generic $b_2$, we can consider
cases where we can solve \refb{keq} in terms of $k_2$ as well as $k_3$, and we
expect a symmetry in $D_2\leftrightarrow D_3$. Indeed, then 
$D_2^{-1}D_1$ and $D_2^{-1}D_3$ are integral matrices too. The only way for $D_2^{-1} D_3$ and $D_3^{-1} D_2$ both to be integral is to satisfy $|\det (D_2)| = |\det (D_3)|$, in which case \refb{uDdet} reduces to $\det \underline{D} = \pm \det (D_1) \,\det(D_1+D_2+D_3)$, which is symmetric under the exchange of 2 and 3. Similar comments hold of course as well for $k_1$.

Using general formula's for inverses of block matrices \cite{Lu:2002}, we derive that the inverse of $\underline D$ reads
\be
\label{uDinverse}
\begin{split}
 \underline D^{-1}&=\left( \begin{array}{cc} D^{-1}(D_2+D_3)D_1^{-1} & \quad -D^{-1 }\\ -D^{-1} & D^{-1}(D_1+D_3)D_2^{-1} \end{array}\right)\\
&=\left( \begin{array}{cc} D_1^{-1}-D^{-1} & \quad -D^{-1 }\\ -D^{-1} & D_2^{-1}-D^{-1}\end{array}\right),
 \end{split}
\ee  
One can verify that the determinant of  $\underline D^{-1}$ is indeed the inverse of (\ref{uDdet}).
Indeed, if we introduce the two components $\bfmu_1$ and $\bfmu_2$,
\be
\begin{split}
\bfmu_1&=\mu_1- D_1 D_3^{-1}\mu_3,\\
\bfmu_2&=\mu_2-D_2 D_3^{-1}\mu_3,
\end{split}
\ee
the quadratic form $\bfQ^2$ (\ref{bfQk}) for $k_1=k_2=0$, $\bfQ^2=\bfmu^2$, can be written as
\be
\bfmu^2=- (\bfmu_1+\bfmu_2)^2+(\bfmu_1)_1^2+ (\bfmu_2)_2^2=(\bfmu_1,\bfmu_2)\underline D^{-1}(\bfmu_1,\bfmu_2)^T.
\ee
For $k_1,k_2$ non-zero, we have $\bfQ=(\bfQ_1,\bfQ_2)$ with
\begin{eqnarray}\label{bfqq}
\bfQ_1&=\mu_1- D_1 D_3^{-1}\mu_3+(D_1+D_1D_3^{-1}D_1)k_1+D_1 D_3^{-1}D_2k_2,\\ \nn
\bfQ_2&=\mu_2-D_2 D_3^{-1}\mu_3+D_2D_3^{-1}D_1k_1+(D_2+D_2 D_3^{-1}D_2)k_2,
\end{eqnarray}
or equivalently
\be
\begin{split}
\bfQ_1&=Q_1-D_1D_3^{-1}Q_3,\\
\bfQ_2&=Q_2-D_2D_3^{-1}Q_3,
\end{split}
\ee
which we can write more compactly as $\bfQ=\bfmu+\underline D\,\bfk$ with $\bfk=(\bfk_1,\bfk_2)^T$.
Thus an element $\bfmu\in \underline \bfLambda^*/\underline \bfLambda$ is completely determined by
\be
\label{bfmuDef}
\bfmu=\{ (\mu_1,\mu_2,\mu_3,\mu)|\,\,\mu_j\in \Lambda_j^*,\quad \mu_1+\mu_2+\mu_3=\mu\in \Lambda^*\}.
\ee 
\vspace{.2cm}\\
{\it Generic number of constituents (with simplifications)}\\
The analysis for a generic number $n$ of constituents (including $n=2$) follows analogously. The quadratic form reads as in (\ref{bfQ2}), but with 3 replaced by $n$ in the summation, and with constraint $\sum_{j=1}^nQ_j=\mu+P/2$. Similarly to (\ref{keq}), the constraint can be expressed as
\be 
\sum_{j=1}^n D_{jab}k_j^b=0.
\ee 
With the {\it assumption} that $D_n^{-1}D_j$ is an integer matrix for all $j=1,\dots,n$, this can be solved and the quadratic form becomes
\be
\underline D= {\rm diag}(D_1,\dots,D_{n-1})+\left(\begin{array}{c} D_1 \\  \dots \\ D_{n-1} \end{array} \right) \,(D_n^{-1}D_1,\dots,D_n^{-1}D_{n-1}),
\ee
with inverse
\be
\underline D^{-1}= {\rm diag}(D_1^{-1},\dots,D_{n-1}^{-1})-\left(\begin{array}{c} D^{-1} \\  \dots \\ D^{-1} \end{array} \right) \,(1,\dots,1),
\ee
where $D=\sum_{j=1}^n D_j$. Moreover, the determinant of $\underline D$ is
\be
\begin{split}
{\rm det}(\underline D)&=\det(D_{n}^{-1})\,\det(D)\,\prod_{j=1}^{n-1}
\det(D_{j})\\
&=\det(D_{n}^{-1} D)\,\prod_{j=1}^{n-1}
\det(D_{j}),
\end{split}
\ee
from which $N_g$ and $N_q$ are easily determined.
\vspace{.2cm}\\
{\it Characteristic vectors}\\
We briefly discuss here a characteristic vector for the lattice
$\underline \bfLambda$ with $n=3$, which is important for the theta
series of the scaling solutions.
A sign which frequently occured in Section \ref{BHsols} is
$(-1)^{a+b+c}$. Such signs in a theta series are typically written in
terms of a characteristic vector. We therefore express $a+b+c$ as 
\be
a+b+c=\bfK.\bfQ,
\ee
with
\be
\bfK=(P_3-P_2, P_1-P_3).
\ee
This is a characteristic vector of $\underline
\bfLambda$. Indeed,
we have with $\vec P=(P_1,P_2,P_3)$,
\be
\vec k\cdot \vec P+\vec k^2\in 2\mathbb{Z},
\ee
since $P_j$ is a characteristic vector for $\Lambda_j$, $j=1,2,3$.
We can decompose with respect to the lattice decomposition $\underline
\bfLambda\oplus \overline \bfLambda$,
\be
\bfk\cdot \bfK+\bfk^2+k\cdot P+k^2.
\ee
Since $P$ is a characteristic vector for $\Lambda$, this shows that
$\bfK$ is a characteristic vector for $\bfLambda$.

However, if we express $a+b+c+P.Q$ in terms of vectors in
$\bfLambda=\sum_j \Lambda_j$, such that $a+b+c=\vec K.\vec Q \mod 2$, then $\vec
K=(P_3-P_2,P_1-P_3,P_2-P_1)$. Then $(\vec K+\vec P)\cdot \vec
Q=\bfK\cdot \bfQ+\vec P\cdot \vec Q=P\cdot
Q \mod 2$. Moreover, $\vec P$ is a
characteristic vector of $\bfLambda$, and
\be
\vec P^2=P^3+\bfK^3 \mod 4.
\ee

\subsection{Partition functions} 
We consider black hole bound states with three cores. The $j^{\rm th}$
core carries electric and magnetic charges $Q_j$ and $P_j$
respectively. It is natural to work with a mixed ensemble with total
magnetic charge $P$ held fixed. For the present purpose, we shall fix
the total electric D2-brane charge $Q=\sum_j Q_j\in \Lambda^*+P/2$ with
$\mu\in \Lambda^*/\Lambda$ as well.  
We work at the attractor value of the moduli, corresponding to total
charge vector $(P, Q)$. At this point apart from single core black
holes, the only other black holes to survive are the scaling black
holes. A natural question is - for a fixed total charge, how many
scaling black holes are there and what is their contribution to the
index? 

To this end, we define the generating function $h^T_\mu(\tau)$ of
numerical total core invariants $\Omega_T$ in
analogy to the attractor indices,
\be
h^T_{P,\mu}(\tau)=\sum_{Q_{0}} \bar \Omega_T(\gamma)\,q^{\hat Q_{\bar 0}}.
\ee
We can similarly define the partition function of single core indices
$h^S_{P,\mu}(\tau)$, with $\Omega_T$ replaced by $\Omega_S$.
The $\Omega_T(\gamma)$ are determined from the refined ones
(\ref{defOmTy}) using the regularization (\ref{regindex}). For the
3-core case, this gives (\ref{3centerOmT}). If $P$ is
irreducible, i.e. it can not be written as as sum of more than 1
positive magnetic charge, and the three partition functions agree.
 
Based on (\ref{gCformula}), we can express the attractor partition
function $h_{P,\mu}$  in terms of the partition function $h^T_{P,\mu}$. We have schematically
\be
\label{hPhTP}
h_{P,\mu}(\tau)=h^T_{P,\mu}(\tau) + \sum_{n>1}\sum_{\sum_{j=1}^n
  P_j=P\atop \sum_{j=1}^n
  Q_j=Q}  
\frac{g_C(\{\gamma_j\}, \{c_{j}^\lambda\})}{|{\rm
    Aut(\{\gamma_j\})}|} q^{\mu^2/2-\sum_j (Q_j)_j^2/2} \prod_{j=1}^n h^T_{P_j,\mu_j}(\tau).
\ee
Recall that there are no 2-center/core scaling black holes, such that there
is no contribution from $n=2$ on the rhs.

We will proceed by considering the term in (\ref{hPhTP}) with $n=3$. Using the
notation introduced in Section \ref{BoStaLat}, we can enumerate the
number of three-core scaling black holes as 
\begin{equation} 
\label{defhs}
h^{3T}_{\{P_j\},\mu}(\tau)=\sum_{\mu_j\in \Lambda^*_j/\Lambda_j,\,\,j=1,2,3, \atop{\mu_1 + \mu_2 + \mu_3 = \mu}}
h^T_{P_1,\mu_1}(\tau)\, h^T_{P_2,\mu_2}(\tau)\,h^T_{P_3,\mu_3}(\tau) \, \Psi_{\bfmu}(\tau) \, , 
\end{equation}
with $\bfmu$ as in (\ref{bfmuDef}), and where $\Psi_\bfmu$ is the indefinite theta series
\begin{equation} 
\label{DefPsi} 
\Psi_{\bfmu}(\tau) = \sum_{\bfQ \in \underline {\boldsymbol
    \Lambda}^*_{\bfmu}}  g_{C}(\{\gamma_j\},\{c_j^\lambda\}) \,q^{ -
  \bfQ^2/2}.
\end{equation} 
With $y=e^{2\pi iz}$, we define the refined series as
\begin{equation} 
\label{DefPsiy} 
\Psi_{\bfmu}(\tau,z) = (y-y^{-1})^2\sum_{\bfQ \in \underline {\boldsymbol \Lambda}^*_{\bfmu}}  g_{C}(\{\gamma_j\},\{c_j^\lambda\};y) \,q^{ - \bfQ^2/2},
\end{equation} 
with $g_{C}(\{\gamma_j\},\{c_j^\lambda\})$ as in (\ref{defgC}) and
$g_{C}(\{\gamma_j\},\{c_j^\lambda\};y)$ as in (\ref{defgCy}). Note
that $\Psi_\bfmu$ is
symmetric as function of $z$,
\be
\Psi_{\bfmu}(\tau,-z)=\Psi_{\bfmu}(\tau,z).
\ee
The  two functions are related by (\ref{regindex}),
\be\label{derPsi}
\Psi_{\bfmu}(\tau)=\left(\frac{1}{4\pi i} \frac{\partial}{\partial
    z}\right)^2 \left.\Psi_{\bfmu}(\tau,z)\right|_{z=0}.
\ee

The kernel $g_{C}(\{\gamma_j\},\{c_j^\lambda\})$ and therefore $\Psi_{\bfmu}$
is unchanged under a symplectic transformation
\eqref{spectransimple}, such that $h^{T}_{\{P_j\},\mu}$ is invariant under
spectral flow as required. The number of terms in the sum over $\mu_j$
in (\ref{defhs}) is given by $N_q$ (\ref{Nq}). 

To determine the modular properties of $\Psi_\bfmu$, we consider first 
the generating function of $f_C$ (\ref{deff}), which enumerates the number of
scaling charge configurations for a given total charge. We define this function
$\Phi_{\bfmu}$ as the following theta series,  
\be
\label{DefPhi}
\begin{split}
\Phi_{\bfmu}(\tau) &= \sum_{\bfQ \in \underline {\boldsymbol
    \Lambda}^*_{\bfmu}} f_C(\{\gamma_j\},\{c_j^\lambda\})\,q^{ -\bfQ^2/2},
\end{split}
\ee
with $f_C(\{\gamma_j\},\{c_j^\lambda\})$ as in (\ref{deff}).

The following subsections will demonstrate that $\Phi_{\bfmu}(\tau)$
is a convergent $q$-series, which can be
completed to a function $\widehat \Phi_\bfmu$ which transform as a
vector-valued modular form. The transformation properties under the $S$ and $T$ transformations are
\be 
\begin{split}
\widehat \Phi_{\bfmu}(-1/\tau,-1/\bar \tau) & =
-\frac{(-i\tau)^{b_2}}{\sqrt{|\underline \bfLambda^*/\underline \bfLambda|}}\, e^{\pi i \bfK^2/2}\sum_{\bfnu \in \underline
  \bfLambda^*/\underline \bfLambda} e^{2\pi i \bfmu.\bfnu}\, \widehat \Phi_{\bfnu}(\tau,\bar\tau),\\
\widehat \Phi_{\bfmu}(\tau+1,\bar \tau+1) &= e^{\pi i (\bfmu+\bfK/2)^2}\,\widehat \Phi_{\bfmu}(\tau,\bar \tau).
\end{split}
\ee

The partition function $\widehat
\Psi_{\bfmu}(\tau)$ can be obtained by introducing a suitable
elliptic variable in $\widehat \Phi_{\bfmu}$ and
subsequently differentiating twice to this variable.
As a result, the modular transformations of the completed function
$\widehat \Phi_{\bfmu}$ equal those of the completed
$\widehat \Psi_{\bfmu}$ except that the weight of $\widehat
\Psi_{\bfmu}$ is increased by two compared to $\widehat
\Phi_{\bfmu}$. The weight of $\widehat \Psi_{\bfmu}$ is thus $b_2+2$.
The non-holomorphic terms are determined in this way in Section \ref{modcompPsi}, specifically Eq. (\ref{completePsi}).
The end result is that $\widehat \Psi_{\bfmu}$ transforms as 
\be 
\begin{split} 
\widehat \Psi_{\bfmu}(-1/\tau,-1/\bar \tau) & =
\frac{(-i\tau)^{b_2+2}}{\sqrt{|\underline \bfLambda^*/\underline \bfLambda|}}\, e^{\pi i \bfK^2/2}\sum_{\bfnu \in \underline
  \bfLambda^*/\underline \bfLambda} e^{2\pi i \bfmu.\bfnu}\, \widehat \Psi_{\bfnu}(\tau,\bar\tau),\\
\widehat \Psi_{\bfmu}(\tau+1,\bar \tau+1) &= e^{\pi i (\bfmu+\bfK/2)^2}\,\widehat \Psi_{\bfmu}(\tau,\bar \tau).
\end{split}
\ee

Therefore, the completion of $h^{3T}_{\{P_j\},\mu}$ (\ref{defhs}),
\be
\widehat h^{3T}_{\{P_j\},\mu}(\tau,\bar \tau)=\sum_{\mu_j\in \Lambda^*_j/\Lambda_j,\,\,j=1,2,3, \atop{\mu_1 + \mu_2 + \mu_3 = \mu}}
\widehat h^T_{P_1,\mu_1}(\tau)\, \widehat h^T_{P_2,\mu_2}(\tau)\,\widehat h^T_{P_3,\mu_3}(\tau) \, \widehat\Psi_{\bfmu}(\tau) \, , 
\ee
 transforms as $\widehat h_{P,\mu}$ (\ref{STh}) as we aimed to
 show. We can furthermore combine $\widehat h_{P,\mu}$ with the theta
 series $\Theta_\mu$, 
\be
\widehat \CZ^{3T}_{P}(\tau,C,t)=\sum_{\sum_{j=1}^3 P_j=P}\sum_{\mu\in \Lambda^*/\Lambda}
\widehat h^{3T}_{\{P_j\},\mu}(\tau,\bar \tau)\, \Theta_\mu(\tau,\bar \tau, C, B).
\ee
We can then decompose the attractor partition $\widehat \CZ^\lambda_P$ in terms of the
multi-core partition functions $\widehat \CZ^{nT}_{P}$,
\be 
\label{AttrPDecomp}
\widehat \CZ^\lambda_P(\tau,C,t)=\widehat \CZ^T_P(\tau,C,t)+\widehat \CZ^{3T}_{P}(\tau,C,t)+\dots.
\ee
Since the partition functions transform the same way, this raises the
question which terms are captured by the MSW conformal field
theory. As mentioned in the introduction, it will also be interesting
to deduce the non-holomorphic terms of $\widehat \CZ^T_P$ using those
determined for $\widehat \CZ^{3T}_{P}$ in this paper, and those for
$\widehat \CZ^\lambda_P$ in \cite{Alexandrov:2018lgp}.

\subsection{Convergence}
A crucial aspect of $\Phi_\bfmu$ (and $\Psi_\bfmu$) is whether the sum on
the rhs of \eqref{DefPhi} (and \eqref{DefPsi}) is convergent. If $\bfQ^2$
would be negative definite, convergence of these series would be guaranteed. However this is
not the case since the electric charge lattice has
signature $(2, 2b_2-2)$, i.e. has 2 positive directions.  

To prove the convergence, we first introduce a theta series
$\Theta_\mu[\CK](\tau)$ with kernel $\CK$ for a generic indefinite theta
lattice $L$ and $\mu\in L^*$,  
\begin{align} 
\Theta_\mu[\CK](\tau;L) &= \sum_{x \in L + \mu} \mathcal{K}(x)\, q^{- B(x)/2} \, , \label{thetajan}
\end{align}
with integral quadratic form $B$. If $L$ is negative definite,
we also use
\begin{align}
\label{NormalTheta} 
\theta_\mu(\tau;L) &= \Theta_\mu[1](\tau;L) .
\end{align}

For an indefinite lattice the kernel $\CK(x)=\CK(x,\CV)$ depends on a collection
$$\CV=\{V_1,V_2,\dots,V_N\},$$ of positive vectors. For signature
$(2,2b_2-2)$, $\CK(x,\CV)$ can be expressed as \cite{Alexandrov:2016enp, Alexandrov:2017qhn, FunkeNotes:2018}
\be
\label{sumjan}
\mathcal{K}(x,\CV) = \frac{1}{4} \left(  w(\CV)  + \sum_{j=1}^N \sgn(B(x,V_j)) \sgn(B(x,V_{j+1})) \right) \, , 
\ee 
where for any strictly positive vector $v\in L$, $v^2>0$,
\be
w(\CV) =   - \sum_{j=1}^N \sgn(B(v,V_j))
  \sgn(B(v,V_{j+1})) ,
\ee
which is independent of the choice of positive vector $v$ \cite{FunkeNotes:2018}. There are various sufficient conditions for
convergence put forward in the literature \cite{Alexandrov:2017qhn, Alexandrov:2016enp,
  kudla2016theta, funke_kudla_2019, FunkeNotes:2018}.
We will consider here the following $N$-gon conditions put forward in
\cite{Alexandrov:2017qhn, Alexandrov:2016enp, FunkeNotes:2018}, which read
\be
\begin{split} 
& B(V_j,V_j)  > 0, \\ 
& B(V_j,V_j)\,B(V_{j+1},V_{j+1}) - B(V_j,V_{j+1})^2 > 0, \\
& B(V_j,V_j)\,B(V_{j-1},V_{j+1}) - B(V_j,V_{j-1})\,B(V_j,V_{j+1})  <0.
   \label{FK}
\end{split}
\ee
Now let us return to the sum (\ref{DefPhi}) at hand. It comprises of six individual
sums, which are each of the form
\begin{align}
\begin{split}
&\sum_{Q_i \in \mu_i + \Lambda_i + P_i /2 \atop {Q_1+Q_2 +
    Q_3=\mu+P/2}} F_\ell(a,b,c)\,(-1)^{a+b+c}\,q^{ Q^2/2- \sum_i (Q_i)_i^2/2}  =\sum_{\bfQ \in \underline {\boldsymbol
    \Lambda}^*_{\bfmu}} F_\ell(a,b,c)\,(-1)^{a+b+c}   \,q^{ -\bfQ^2/2},
\end{split}
\end{align}
with $\ell=1, \dots , 6$.

The simplification $F_{\rm total}(a,b,c)$ for $\sum_{j=1}^6 F_j(a,b,c)$ put forward in \refb{F0fin} is precisely of the form (\ref{sumjan}) with $N=3$. 
To present the vectors, we define
\be
\label{Cabc}
\begin{split}
&C_a=(-P_2,P_1,0),\\
&C_b=(0,-P_3,P_2),\\
&C_c=(P_3,0,-P_1),
\end{split}
\ee
such that $C_a.\bfQ=a$, $C_b.\bfQ=b$ and $C_c.\bfQ=c$, with $a,b$ and
$c$ as in \eqref{abcPQ}. The $V_j$ are then identified with $C_j\in \underline\bfLambda$ with the $C_j$
given by,
\be
\begin{split}
&C_1=C_a+C_b-C_c=(-P_2 - P_3,\, P_1 - P_3, P_1 + P_2),\\
&C_2=C_a-C_b+C_c=(-P_2 + P_3, \, P_1+P_3,\, -P_1-P_2),\\
&C_3=-C_a+C_b+C_c=(P_2 +P_3, \, -P_1 -P_3, \, -P_1+P_2) \, . \label{v123new}
\end{split}
\ee
If we assume that $P_j$ is an ample divisor for each $j\in 1,2,3$,
triple intersections $P_iP_jP_k>0$ for all $i,j,k\in \{1,2,3 \}$. The
conditions for convergence (\ref{FK}) are then satisfied. 

It is also useful to consider the convergence for the kernel
due to a single permutation separately, since the
different permuations are weighted by a different factor in $g_C$. The
vectors $V_j$ for the kernel $F^*(123)$ (\ref{F0123}) can be chosen as
$C_j^{(123)}$,
\be
\label{Cj123}
\begin{split}
C_1^{(123)}&= C_a-C_c=(C_1-C_2)/2,\\
C_2^{(123)}&= C_b-C_c=(C_1-C_3)/2,\\
C_3^{(123)}&= C_c-C_a-C_b=-C_1.
\end{split}
\ee
Again one may verify that with the assumption $P_iP_jP_k>0$ for all
$i,j,k\in \{1,2,3 \}$, these vectors satisfy the conditions for convergence (\ref{FK}). 
%
\subsection{Modular completion of $\Phi_\bfmu$} \label{ssmodcom} 
Having discussed the convergence of $\Phi_\bfmu$, we proceed in this
Section to discuss its modularity. Since $\Phi_\bfmu$ is a sum over a
subset of an indefinite lattice, the function is not modular in the
classical sense. Our task is to determine a modular completion
$\widehat \Phi_\bfmu$, which differes from $\Phi_\bfmu$ by subleading
non-holomorphic terms, and which does transform as a modular form. 
Essentially, products of
sgn-functions are replaced by a generalized error function \cite{Alexandrov:2016enp}.
Following this approach, we will demonstrate that the difference between $\widehat \Phi_\bfmu$ and $\Phi_\bfmu$, is
given by iterated integrals of modular forms.
  
Such non-holomorphic contributions have appeared in similar contexts.
In specific cases, the non-holomorphic contributions are derived from
different physical points of view, 
for example the continuum of multi-particle states in $\mathbb{R}^4$ \cite{Pioline:2015wza}, or the quantum field theory on the world volume of the
D-branes \cite{Dabholkar:2020fde, Bonelli:2020xps, Manschot:2021qqe}, or the perspective of
D3-instantons in the hypermultiplet moduli space
\cite{Alexandrov:2016tnf}. 

To this end let us consider $F_{\rm total}(a,b,c)$ in \refb{F0fin}. Under modular completion, one adds certain extra terms to $\sgn(V_1,x)
\sgn(V_2,x)$, therefore replacing $\sgn(V_1,x) \sgn(V_2,x)$  with the double
error function $E_2(\alpha,u_1, u_2)$: 
\be 
\sgn(V_1,x) \sgn(V_2,x)\to E_2(\alpha,\sqrt{2\tau_2}\,\bfu),
\ee
with \cite{Alexandrov:2016enp} 
\be
\label{E2def1}
E_2(\alpha; \bfu)=\int_{\mathbb{R}^2} e^{-\pi (u_1-u_1')^2-\pi (u_2-u_2')^2}\,\sgn(u_2')\,\sgn(u_1'+\alpha\,u_2')\,du_1'\,du_2',
\ee
whose arguments are given in terms of $V_1$, $V_2$ and $x$ by
\be
\label{alphauC}
\begin{aligned}  
&\alpha=\alpha(V_1,V_2) =\frac{(V_1,V_2)}{\sqrt{V_1^2\,V_2^2-(V_1,V_2)^2}}, \\
&\bfu=\bfu(V_1,V_2;x)=(u_1(V_1,V_2;x),u_2(V_1,V_2;x)),
\end{aligned}
\ee
with
\begin{align}
&u_1(V_1,V_2;x)= \frac{(V_{1 \perp 2}, x )}{\sqrt{ (V_{1 \perp 2}, V_{1 \perp
      2})}} \, ,\\
&u_2(V_1,V_2;x) =  \frac{(V_2, x )}{\sqrt{ (V_2, V_2) }} \, 
\end{align}
and $V_{1\perp 2}$ the component of $V_1$ orthogonal to $V_2$, 
\be
V_{1\perp 2}=V_1-\frac{(V_1,V_2)}{(V_2,V_2)}\,V_2.
\ee
To stress the dependance of $E_2$ on the vectors $V_1$,
$V_2$ and $x$, we will also use $E_2$ with alternative arguments,
\be
E_2(\alpha,\bfu)\equiv E_2(V_1,V_2;x),
\ee
with the identifications as in (\ref{alphauC}).

In addition to (\ref{E2def1}), another (equivalent) expression for
$E_2$ is in terms of Eichler integrals. To this end,  we first define the Eichler integrals $M_1$, 
\be
M_1(u) =\left\{ \begin{array}{cc} \frac{i u}{\sqrt{2 \tau_2}} q^{\frac{u^2}{4\tau_2}} \int_{-
    \bar{\tau}}^{i\infty}  \frac{e^{\frac{i \pi u^2 w}{2\tau_2}}}{\sqrt{-i
      (w+\tau)}} dw, & \qquad u\neq 0, \\ 0, & \qquad u=0. \end{array}\right. \\
\ee
and the (iterated) Eichler integral $M_2$
and $m_2$, for $u_1 \neq 0$, and  $u_2 - \alpha u_1 \neq 0$,
\be
\begin{split}
m_2(u_1, u_2) &= \left\{ \begin{array}{cc}\frac{u_1 u_2}{2\tau_2}
    q^{\frac{u_1^2}{4\tau_2} + \frac{u_2^2}{4\tau_2}} \int_{- \bar{\tau}}^{
      i\infty} dw_2 \int_{w_2}^{i\infty} dw_1 ~\frac{e^{\frac{\pi i
          u_1^2 w_1}{2\tau_2} + \frac{\pi i u_2^2 w_2}{2\tau_2}
      }}{\sqrt{-(w_1+\tau) (w_2+\tau)}}, &\qquad u_1\neq 0\\ 0,&
    \qquad u_1=0. \end{array}\right.  \, \\ 
M_2(\alpha;u_1,u_2)  &= 
\left\{ 
\begin{array}{rl}
&- m_2 (u_1, u_2) - m_2 \left( \frac{u_2 - \alpha u_1}{\sqrt{1+\alpha^2}} , \frac{u_1 + \alpha u_2}{\sqrt{1+\alpha^2}} \right) \, \quad u_1 \neq 0, u_2 - \alpha u_1 \neq 0 \, ,\\
& - m_2 \left( \frac{u_2 - \alpha u_1}{\sqrt{1+\alpha^2}} , \frac{u_1 + \alpha u_2}{\sqrt{1+\alpha^2}} \right) \,  \hspace{27mm} u_1=0, u_2 \neq 0 \, , \\
& - m_2 (u_1, u_2) \, \hspace{45mm} u_1 \neq 0, u_2 - \alpha u_1 =0 \, , \\
& \frac{2}{\pi} \arctan \alpha \,  \hspace{49mm} u_1 = u_2 = 0 . \end{array}\ \right. \\
\label{integralE2}
\end{split}
\ee

With $\bfu=(u_1,u_2)$ as before, the double error function $E_2$ is
then defined as a linear combination of $M_1$ and $M_2$
\cite{Alexandrov:2016enp, bringmann2018higher}
\be
\label{E2defn}
\begin{split}
E_2(\alpha; \bfu) &= \sgn(u_2) \sgn(u_1 + \alpha u_2) + \sgn(u_1) M_1 (u_2) \\
&\quad + \sgn(u_2 - \alpha u_1)\, M_1\!\left(\frac{u_1 + \alpha u_2}{\sqrt{1 + \alpha^2}}\right) + M_2(\alpha; u_1 , u_2) \, .
\end{split}
\ee
See \cite{Alexandrov:2016enp} for other representations of $E_2$. 

Thus $E_2$ consists of  the original $\sgn(V_1,x) \sgn(V_2,x)$ plus
four more terms. 
Noting that 
\begin{align}
\nn
\frac{u_2(V_2,x) - \alpha u_1 (V_1, V_2,x)}{\sqrt{1 + \alpha^2}} &= u_1 (V_2, V_1, x) \, , \\
\frac{u_1(V_1, V_2,x) + \alpha u_2 (V_2,x)}{\sqrt{1 + \alpha^2}} &= u_2(V_1, x) \, , 
\end{align}
we can write
\begin{align}
\nn
E_2(\alpha; \bfu) &= \sgn{} (V_1.x) \, \sgn{} (V_2.x) + \sgn{}
(u_1(V_1, V_2, x)) \, M_1 (u_2 (V_2, x)) \\
&+  \sgn{} (u_1(V_2, V_1, x)) \, M_1 (u_2 (V_1, x)) \\
&\nn- m_2 (u_1 (V_1, V_2,x), \, u_2 (V_2, x)) - m_2 (u_1 (V_2, V_1, x), \, u_2 (V_1, x)) \, .
\label{E2useful}
\end{align}
$E_2$ satisfies an identity similar to (\ref{signid}). This reads
\be
\label{E2iden}
E_2(V_1,V_1+V_2;x)+E_2(V_2,V_1+V_2;x)- E_2(V_1,V_2;x)=1
\ee
and is valid for any choice of the arguments such that the
corresponding $\alpha$, $u_1$ and $u_2$'s are in $\mathbb{R}$.

The sum at hand \refb{F0fin} has the form
\begin{align}
\sgn{} (C_1.x) \, \sgn{} (C_2.x) \, + \, \sgn{} (C_2.x) \, \sgn{} (C_3.x) \, + \, \sgn{} (C_3.x) \, \sgn{} (C_1.x) \, , 
\end{align}
with $C_1, C_2, C_3$ given in \refb{v123new}. The modular completion $\widehat \Phi_\bfmu$ of $\Phi_\bfmu$ (\ref{DefPhi}) follows by the adding to the coefficient $f_C$ (\ref{deff}) the following terms 
\be
\label{compterms}
\frac{1}{4}\sum_{\ell =1,2,3} \left[ E_2(C_\ell, C_{\ell +1};\sqrt{2\tau_2}  x) -
  \sgn{} (C_\ell.x) \, \sgn{} (C_{\ell+1}.x)-A_\ell\,\delta_{(C_\ell.x)}\,\delta_{(C_{\ell+1}.x)} \right].
\ee
This essentially amounts to replacing $f_C$ by a linear combination of
$E_2$'s. Since the latter satisfies Vign\'eras equation, modular
transformation properties are ensured \cite{Vigneras:1977}. 

Our first aim is to determine the value of $A_\ell$ such that the completion is subleading, i.e. it vanishes in the limit that ${\rm Im}(\tau)=\tau_2\to \infty$.
This follows from realizing that the difference $E_2(C_\ell, C_{\ell +1};\sqrt{2\tau_2}  x) -
  \sgn{} (C_\ell.x) \, \sgn{} (C_{\ell+1}.x)$ vanishes in this limit except if $(C_\ell.x)=(C_{\ell+1}.x)=0$, when it equals $\frac{2}{\pi}\arctan(\alpha_\ell)$, 
$\alpha_\ell=\alpha(C_\ell,C_{\ell+1})$ (\ref{alphauC}). Requiring that (\ref{compterms}) vanish, we thus arrive at 
\be
A_\ell=\frac{2}{\pi}\arctan(\alpha_\ell).
\ee
Surprisingly this implies that $A_\ell$ can be irrational, as we will see in the
explicit case studies in Section \ref{sexample}. This of course
obstructs an interpretation as a ``counting'' function for the
coefficients to which $A_\ell$ contribute. On the other hand, since the function
$\Phi_\bfmu$ is not a proper physical partition function summing over
a Hilbert space, we are not very concerned about this.

To proceed with determining the completion, we rearrange the terms and
write (\ref{compterms}) as
\be
\begin{split}
&\frac{1}{4}\sum_{\ell =1,2,3} \left[ \sgn{} (u_1 (C_{\ell +1}, C_\ell, x)) + \sgn{} (u_1 (C_{\ell -1}, C_\ell, x)) \right] M_1 (\sqrt{2\tau_2}\, u_2 (C_\ell, x)) \\
& -m_2 ( \sqrt{2\tau_2} u_1 (C_{\ell+1}, C_\ell,x), \sqrt{2\tau_2} u_2(C_\ell, x)) - m_2 ( \sqrt{2\tau_2} u_1 (C_{\ell-1}, C_\ell,x), \sqrt{2\tau_2} u_2(C_\ell, x)).
\end{split}
\ee
We determine now the $A_\ell$ by requiring that this expression
vanishes in the limit $y\to \infty$. 

Using these expressions, we thus naturally separate the holomorphic part $\Phi_\bfmu$ from the completion $\widehat{\Phi}_\bfmu$, 
\be
\label{defhatPhinew}
\widehat{\Phi}_{\bfmu}(\tau,\bar \tau) = \Phi_\bfmu(\tau)
+R^{\Phi}_\bfmu(\tau,\bar \tau),
\ee
with the non-holomorphic completion $R^{\Phi}_\bfmu$ defined by
\be
\label{Phihatnew}
\begin{split}
R^{\Phi}_\bfmu(\tau,\bar \tau)  = &\sum_{\bfQ\in \underline {\boldsymbol \Lambda}^*_\mu
}\,\, 
\sum_{\ell=1,2,3} \\
&\left[
\left[ \sgn{} (u_1 (C_{\ell +1}, C_\ell, x)) + \sgn{} (u_1 (C_{\ell -1}, C_\ell, x)) \right] M_1 (\sqrt{2\tau_2} u_2 (C_\ell, x)) \right. \\
&\left.- m_2 ( \sqrt{2\tau_2} u_1 (C_{\ell+1}, C_\ell,x),
  \sqrt{2\tau_2} u_2(C_\ell, x))\right.\\
& \left.- m_2 ( \sqrt{2\tau_2} u_1
  (C_{\ell-1}, C_\ell,x), \sqrt{2\tau_2} u_2(C_\ell, x)) \right] \, \\
&\times (-1)^{\bfK.\bfQ} q^{-\bfQ^2/2}. 
\end{split}
\ee
Our next aim is to write the non-holomorphic part as an (iterated)
integral over modular forms. This makes the modular properties of the
holomorphic $q$-series manifest, since the modular properties of
integrals of modular forms are readily determined. Moreover, it is
straightforward to determine the non-holomorphic anomaly.

To determine this form of the non-holomorphic part, we write $R^{\Phi}_\bfmu$ as 
\begin{align}
R^{\Phi}_\bfmu(\tau,\bar \tau)
&=
\sum_{\ell=1,2,3} \left[ R_{\bfmu, 1, \ell} (\tau,\bar \tau) + R_{\bfmu, 2, (\ell -1, \ell)} (\tau,\bar \tau) + R_{\bfmu, 2, (\ell + 1, \ell)} (\tau,\bar \tau) \right] \label{RPhi12}
\end{align}
and $R_{\bfmu, 1, \ell}, R_{\bfmu, 2, (k, \ell)}$ are defined as 
\begin{align}
\nn
R_{\bfmu, 1, \ell} (\tau,\bar \tau) &= 
\sum_{\bfQ\in \underline {\boldsymbol \Lambda}^*_\mu
}\,\, 
\left[ \sgn{} (u_{1, (\ell+1, \ell)} + \sgn{} (u_{1,(\ell-1, \ell)}) \right] M_1 (\sqrt{2y} u_{2, \ell})   (-1)^{\bfK.\bfQ} q^{-\bfQ^2/2} \, , \\
R_{\bfmu, 2, (k, \ell)} (\tau,\bar \tau) &=  - \sum_{\bfQ\in \underline {\boldsymbol \Lambda}^*_\mu
}\,\, 
m_2 ( \sqrt{2\tau_2} u_{1,(k,l)},\, \sqrt{2\tau_2} u_{2, \ell}) \,  (-1)^{\bfK.\bfQ} q^{-\bfQ^2/2} \, , \label{Rmu1mu2}
\end{align}
where we have used the abbreviations $u_{1,(k,\ell)} = u_1 (C_k, C_\ell, x), \, u_{2, \ell} = u_2 (C_\ell, x)$.

To evaluate the sums \refb{Rmu1mu2}, it is useful to split $\bfQ^2$ as follows
\begin{align}
\bfQ^2 &= u_{1,(k,\ell)}^2 + u_{2,\ell}^2 + (\bfQ^2 - u_{1,(k,\ell)}^2 - u_{2,\ell}^2) \, , \label{Qsplitnew}
\end{align}
$u_{1,(k,\ell)}^2, \, u_{2,\ell}^2$ and $(\bfQ^2 - u_{1,(k,\ell)}^2 - u_{2,\ell}^2)$ are naturally associated with quadratic forms and lattices as described in Table \ref{table2}.
\begin{table}[h]
\begin{centering}
\begin{tabular}{|c|c|c|c|} 
\hline 
term & form & dual lattice & signature\\
\hline \hline
$\bfQ^2$ & $\underline D$ & $\underline \bfLambda^*_\bfmu$ & $(2,2b_2-2)$ \\
\hline
$u_{2,\ell}^2$ & $\underline D_\ell =|C_\ell|^2$ & $(L_{\ell})^*_\bfmu \subset \underline\bfLambda_\bfmu^*$ & (1,0) \\
\hline
$\bfQ^2 - u_{2,\ell}^2$ & $\underline D_{\perp \ell}$ & $(L_{\ell}^{\perp})^*_\bfmu \subset \underline\bfLambda_\bfmu^*$ & $(1,2b_2-2)$\\
\hline
$u_{1,(k,l)}^2$ & $\underline D_{(k,l)} =|C_{k \perp \ell}|^2$ & $(L_{k \ell})^*_\bfmu \subset (L_{\ell}^{\perp})^*_\bfmu$ & (1,0)\\
\hline 
$\bfQ^2 - u_{1,(k,l)}^2 - u_{2,\ell}^2$ & $\underline D^\perp_{(k \ell)}$ & $ (L_{k \ell}^\perp)^*_\bfmu  \subset (L_{\ell}^{\perp})^*_\bfmu$ & $(0,2b_2 -2)$\\
\hline
\end{tabular}
\caption{Lattices and quadratic forms associated to the splitting in
  Equation (\ref{Qsplitnew}). } \label{table2}
\end{centering} 
\end{table}

We let $L_{\ell}$ be the 1-dimensional lattice spanned by
$C_\ell$. The quadratic form is a number in this case, $D_\ell=|C_\ell|^2$. We
denote the dual lattice with quadratic form $|C_\ell|^{-2}$ by $(L_\ell)^*$. The
projection of $\bfmu$ to $(L_\ell)^*$ is $\mu_\ell=(\bfmu.C_\ell)\, C_\ell \in
\mathbb{Z}^{2b_2}$. 
To express a generic vector in $\underline \bfLambda_\bfmu^*$ as an
element of $(L_{\ell})^* \oplus (L_{\ell}^\perp)^*$, we introduce glue
vectors $\rho$. We denote by $(L_\ell)^*_\bfmu$ the set of vectors $\mu_\ell
\mod L_\ell \in (L_\ell)^*$, and by  $(L_\ell)^*_{\bfmu+\rho}$ the set of
vectors $\mu_\ell +\rho\mod L_\ell \in (L_\ell)^*$. We introduce similar
notation for $L_\ell^\perp$. Using this notation, the direct sum $(L_{\ell})^*_\bfmu \oplus (L_{\ell}^\perp)^*_\bfmu$ is a subset of the lattice
$\underline \bfLambda_\bfmu^*$, $(L_{\ell})^*_\bfmu \oplus
(L_{\ell}^\perp)^*_\bfmu  \subset \underline \bfLambda_\bfmu^*$.

A generic element $\bfk$ of $\underline \bfLambda_\bfmu^*$ can be written as a sum,
$\bfk=\bfl_\ell + \bfl_{\ell}^\perp\in (L_{\ell})^*_{\bfmu,\rho}\oplus
(L_{\ell}^\perp)^*_{\bfmu,\rho}$ for some $\rho$, where the projection $\bfl_\ell$ to
$(L_{\ell}^\perp)^*$ vanishes, and similarly for the projection of
$\bfl_\ell^\perp$ to $(L_{\ell})^*$. We have a similar decomposition of
vectors in $(L_{\ell})^\perp$ with respect to the decomposition
$(L_{k \ell})^*\oplus (L_{k \ell}^\perp)^*$. The representatives
of minimal length in $(L_{\ell})^* \oplus (L_{\ell}^\perp)^*$ appearing in such
splits are called ``glue vectors''.  We use glue vectors $\rho$ for 
the splitting $(L_{\ell})^*_{\bfmu,\rho}\oplus
(L_{\ell}^\perp)^*_{\bfmu,\rho}$ and $\nu$ for the splitting $(L_{k \ell})^*\oplus (L_{k \ell}^\perp)^*$.
 We then have,
\be
\begin{split}
&\underline \bfLambda_{\bfmu}^*= \sum_\rho (L_{\ell})^*_{\bfmu+\rho} \oplus (L_{\ell}^\perp)^*_{\bfmu+\rho}, \\
&(L ^\perp_{\ell})^*_\rho=\sum_\nu \subset (L_k)^*_{\rho+\nu} \oplus (L_{k \ell}^\perp)^*_{\rho+\nu}.
\end{split}
\ee 
Now let us evaluate the sum $R_{\bfmu,1}$ and $R_{\bfmu,2}$
\refb{Rmu1mu2}. The embedding of $L_\ell$ in $\underline \bfLambda_\bfmu \otimes \mathbb{Q}$ is spanned by $k\,C_\ell$ with 
$k\in \mathbb{Z}+\rho$ where $\rho\in \mathbb{Q}$ is (the projection of) the glue vector. Then, $u_{2,\ell} = |C_\ell| k$, such that $R_{\bfmu,1, \ell}$ reads
\be
\begin{aligned}
R_{\bfmu,1, \ell}(\tau,\bar \tau)&=\sum_\rho\sum_{\bfl_{\ell}^\perp \in (L_{\ell}^\perp)^*_{\bfmu+\rho} \atop k\in \mathbb{Z}+\bfmu+\rho} (\sgn(u_{1,(\ell-1, \ell)})+\sgn(u_{1,(\ell+1,\ell)})) \\
&\qquad \times (-1)^{\bfK.\bfl_{\ell}^\perp+\bfK.C_{\ell}\,k}  \,
M_1\!\left(\sqrt{2\tau_2}\, |C_{\ell}|\,k\right)
q^{-(\bfl_{\ell}^\perp)^2/2- C_{\ell}^2\,k^2/2}. 
\end{aligned}
\ee
To write this more compactly, we define for a generic lattice $L$ of signature
$(1,\dim(L)-1)$ and characteristic vector $K$, 
\begin{align}
\Theta_\alpha(\tau; L, \{V, V'\} ) &:= \sum_{x\in L + \alpha}
\left( \sgn( V, x) - \sgn(V',x) \right) (-1)^{K.x} q^{-x^2/2} \, . 
\end{align}
Convergence of $\Theta_\alpha$ requires \cite{ZwegersThesis} 
\begin{align}
V^2 &>0 \, , \, (V')^2 >0 \, , \, (V,V')> 0 \, ,
\end{align}
Moreover, we define the unary theta series $\Upsilon_\alpha$,
\be
\label{Upsilon}
\Upsilon_\alpha(\tau,M,N)=\sum_{x\in \mathbb{Z}+\alpha} x\,(-1)^{Nx}\,e^{\pi i
  \tau Mx^2}.
\ee 
For $\sigma\in \bar{\mathbb{H}}$, we introduce the period integral
\be
\begin{split}
R_\alpha(\tau, \sigma; M,N) &:= i \sum_{x\in \mathbb{Z}+\alpha}
(-1)^{Nx} x \int_{-\sigma}^{i \infty} dw \, \frac{e^{\pi i w M x^2}}{\sqrt{ -i (w + \tau)}} \, \\
 &= i \int_{-\sigma}^{i \infty} dw \, \frac{\Upsilon_\alpha(w,M,N)}{\sqrt{ -i (w + \tau)}} \, .
\end{split}
\ee 
The non-holomorphic modular completion $\widehat \Theta_\alpha$ of $\Theta_\alpha$ is expressed in terms of $R_\alpha$ as
\be
\label{whThetanew}
\begin{split}
\widehat \Theta_\alpha(\tau,\bar \tau; L, \{V, V'\}
)&=\Theta_\alpha(\tau; L, V, V'\} )\\
&\quad +\sum_{\nu} \theta_{\alpha+\nu}(\tau;
L^\perp_{V})\, R_{\alpha+\nu}(\tau,\bar \tau;V^2,K_L.V) \\
&\quad  - \sum_{\nu'} \theta_{\alpha+\nu'}(\tau; L^\perp_{V'})\,
R_{\alpha+\nu'}(\tau,\bar \tau; (V')^2,K_L.V'),
\end{split}
\ee
with $\theta_\alpha$ as in (\ref{NormalTheta}). 

Returning to $R_{\bfmu,1, \ell}$, we can now write this as
\be
\label{Rmurhonew}
R_{\bfmu,1, \ell}(\tau,\bar \tau)=\sum_\rho \Theta_{\bfmu+\rho}(\tau, L_\ell^\perp,\{C_{\ell-1}, \, C_{\ell+1}\})\,R_{\bfmu+\rho}(\tau, \bar \tau; C_{\ell}^2,\bfK.C_{\ell}).
\ee
The relations for convergence of $\Theta_{\bfmu+\rho}$ are indeed satisfied for the vectors given in \refb{v123new}. We will continue to demonstrate that these are captured by the term $R_{\bfmu,2}$
in \refb{RPhi12}. We have
\be
\begin{split}
R_{\bfmu,2, (k, \ell)}(\tau,\bar \tau)&=-\sum_{\bfQ\in \underline
  \bfLambda_\bfmu^*} m_2(\sqrt{2\tau_2}\,u_1 (C_k, C_\ell, \bfQ),
\sqrt{2\tau_2}\,u_2 (C_\ell, \bfQ))\,(-1)^{\bfK\cdot \bfQ}\,q^{-\bfQ^2/2}\\
&=-\sum_{\bfQ\in \underline
  \bfLambda_\bfmu^*} u_1 (C_k, C_\ell, \bfQ) u_2(C_\ell, \bfQ) \,(-1)^{\bfK\cdot
  \bfQ}\,q^{(u_1(C_k, C_\ell, \bfQ)^2+u_2(C_\ell, \bfQ)^2-\bfQ^2)/2}\\
  &\quad \times\int_{-\bar \tau}^{i\infty} dw_2
\int_{w_2}^{i\infty} dw_1 \frac{e^{\pi i (w_1u_1(C_k, C_\ell, \bfQ)^2+w_2
    u_2(C_\ell, \bfQ)^2)}}{\sqrt{-(w_1+\tau)(w_2+\tau)}}. 
\end{split}
\ee
Using the splits of the lattices and $\Upsilon_\alpha$ (\ref{Upsilon}), we can express this as
\be
\label{Rmu2new}
\begin{split}
&R_{\bfmu,2, (k, \ell)}(\tau,\bar \tau)
=-\sum_\rho\theta_{\bfmu+\rho+\nu}(\tau;L_{k \ell}^\perp)\\
&\quad \times \,\int_{-\bar \tau}^{i\infty}
dw_2 \int_{w_2}^{i\infty}
dw_1\frac{\Upsilon_{\bfmu+\rho}(w_2; C_\ell^2,\bfK\cdot
  C_\ell)\,\Upsilon_{\bfmu+\rho+\nu}(w_1;C_{k \perp \ell}^2,\bfK\cdot
  C_{k \perp \ell})}{\sqrt{-(w_1+\tau)(w_2+\tau)}}.
\end{split}
\ee
Next, we aim to combine the sums in (\ref{Phihatnew}).
We recognize the two $\Upsilon$'s in the integrand in
(\ref{Rmu2new}). The first, $\Upsilon_{\bfmu+\rho}$, appears in the
integrand of $R_{\bfmu+\rho}$ on the rhs of \refb{Rmurhonew}, whereas
the second $\Upsilon_{\bfmu+\rho+\nu}$ matches with one of the
non-holomorphic terms on the rhs of (\ref{whThetanew}). More precisely
and concisely, we arrive at
\be
R^\Phi_\bfmu(\tau,\bar \tau)=\sum_{\ell=1,2, 3} i\int_{-\bar
  \tau}^{i\infty} dw \frac{\sum_\rho\widehat
  \Theta_{\bfmu+\rho}(\tau,-w; L^\perp_{\ell},\{C_{\ell-1}C_{\ell+1}\})\,\Upsilon_{\bfmu+\rho}(w; C_{\ell}^2,\bfK\cdot C_{\ell})}{\sqrt{-i(w+\tau)}}. \label{Rphifinalnew}
\ee
We thus have succeeded to express $R^\Phi_\bfmu$ in (\ref{defhatPhinew}) as an iterated
integral of modular forms.    
 
Note that for $\ell=1$, there is a symmetry exchanging $P_1\leftrightarrow
P_2$, and there are similar symmetries for $\ell=2$ and $\ell=3$. 
The form (\ref{Rphifinalnew}) makes the determination of the
anti-holomorphic derivative of $\widehat \Phi_\bfmu$ immediate. We
have,
\be 
\frac{\partial \widehat \Phi_\bfmu(\tau,\bar \tau)}{\partial \bar
  \tau}=\sum_{\ell=1,2,3} \sum_\rho\frac{\widehat
  \Theta_{\bfmu+\rho}(\tau,\bar \tau;
  L^\perp_{\ell},\{C_{\ell -1},C_{\ell+1}\})\,\Upsilon_{\bfmu+\rho}(-\bar
  \tau; C_{\ell}^2,\bfK\cdot C_{\ell})}{\sqrt{2\tau_2}}.
\ee
    
\subsection{Modular completion of $\Psi_\bfmu$}
\label{modcompPsi}
We will treat in this subsection the completion of
$\Psi_\bfmu$. Eq. (\ref{derPsi}) related the holomorphic $q$-series
$\Psi_\bfmu(\tau)$ to that the second derivative of the function
$\Psi_\bfmu(\tau,z)$. To determine the non-holomorphic completion
$\widehat \Psi_\bfmu(\tau,\bar \tau)$, we consider the non-holomorphic differential operator 
\be
\label{Jacdiffoper}
-\frac{1}{4\pi^2} \left(\frac{\partial^2}{\partial z^2} +\frac{2\pi m}{\tau_2}\right),
\ee
on the completion $\widehat \Psi_\bfmu(\tau,\bar \tau,z,\bar z)$ of
$\Psi_\bfmu(\tau,z)$ (\ref{DefPsiy}). The differential operator maps a Jacobi form of weight $k$ and index $m$ to a Jacobi form
of weight $k+2$ and index $m$. The second non-holomorphic term in (\ref{Jacdiffoper}) is
required for modularity but does is not relevant for the holomorphic part.

To determine the completion $\widehat \Psi_\bfmu(\tau,\bar \tau,z,\bar
z)$, we set $z=\beta \tau +\delta$ with $\beta, \delta\in \mathbb{R}$, so that $\beta={\rm
  Im}(z)/\tau_2$. We first note that completing the square gives for a
generic vector $V\in \underline \bfLambda$,
\be
y^{V.\bfQ}\,q^{-\frac{1}{2}\bfQ^2} = q^{\frac{\beta^2}{2}V^2-\frac{1}{2}(\bfQ-\beta V)^2} e^{2\pi i(V.\bfQ)\delta},
\ee
for an arbitrary vector $V$. To write the modular completion, we can treat the three permutations
separately. With $C_1$ as in (\ref{v123new}), and $C^{(123)}_\ell$ as
in \eqref{Cj123}, we find that the kernel $(-1)^{a+b-c} F^*(123)\,(y^{a+b-c}+y^{-a-b+c}) $ is to
be completed to
\be
\widehat F^*(123,y)=\frac{(-1)^{a+b-c}}{4}\sum_\pm\left[ 1+ \sum_{\ell=1}^3
  E_2(C^{(123)}_\ell,C^{(123)}_{\ell+1};\sqrt{2\tau_2}\, (\bfQ\mp
  \beta C_1))\right] y^{\pm (a+b-c)}.
\ee
Then the completion $\widehat \Psi_\bfmu(\tau,\bar \tau,z,\bar z)$ reads,
\be
\widehat \Psi_\bfmu(\tau,\bar \tau,z,\bar z)=\sum_{\bfQ\in \underline
  \bfLambda_\bfmu^*} [\widehat F^*(123,y) +\widehat F^*(213,y)+ \widehat F^*(132,y)]\,q^{-\bfQ^2/2}.
\ee
The completion $\widehat \Psi_\bfmu(\tau,\bar \tau,z,\bar z)$
transforms as a Jacobi form of weight $b_2/2$ and index $m_P$ given by,
\be
m_P=-\frac{1}{2}C_1^2=-P_1P_2P_3-\frac{1}{2}\sum_{\ell=1}^3 P_\ell\,(P_{\ell+1}^2+P_{\ell+2}^2).
\ee
This is symmetric under permutations, and thus also equal to $-C_2^2/2=-C_3^2/2$.

To write the modular completion, we define the function
\be
\label{delz2}
\begin{split}
G(V_1,V_2,V_3;\bfQ,\tau_2) &= -\frac{1}{4\pi^2 }\partial_z^2
\left(E_2(V_1,V_2; \sqrt{2\tau_2}\,(\bfQ-\beta
  V_3))\,y^{V_3.\bfQ}\right)|_{z=0} \\
&= (V_3.\bfQ)^2 E_2(V_1,V_2; \sqrt{2\tau_2}\,\bfQ)\\
&\quad - \frac{1}{4\pi^2}\partial_z^2 E_2(V_1,V_2;\sqrt{2\tau_2}\,(\bfQ-\beta V_3))|_{z=0}\\ 
&\quad +\frac{1}{2\pi i} (V_3.\bfQ)\,\partial_z E_2(V_1,V_2;\sqrt{2\tau_2}\, (\bfQ-\beta V_3))|_{z=0}.
\end{split}
\ee 
The limit $\tau_2\to \infty$ is determined by the first term,
\be
\lim_{\tau_2\to \infty}
G(V_1,V_2,V_3;\bfQ,\tau_2)=\left\{ \begin{array}{rl}
    (V_3.\bfQ)^2\,\arctan(\alpha), & \quad {\rm if}\,\, V_1.\bfQ=V_2.\bfQ=0,\\ 
    (V_3.\bfQ)^2\,\sgn(V_1.\bfQ)\,\sgn(V_2.\bfQ), & \quad {\rm otherwise}, \end{array} \right.
\ee
with $\alpha$ as in (\ref{alphauC}). 

To write the modular completion, we can treat the three permutations
separately. We set
\be
\widehat G^*(123)=\frac{(-1)^{a+b-c}}{4}\left((a+b-c)^2+ \sum_{\ell=1}^3\,G(C^{(123)}_\ell,C^{(123)}_{\ell+1},C_1;\bfQ,\tau_2)\right),
\ee
with $C_1$ as in (\ref{v123new}), and $C^{(123)}_\ell$ as in \eqref{Cj123}
The other permutations $(213)$ and $(132)$ give similarly rise to
$\widehat F^*(213)$ and $\widehat F^*(132)$. We define, 
\be
\begin{split}
\widehat g_C(\{\gamma_j\};\{c_j^*\})&=\frac{1}{4}\left[\widehat
  G^*(123)+\widehat G^*(213)  + \widehat G^*(132)\right].
\end{split}
\ee
The modular completion $\widehat \Psi_\bfmu$ now reads
\be
\label{completePsi}
\widehat \Psi_\bfmu(\tau,\bar \tau)=\sum_{\bfQ\in \underline
  \bfLambda_\bfmu^*}\widehat
g_C(\{\gamma_j\};\{c_j^*\})\,q^{-\bfQ^2/2}-\frac{m_P}{2\pi
  \tau_2}\,\widehat \Phi_\bfmu(\tau,\bar \tau).
\ee
where the last term is due to the $m$-dependent term in
(\ref{Jacdiffoper}). We let this expression be our final form for
$\widehat \Psi_{\bfmu}(\tau,\bar \tau)$.
 
Our last task is to determine the constant $A$ in
$g_C(\{\gamma_j\};\{c_j^*\})$  (\ref{defgC}). To this end, we consider
the $\tau_2 \to \infty$ limit of $\widehat
g_C(\{\gamma_j\};\{c_j^*\})$ and require that it reduces to
$g_C(\{\gamma_j\};\{c_j^*\})$. If we 
consider the term $F^*(123)\,(a+b-c)^2$, we see that only the
completion of the term $\sgn(a-c)\,\sgn(b-c)$ in $g_C$ can contribute a non-vanishing
difference. Indeed, if we include the other permutations, the only remaining terms in the difference are the equilateral cases with
$a=b=c$,\footnote{We stress that the equilateral condition $a=b=c$, is
 the condition on $\bfQ$ to satisfy $C_a.\bfQ=C_b.\bfQ=C_c.\bfQ$, and does {\it not} imply
  equalities among $C_a$, $C_b$ and $C_c$.}
\be
\begin{split}
&\lim_{\tau_2\to \infty}\widehat g_C(\{\gamma_j\};\{c_j^*\})-g_C(\{\gamma_j\};\{c_j^*\})=\frac{(-1)^a}{4}\,a^2\,\delta_{a,b}\,\delta_{b,c} \\
&\qquad \times \lim_{\tau_2\to \infty} \left(E_2(C_a-C_c, C_b-C_c;\sqrt{2\tau_2}\,\bfQ)+E_2(C_a-C_b,
C_c-C_b;\sqrt{2\tau_2}\,\bfQ)\right.\\
&\qquad \left.+E_2(C_c-C_a, C_b-C_a;\sqrt{2\tau_2}\,\bfQ)-A)\right).
\end{split}
\ee
Now the sum of $E_2$'s is precisely of the form (\ref{E2iden}), and thus equals
1. As a result, we find that with $A=1$ the limit vanishes for any choice of Calabi-Yau and
charge configurations. This matches perfectly with the physical
expectation. Note that the individual values for $E_2$ are given by an
arctan, and are generically irrational, but that this combination adds
up to 1.

It is quite striking that the values of $E_2$'s precisely confirm the physical expectation.
Also in other cases \cite{Manschot:2017xcr, Alexandrov:2019rth}, the values of the generalized error
functions for vanishing arguments has matched with the expectations
based on BPS invariants.
  
\section{Relation to M5-branes and AdS$_3$/CFT$_2$}
\label{sec:Mtheory}
In this brief section, we discuss our findings from the point of view
of M-theory and the MSW CFT. We discuss how the partition functions $\widehat
Z_P^{nT}$, $n>2$, may arise from the 2-dimensional perspective.
  
The uplift of the D4-branes to M5-branes in
M-theory is useful to understand the modular properties of the partition functions \cite{Denef:2007vg, Maldacena:1997de, Gaiotto:2006wm, deBoer:2006vg}. The spatial dimensions of the M5-brane are $P \times S^1_M$, where
$P \in H_4(X, \mathbb{Z})$ is a four-cycle in the Calabi-Yau threefold and $S^1_M$ is the
M-theory circle. The D2 branes of IIA string theory are realized as excitations of the
self-dual 2-form field on M5 brane world volume, while the D0-branes
are realized as momenta of the brane system along $S^1_M$ with radius
$R=g_s\,\ell_s=\ell_{11}^3/\ell_s^2$ in terms of the string coupling
$g_s$, string length $\ell_s$, and eleven dimensional Planck length $\ell_{11}$. The
world volume theory of the M5 brane gives a low energy description of
the system, provided gravitational effects can be ignored. This can be
ensured by taking the volume of the CY$_3$ to be large, namely $V_X/\ell_{11}^6$
large but fixed. Supersymmetry of the M5 world volume theory implies
that the MSW CFT has (0,4) supersymmetry, and
is dual to the near horizon AdS$_3$ geometry \cite{Maldacena:1997de, Minasian:1999qn}. Bosons
of this CFT include, moduli of the divisor $P$ inside $X$,
translations along flat $\mathbb{R}^3$ and (anti)-chiral scalars
coming from reduction of self-dual two-form field in the M5 brane
world volume. The number of fields can be determined using geometric
data of $X$, when $P$ is a very ample divisor. Consequently the
central charges can be determined. The central charge of the left-moving, non-supersymmetric sector is
\be
\label{CFTccharge}
c_L=P^3+c_2(X)\cdot P,
\ee
where $c_2(X)$ is the second Chern class of the Calabi-Yau three-fold $X$.
Using Cardy's formula, Reference \cite{Maldacena:1997de} demonstrated that microscopic entropy is in
agreement with one loop corrected Bekenstein-Hawking entropy. 

As for any CFT, a key feature of the MSW CFT is the modular
symmetry. When time direction is
compactified, the CFT lives on a 2-torus $T^2=S^1_M \times S^1_t$. The
linear fractional transformation of the complex structure modulus
$\tau$ of $T^2$ by an
element of $SL(2, \mathbb{Z})$ corresponds to the same torus $T^2$. Thus the $SL(2, \mathbb{Z})$ symmetry, or
modular symmetry, appears as a symmetry of the CFT. This is a strong
constraint on the degeneracies. Another property of the CFT is the
spectral flow of the $U(1)^{b_2}$ current algebra. This leads to a
symmetry of degeneracies as function of the charges, such that the
partition function can be decomposed as a finite sum of theta series
$\times$ modular functions, much as is the case for the attractor
partition function in Section \ref{AttrPart}.

We proceed by considering the AdS$_3$ dual to the MSW
CFT following \cite{deBoer:2008fk}. The relation of $\lambda$ in 
(\ref{LVattractor}) to the
five-dimensional quantities can be understood from four-dimensional
Newton's constant. From the IIA perspective, we have
\be
G_4=\frac{g_s^2 \,\ell_s^8}{V_X}=\frac{g_s^2\,\ell_s^2}{\lambda^3},
\ee
while in terms of the five-dimensional quantities,
\be
G_4=\frac{\ell_5^3}{R}=g_s^2\,\ell_s^2\frac{\ell_5^3}{R^3},
\ee
such that one has,
\be
\label{lambdaell}
\lambda=\frac{R}{\ell_5}.
\ee
Starting from five dimensional asymptotically flat geometries
with appropriate charges, one takes the decoupling limit as follows:  
size of M-theory circle $R$ is kept fixed (in absolute units),
Calabi-Yau volume is kept finite in units of five dimensional Planck
length $\ell_5$. In the decoupling limit
to AdS$_3$, $\ell_5 \rightarrow 0$, hence $\lambda \to \infty$ by
(\ref{lambdaell}).
Multi-centered
geometries where centers have mutual distance of the order of
$\ell_5^3\sim \lambda^{-3}$ or less, survive this limit and go over to asymptotically
AdS$_3 \times S^2$ geometries, with asymptotic moduli fixed to their
attractor values. These are the $\lambda$-core geometries mentioned in
the introduction, and include centers with non-vanishing
individual D6 brane charges, which add up to zero. On the other hand,
if the distances between the centers is larger than $\sim\ell_5^3/R^2$ for
$\ell_5\to 0$, the bound states decouple from the spectrum. As we can see from the Denef
equations (\ref{Deqs}), this is for example the case
if the centers carry a non-vanishing D4-brane charge with
vanishing D6-brane charge. Then the distances between the centers
scale as $\ell_5$ for non-scaling bound states, and the centers give rise to disconnected AdS$_3$
throats in the decoupling limit. On the other hand for scaling solutions, the distance between the centers contains a regime where the centers can come
arbitrarily close.
 
As explained in Section \ref{BHsols}, the BPS index can be determined using localization with respect to
rotation around the $z$-axis, and we can therefore concentrate on
collinear solutions to Denef's equations \cite{Manschot:2011xc}. These collinear solutions admit two branches, one corresponding to centers at
finite distances, and the other one is when the centers are nearly coincident. The
second branch, sometimes referred to as ``deep scaling regime"
reproduces pure Higgs states and goes to a single throat in $\ell_5
\rightarrow 0$ limit \cite{Beaujard:2021fsk}.   
This is in accordance with the expectation that MSW CFT captures the
near coincident regime of scaling solutions since these are smoothly
connected to the single center black hole. The separation between the
centers at the collinear fixed point is of order $\lambda^{-1}\sim \ell_5$, and
these therefore decouple from the AdS$_3$ geometry. As a result, these do not appear to be captured by the CFT. This leads us to
speculate that the first term on the rhs of (\ref{CZlambdaDec}),
$\widehat \CZ_P^T$, corresponds to the AdS$_3$/CFT$_2$ partition
function while the other terms on the rhs do not.    
 
It is intriguing that the terms due to scaling solutions $\widehat
\CZ_P^{nT}$ with $n\geq 3$ do satisfy the restrictive modular
transformations as well as the spectral flow symmetry, and it is desirable to understand the origin. We
think that these terms can appear after
turning on an irrelevant deformation in the $(0,4)$ CFT, away from the conformal fixed
point and reversing the attractor
flow and $\ell_5\to 0$ limit. While such a deformation
does not lead to a finite change in the partition function for $(4,4)$ CFT
\cite{deBoer:2008ss}, it seems plausible to us that this can happen with reduced
supersymmetry. 
This deformation is distinguished among other deformations spanning
 the space of attractor moduli, since this deformation does preserve the spectral
flow symmetry which is in general not the case for variations of the
K\"ahler moduli orthogonal to $P$ \cite{Manschot:2009ia}. We leave a further exploration of these interesting
aspects for future work.

\section{Case studies} \label{sexample}
To make a suitable choice of a Calabi-Yau threefold, we note that the
lattices involved have dimensions linear in $b_2$, the second Betti
number. Thus the computations are less involved for smaller $b_2$. However for the simplest case $b_2=1$, there are no
cyclic quivers and hence no scaling solutions.\footnote{Consider triangular quiver to start with. For
 $b_2=1$, electric and magnetic charges are numbers and satisfy the identity $a P_3 + b P_1 + c P_2 =0$. Since $P_i >0$ this implies $a,b,c$ can not all have the same sign hence the quiver can not be cyclic. This is easily generalized to any cyclic quiver.} Also, the lattices concerned are positive
definite, and therefore do not give rise to mock modular forms. So we settle for the next
simplest case $b_2=2$. 

In order to define quadratic forms on the lattices, one needs the
intersection numbers. For a Calabi Yau threefold with $b_2=2$, there
are $2^3=8$ intersection numbers of which only 4 are independent. For the sake of simplicity we shall
choose a Calabi Yau threefold $X$, for which all intersection numbers are $\in
2\mathbb{Z}$ such that Simplification 2 of Section
\ref{BoStaLat} holds, and for which most of these numbers
vanish. To be specific, we choose $X$ to be a $K3$ fibration
over $\mathbb{P}^1$.  
This Calabi-Yau manifold corresponds to Polytope ID \# =14 in the online database \cite{CYdatabase}.
$X$ can be realized as a divisor in an ambient toric variety
$\mathbb{P}^3\times \mathbb{P}^1$, specified by the weight matrix
\begin{align}
W &=  
\begin{pmatrix}
0 & 0 & 0 & 1 & 1 & 0 \\
1 & 1 & 1 & 0 & 0 & 1
\end{pmatrix} \, .
\end{align}
$W$ defines the toric variety through the identification on $\mathbb{C}^6/\{0\}$
\begin{align}
(z_1, \dots , z_6) &\sim (\lambda^{W_{i1}} z_1, \dots, \lambda^{W_{i6}} z_6) , \, \, \lambda \in \mathbb{C}^\star, \, \, i=1,2 \, ,
\end{align}
where $z_1, \dots , z_6$ are coordinates of $\mathbb{C}^6$.
Anti-canonical divisor of this toric variety defines the Calabi Yau threefold under consideration. It has the following Hodge numbers
\begin{align}
h^{1,1} &= 2,\quad  \, h^{2,1}= 86 \, 
\end{align}
and consequently the Euler number is $\chi = 2 (h^{1,1} - h^{2,1}) = - 168$.
Intersection polynomial and second Chern class of $X$ are given respectively by  
\be
\begin{split}
\text{intersection polynomial} &= 4 J_1 J_2^2 + 2 J_2^3 , \, \\
\text{second Chern class} &= 8J_1  J_2+6 J_2^2 \, , 
\end{split}
\ee
where $J_i$'s are a basis of 2-forms on $X$ and generators of the
K\"ahler cone.  $X$ is favorable, meaning that K\"{a}hler forms on $X$ descend from the ambient toric variety.
In our notation, the intersection numbers $d_{abc}$ and second Chern class $c_{2,a}$ read
\be
\begin{split}
& d_{111}=d_{112}=0,\qquad d_{122}=4,\qquad d_{222}=2,\\
& c_{2,1}=24,\qquad c_{2,2}=44
\end{split}
\ee
and the $d_{abc}$ for other indices follow by permutation.

\subsection{Charge configuration 1: $P_1=P_2=(0,1),\; P_3=(1,1)$}
\label{CC1}
In this example we choose the total magnetic charge $P=(1,3)$ magnetic charges split into 3 centers as
\begin{eqnarray}
P_1=P_2=(0,1),\quad P_3=(1,1).
\end{eqnarray}
We note that $P_{1,2}$ are irreducible while $P_3$ is
reducible. Therefore $h_{P_{1,2},\mu}$ is a weakly holomorphic modular
form, while $h_{P_{3},\mu}$ is a mock modular form. The corresponding central charges of the left-moving sector of the CFT (\ref{CFTccharge}) are,
\be
c_L(P_{1,2})=46,\qquad c_L(P_{3})=92,\qquad c_L(P)=318.
\ee
For the $j$-th center, we have for the innerproduct $D_j=d_{abc}P_j^c$,
\begin{eqnarray}
D_1=D_2=\begin{pmatrix}
0 & 4 \\
4 & 2
\end{pmatrix},\quad D_3=\begin{pmatrix}
0 & 4\\
4 & 6
\end{pmatrix}.
\end{eqnarray}
For this choice, $D_3^{-1}D_1=D_3^{-1}D_2$ is an integral matrix, such
that Simplication 1. of Section \ref{BoStaLat} is satisfied, and we
can use the results of that section. We have,
$D_1D_3^{-1}D_1=\begin{pmatrix}
0 & 4\\
4 & -2
\end{pmatrix}$
which produces the quadratic form of $\underline \bfLambda$ (\ref{underD}),
\begin{eqnarray}
\underline D &=& \begin{pmatrix*}[c]
0 & 8 & 0 & 4\\
8 & 0 & 4 & -2\\ 
0 & 4 & 0 & 8\\
4 & -2 & 8 & 0
\end{pmatrix*},\quad \det{\underline D}=2304.
\end{eqnarray}
The quadratic form on the lattice of electric charges $\underline{D}^{-1}$ is given by,
\begin{eqnarray}\label{dinv}
\underline{D}^{-1}=\left(
\begin{array}{cccc}
 -\frac{1}{18} & \frac{1}{6} & \frac{5}{72} & -\frac{1}{12} \\
 \frac{1}{6} & 0 & -\frac{1}{12} & 0 \\
 \frac{5}{72} & -\frac{1}{12} & -\frac{1}{18} & \frac{1}{6} \\
 -\frac{1}{12} & 0 & \frac{1}{6} & 0 \\
\end{array}
\right).
\end{eqnarray}
Electric charge vectors $Q_j\in \Lambda_j^*$ have the following form
\[Q_j= \begin{pmatrix} q_{j1} \\ q_{j2} \end{pmatrix}.\] There are $\det(D_1)\,\det(D_2)=256$ conjugacy classes of the form
\begin{eqnarray}
q_{11}&=&4k_{12}+\mu_{11},\quad q_{12}=4k_{11}+2k_{12}+\mu_{12},\\ \nn
q_{21}&=&4k_{22}+\mu_{21},\quad q_{22}=4k_{21}+2k_{22}+\mu_{22},
\end{eqnarray}
where $\mu_{ij}\in\{0,1,2,3\}$, $k_{ij}\in\mathbb{Z}$. These conjugacy
classes have the exchange symmetry between $\mu_1,\mu_2$. Since $D_3$
divides $D_1$ and $D_2$, $N_q=\det(D_1)\,\det(D_2)$ (\ref{SimpNgNq}) is the
number of conjugacy classes of $\underline \bfLambda^*/\underline \bfLambda$ for fixed $\mu_1+\mu_2+\mu_3=\mu\in
\Lambda^*/\Lambda$, since the class $\mu_3\in 
\Lambda_3^*/\Lambda_3$ by the simplification. Using the equations in (\ref{bfqq}), we can write
\begin{eqnarray}
\bfQ_1 &=& \begin{pmatrix}
8k_{12}+4k_{22}+\mu_{11}-\mu_{31}\\
8k_{11}+4k_{21}-2k_{22}+\mu_{12}+\mu_{31}-\mu_{32}
\end{pmatrix},\\ \nn
\bfQ_2 &=& \begin{pmatrix}
8k_{22}+4k_{12}+\mu_{21}-\mu_{31}\\
8k_{21}+4k_{11}-2k_{12}+\mu_{22}+\mu_{31}-\mu_{32}
\end{pmatrix}.
\end{eqnarray}

We restrict to $\mu=0=\mu_1+\mu_2+\mu_3$, such that
\begin{eqnarray}\nn
\bfQ_1 &=& \begin{pmatrix}
8k_{12}+4k_{22}+2\mu_{11}+\mu_{21}\\
8k_{11}+4k_{21}-2k_{22}+2\mu_{12}+\mu_{22}-\mu_{11}-\mu_{21}
\end{pmatrix}, \quad \bfmu_1=\begin{pmatrix}
2\mu_{11}+\mu_{21}\\ \nn
2\mu_{12}+\mu_{22}-\mu_{11}-\mu_{21}
\end{pmatrix},\\
\bfQ_2 &=& \begin{pmatrix}
8k_{22}+4k_{12}+2\mu_{21}+\mu_{11}\\
8k_{21}+4k_{11}-2k_{12}+2\mu_{22}+\mu_{12}-\mu_{11}-\mu_{21}\end{pmatrix}, \quad \bfmu_2=\begin{pmatrix}
2\mu_{21}+\mu_{11}\\ \nn
2\mu_{22}+\mu_{12}-\mu_{11}-\mu_{21}
\end{pmatrix}.
\end{eqnarray}

\subsection*{List of $\Phi_\bfmu$}
We next present explicit $q$-expansions of $\Phi_\bfmu$ for a various
choices of $\bfmu$. All the convergence conditions of the types
$C_i^2C_{i+1}^2-(C_i.C_{i+1})^2>0$ and
$C_i^2(C_{i-1}.C_{i+1})-(C_i.C_{i-1}).(C_i.C_{i+1})<0$ are satisfied
in this example along with $C_j^2>0$. 
As in explained in Section \ref{BoStaLat}, $\bfmu$ is fully determined by specifying $(\mu_1,\mu_2,\mu_3)$ and $\mu=\mu_1+\mu_2+\mu_3$. 
In the following example, we fix $\mu=\mu_1=0$ and $\mu_3=-\mu_2$, and list $\Phi_\bfmu$ for 16 different values of $\mu_2$. Out of this set, 10 give rise to different $q$-series. The overall sign of the $q$-expansion is given by
$(-1)^{(a+b+c)}=(-1)^{\mu_{11}-\mu_{21}}=(-1)^{\mu_{21}}$. Furthermore,
the growth of the coefficients is moderate, $|c(n)| \leq C\,n^\alpha$ for
some positive constants $\alpha$ and $C$, which is expected for a
holomorphic (mock) modular form of weight 2.

It can be shown that in most of the sectors we study ($\mu_1=(0,0)$) there is no contribution to the holomorphic part when the arguments of $E_2$ are both zero.
When $\mu_2=(1,j)$ or $\mu_2=(3,j)$ $b$ is odd while $a,c$ are both
odd or both even. Under these conditions at least two of $(a+b-c),\;
(a-b+c),\; (c+b-a)$ cannot vanish. Only in the conjugacy classes
$\mu_2=(0,0),(0,2),(2,3)$ we find integral
$k_{11},k_{12},k_{21},k_{22}$ to satisfy at least two of $(a+b-c),\;
(a-b+c),\; (c+b-a)$ to vanish. The contributions $A_\ell$ are given by
\be
\begin{split}
A_1&= \frac{2}{\pi}\,\arctan(-5/\sqrt{11}),\\
A_2&=A_3= \frac{2}{\pi}\,\arctan(-1/\sqrt{8}),
\end{split}
\ee
which are all irrational numbers.

For $\mu_2=(0,0)$, we have:
\begin{eqnarray}\nn
\Phi_\bfmu &=& 2q^6+4q^{20}+6q^{24}+4q^{30}+4q^{44}+4q^{50}+8q^{52}+2q^{54}+12q^{56}+4q^{60}\\ \nn
&+& 4q^{64}+4q^{68}+4q^{70}+12q^{80}+2q^{88}+8q^{90}+8q^{92}+8q^{94}
+14q^{96}+16q^{100}\\ \nn
&&+\frac{(A_1+A_2+A_3+1)}{2}+ (A_1+A_2+A_3+3)(q^8+q^{32}+q^{72})+\dots
\end{eqnarray}
For $\mu_2=(0,1)$, we have:
\begin{eqnarray}\nn
\Phi_\bfmu &=& q^5\left(1+q+2q^{12}+2q^{15}+2q^{17}+2q^{20}+q^{22}+2q^{23}+q^{27}+3q^{28}+q^{30}+2q^{35}+2q^{38}\right. \\ \nn
&+& \left. 3q^{40}+ 2q^{42}+ 4q^{45}+3q^{46}+3q^{49}+4q^{51}+2q^{52}+2q^{53}+3q^{54}+2q^{55}+2q^{57}+
3q^{60}\right.\\ \nn
&+& \left. 2q^{61}+ 2q^{63}+ q^{64}+ 2q^{65}+5q^{69}+6q^{70}+2q^{71}+2q^{73}+3q^{75}+2q^{78}+2q^{79}+2q^{81}\right. \\ \nn
&+& \left. q^{82} + 2q^{83}+6q^{84}+ 2q^{86}+3q^{87}+4q^{88}+5q^{90}+4q^{91}+2q^{92}+6q^{93}+2q^{94}+2q^{95}\right)+\dots\\ \nn
\end{eqnarray}
For $\mu_2=(0,2)$, we have:
\begin{eqnarray}\nn
\Phi_{\bfmu} &=& 2q^{14}+2q^{20}+2q^{24}+2q^{26}+2q^{32}+6q^{36}+4q^{40}+4q^{42}+2q^{44}+4q^{48}\\ \nn
&+& 10q^{56}+6q^{60}+4q^{62}++4q^{68}+4q^{70}+8q^{72}+8q^{76}+10q^{80}+4q^{82}+4q^{86}\\ \nn
&+& 8q^{88}+4q^{92}+4q^{96}+6q^{98}+8q^{100}+(A_1+1)(q^{50}+q^{54}+q^{62}+q^{74}+q^{90})+\dots
\end{eqnarray}
For $\mu_2=(1,0)$, we have:
\begin{eqnarray}\nn
\Phi_{\bfmu} &=& -2q^{29/4}(1+q^4+q^{12}+q^{17}+q^{18}+q^{19}+2q^{21}+q^{25}+q^{31}+q^{37}+2q^{40}+q^{42}+q^{43}\\ \nn
&+&  2q^{45}+q^{48} +2q^{49}+3q^{50}+q^{52}+q^{53}+2q^{54}+q^{55}+q^{56}+q^{57}+2q^{60}\\ \nn
&+& q^{66}+2^{71}+2q^{74}+q^{75}+q^{76}+3q^{79}+q^{80}+2q^{81}+q^{82}+3q^{84}+q^{85}+q^{86}\\ \nn
&+& 3q^{87}+2q^{90}+3q^{91}+3q^{92})+\dots
\end{eqnarray}
For $\mu_2=(1,1)$, we have:
\begin{eqnarray}\nn
\Phi_{\bfmu} &=&-2 q^{83/4}\left(1 +q^3+q^5+q^8+q^9+q^{10}+q^{13}+q^{16}+q^{22}+q^{23}+q^{24}
+q^{26}+3q^{31}\right.\\ \nn
&+& \left. 2q^{33}+q^{34}+q^{35}+ q^{36}+2q^{38}+q^{39}+q^{40}+q^{41}+2q^{42}+q^{44}+q^{45}+q^{46}+q^{49}
+2q^{51}\right.\\ \nn
&+&\left. 2q^{54}+ q^{57}+q^{58}+2q^{59}+2q^{60}+2q^{61}+ q^{68}+q^{69}+q^{70}+2q^{71}+4q^{72}+3q^{72}+ q^{74}\right.\\ \nn
&+&\left.+2q^{75} + q^{76}+ q^{77}+ 3q^{78}+ q^{79}\right)+\dots
\end{eqnarray}
For $\mu_2=(1,2)$, we have:
\begin{eqnarray}\nn
\Phi_{\bfmu} &=& -2 q^{45/4} (1+2q^8+q^{15}+q^{18}+q^{21}+q^{23}+2q^{27}+q^{28}+q^{29}+q^{32}+2q^{35}+q^{39}+q^{41}\\ \nn
&+& 2q^{43}+q^{44}+ q^{46}+2q^{48}+q^{50}+q^{52}+q^{54}+q^{56}+3q^{57}+q^{58}
+2q^{62}+3q^{63}+3q^{65}\\ \nn
&+&  q^{66}+ q^{68}+q^{70}+2q^{72}+q^{73}+2q^{74}+q^{75}+3q^{80}
+q^{82}+2q^{85}+q^{86}+  q^{87}+ q^{88})+\dots\\ \nn
\end{eqnarray}
For $\mu_2=(1,3)$, we have:
\begin{eqnarray}\nn
\Phi_{\bfmu} &=& -2 q^{15/4} (1+q+q^{10}+q^{13}+q^{17}+q^{18}+q^{19}+q^{21}+q^{22}+2q^{32}+q^{35}
+q^{39}\\ \nn
&+&  2q^{40}+ q^{42}+q^{45}+q^{47}+q^{50}+2q^{51}+q^{53}+3q^{54}+q^{56}+2q^{59}
+q^{63}+q^{65}+2q^{66}\\ \nn
&+&  2q^{68}+q^{71}+q^{72}+q^{74}+q^{75}+q^{76}+3q^{77}+q^{79}+2q^{80}
+2q^{83}+q^{84}+2q^{85}+q^{86}\\ \nn
&+&  q^{87}+q^{88}+2q^{90}+2q^{92}+3q^{93}+2q^{95})+\dots\\ \nn
\end{eqnarray}
For $\mu_2=(2,0)$, we have:
\begin{eqnarray}\nn
\Phi_{\bfmu} &=& 2(q^9+q^{15}+q^{19}+2q^{27}+2q^{29}+q^{33}+q^{35}+q^{43}+4q^{45}+4q^{53}+2q^{55}+q^{57}+q^{59}\\ \nn
&+& 2q^{61}+2q^{63}+3q^{65}+q^{67}+2q^{71}+q^{75}+4q^{77}+2q^{81}+3q^{83}+2q^{87}+q^{89}+5q^{91}\\ \nn
&+& q^{93}+6q^{95}+q^{97}+2q^{99}) +\dots
\end{eqnarray}
For $\mu_2=(2,1)$, we have:
\begin{eqnarray}\nn
\Phi_{\bfmu} &=& 2q^{25}\left(2+2q+q^2+q^4+2q^{15}+q^{16}+2q^{18}+q^{22}+2q^{25}+2q^{29}+4q^{32}+3q^{34}+4q^{37}\right.\\ \nn
&+&\left. 2q^{39}+ q^{40}+2q^{41}+q^{44}+q^{46}+2q^{51}+2q^{55}+q^{56}+2q^{59}+4q^{61}+3q^{62}+2q^{65}+2q^{66}\right.\\ \nn
&+& \left. 2q^{68}+4q^{70}+2q^{73}+2q^{74}+2q^{75}\right)+\dots
\end{eqnarray}
For $\mu_2=(2,3)$, we have:
\begin{eqnarray}\nn
\Phi_{\bfmu} &=& 2q^5\left(1+q^6+q^9+q^{15}+q^{19}+2q^{20}+q^{27}+q^{28}+q^{33}+2q^{34}+q^{37}+q^{40}+q^{42}+2q^{47}+3q^{51}\right.\\ \nn
&+& \left. q^{54}+2q^{55}+2q^{56}+q^{60}+q^{63}+2q^{64}+q^{66}+q^{69}+2q^{70}+4q^{71}+2q^{73}+q^{75}+2q^{78}+3q^{81}\right.\\ \nn
&+& \left. 2q^{84}+2q^{85}+2q^{88}+4q^{90}+3q^{91}+2q^{92}+3q^{94}\right)+(A_3+1)\; q^2(1+q^{16}+q^{48}+q^{96})+\dots
\end{eqnarray}

\subsection*{List of $\Psi_{\bfmu}$}
Similarly to the $\Phi_{\bfmu}$, we list the $q$-expansion of $\Psi_\bfmu$ for a various choices of $\bfmu$. In these examples, we keep $\mu=\mu_1=0$ and $\mu_3=-\mu_2$, and list $\Psi_\bfmu$ for 16 different values of $\mu_2$. Out of these 16 only 10 give rise to different $q$-series. The factor
$(-1)^{(a+b+c)}=(-1)^{\mu_{11}-\mu_{21}}=(-1)^{\mu_{21}}$ fixes the
overall sign of the whole series. Comparison of the $\Phi_{\bfmu}$ and
$\Psi_{\bfmu}$ demonstrates that for small powers of $q$, scaling
solutions can be present kinematically, but that the number of bound
states vanishes quantum mechanically. We again observe a moderate,
polynomial growth of the coefficients, as expected for a (mock)
modular form of weight 4.

For $\mu_2=(0,0)$, we have:
\begin{eqnarray}\nn
\Psi_{\bfmu} &=& 16q^{30}\left(1+2q^{22}+4q^{34}+q^{40}+2q^{60}+8q^{62}+2q^{64}+4q^{70}\right)+\dots
\end{eqnarray}
For $\mu_2=(0,1)$, we have:
\begin{eqnarray}\nn
\Psi_{\bfmu} &=& 2q^{17}\left(1+q^5+q^8+q^{11}+q^{16}+q^{23}+q^{26}+q^{28}+4q^{30}+q^{33}+q^{34}+9q^{37}+5q^{39}+q^{40}\right.\\ \nn
&+&\left. q^{41}+q^{42}+4q^{43}+ 4q^{45}+4q^{48}+q^{49}+9q^{51}+q^{53}+q^{57}+6q^{58}+q^{59}+q^{61}+q^{63}+9q^{66}\right.\\ \nn
&+&\left. 4q^{67}+ 16q^{69}+q^{71}+ 5q^{72}+16q^{74}+q^{75}+13q^{76}+26q^{78}+q^{79}+q^{80}+13q^{81}+q^{82}+q^{83}\right)\\ \nn
&&+\dots\\ \nn
\end{eqnarray}
For $\mu_2=(0,2)$, we have:
\begin{eqnarray}\nn
\Psi_{\bfmu} &=& 16q^{36}(1+q^4+q^{12}+2q^{20}+q^{24}+q^{32}+q^{34}+2q^{36}+2q^{40}+q^{44}+q^{46}+2q^{52}+4q^{56}\\ \nn
&+& 4q^{62}+2q^{64})+\dots
\end{eqnarray}
For $\mu_2=(1,0)$, we have:
\begin{eqnarray}\nn
\Psi_{\bfmu} &=& -\frac{q^{29/4}}{2}\left(1+q^4+q^{12}+q^{17}+q^{18}+q^{19}+10q^{21}+9q^{25}+q^{31}+q^{37}+10q^{40}+q^{42}+q^{43} \right.\\ \nn
&+& \left. 34q^{45}+q^{48}+2q^{49}+27q^{50} +
 q^{52}+q^{53}+50q^{54}+q^{55}+q^{56}+q^{57}+58q^{60}+9q^{66}+2q^{71}\right.\\ \nn
 &+& \left. q^{74}+9q^{75}+q^{78}+ 51q^{79}+25q^{80}+10q^{81}+q^{82}+107q^{84}+25q^{85}+49q^{86}+43q^{87}+26q^{90}\right. \\ \nn
&+& \left. 107q^{91}+51q^{92}\right)+\dots
\end{eqnarray}
For $\mu_2=(1,1)$, we have:
\begin{eqnarray}\nn
\Psi_{\bfmu} &=& -\frac{q^{83/4}}{2}\left(9+9q^3+q^5+q^8+q^9+q^{10}+9q^{13}+q^{16}+9q^{22}+q^{23}+9q^{24}+q^{26}+19q^{31}\right. \\ \nn
&+& \left. 34q^{33}+9q^{34}+25q^{35}+9q^{36}+26q^{38}+9q^{39}+25q^{40}+q^{41}+10q^{42} +25q^{44}+9q^{45}\right. \\ \nn
&+& \left. 25q^{46}+q^{49}+2q^{51}+10q^{54}+9q^{57}+9q^{58}+2q^{59}+18q^{60}+34q^{61}+q^{68}+9q^{69}+49q^{70}\right.\\ \nn
&+&\left. 74q^{71}+140q^{72}+131q^{73}+q^{74}+18q^{75}+q^{76}+49q^{77}+43q^{78}+25q^{79}\right)+\dots
\end{eqnarray}
For $\mu_2=(1,2)$, we have:
\begin{eqnarray}\nn
\Psi_{\bfmu} &=& -\frac{q^{45/4}}{2}\left(1+2q^8+q^{15}+q^{18}+9q^{21}+9q^{23}+2q^{27}+q^{28}+9q^{29}+q^{32}+18q^{35}+9q^{39} \right. \\ \nn
&+& \left. 9q^{41}+26q^{43}+ 9q^{44}+25q^{46}+10q^{47}+9q^{50}+q^{52}+9q^{54}+9q^{56}+19q^{57}+25q^{58}+10q^{62}\right. \\ \nn
&+& \left. 75q^{64}+43q^{65}+9q^{66}+9q^{68}+q^{70}+10q^{72}+q^{73}+18q^{74}+q^{75}+131q^{80}+49q^{82}\right. \\ \nn
&+& \left. 50q^{85}+44q^{86}+49q^{87}+25q^{88}\right)+\dots
\end{eqnarray}
For $\mu_2=(1,3)$, we have:
\begin{eqnarray}\nn
\Psi_{\bfmu} &=& -\frac{q^{15/4}}{2}\left(1+q+q^{10}+q^{13}+q^{17}+q^{18}+q^{19}+9q^{21}+q^{22}+10q^{32}+q^{35}+25q^{39}+2q^{40} \right.\\ \nn
&+& \left. q^{42}+q^{45}+q^{47}+9q^{50}+50q^{51}+q^{53}+27q^{54}+9q^{56}+10q^{59}+q^{63}+9q^{65}+10q^{66}\right. \\ \nn
&+& \left. 50q^{68}+q^{71}+25q^{72}+q^{74}+9q^{75}+q^{76}+83q^{77}+q^{79}+10q^{80}+10q^{83}+25q^{84}\right. \\ \nn
&+& \left. 26q^{85}+49q^{86}+q^{87}+q^{88}+26q^{90}+10q^{92}+139q^{93}+34q^{95}\right)+\dots
\end{eqnarray}
For $\mu_2=(2,0)$, we have:
\begin{eqnarray}\nn
\Psi_{\bfmu} &=& 2q^9(1+q^6+q^{10}+2q^{18}+2q^{20}+q^{24}+q^{26}+q^{34}+4q^{36}+12q^{44}+2q^{46}+q^{48}+9q^{50}+2q^{52}\\ \nn
&+& 10q^{54}+3q^{56}+2q^{58}+10q^{62}+q^{66}+4q^{68}+18q^{72}+11q^{74}+2q^{78}+ q^{80}+29q^{82}+q^{84}\\ \nn
&+& 46q^{86}+q^{88}+10q^{90})+\dots
\end{eqnarray}
For $\mu_2=(2,1)$, we have:
\begin{eqnarray}\nn
\Psi_{\bfmu} &=& 4q^{25}\left(1+q+q^{15}+q^{18}+4q^{25}+4q^{29}+5q^{32}+4q^{34}+8q^{37}+q^{39}+q^{41}+q^{51}+4q^{55}\right.\\ \nn
&+&\left. 4q^{59}+ 5q^{61}+4q^{62}+ 4q^{65}+q^{66}+16q^{68}+16q^{70}+9q^{73}+9q^{74}+4q^{75}\right)+\dots
\end{eqnarray}
For $\mu_2=(2,3)$, we have:
\begin{eqnarray}\nn
\Psi_{\bfmu} &=& 4\left(q^{25}+q^{39}+q^{52}+9q^{56}+q^{61}+4q^{69}+q^{75}+17q^{76}+4q^{83}+9q^{89}+q^{90}+q^{93}\right.\\ \nn
&+&\left. +5q^{95}+q^{96}+4q^{97}+25q^{99}\right)+\dots
\end{eqnarray}

\subsection{Charge configuration 2: $P_1=(1,0),P_2=P_3=(0,1)$}
\label{CC2}
In this example the total magnetic charge is $P=(1,2)$ split into
three different centers as $P_1=(1,0),P_2=P_3=(0,1)$. The charge $P_1$ corresponds to the K3 fiber.
 The CFT central charge (\ref{CFTccharge}) of $P_{2,3}$ is given above, while we have for $P_1$ and $P$,
\be
c_L(P_1)=24,\qquad c_L(P)=176. 
\ee 

In the $j$-th center, the quadratic form $D_j=d_{abc}P^c_j$ is given by
\begin{eqnarray}
D_2=D_3=\begin{pmatrix}
0 & 4 \\
4 & 2
\end{pmatrix},\quad D_1=\begin{pmatrix}
0 & 0\\
0 & 4 
\end{pmatrix}. 
\end{eqnarray}
Note that $D_1$ is a singular matrix, and we only keep the direction with the non-vanishing eigenvalue. 
The partition functions for $P_1$ can be determined using
Noether-Lefshetz theory \cite{Gholampour:2013hfa, Bouchard:2016lfg, 2007arXiv0705.1653M}.
On the other hand, the modular properties are in this case very
restrictive, and determine the partition function with a little
further inpute. First, the lowest exponent of $h_{P,\mu}$ is
$-c_L/24=-1$. Therefore, each of $h_{P,\mu}$ has a term with a
negative coefficient, and the vector space of weakly holomorphic
vector-valued modular forms is 3-dimensional. We can fix the
coefficients following \cite{Gaiotto:2006wm, Gaiotto:2007cd}: the coefficient for $\mu=0$ is the Euler number
of the linear system of the divisor, which is $c_2\cdot
P_1/12=2$. Thus the linear system is a $\mathbb{P}^1$. For $\mu=1$ and
2, we need to require that the divisor contains a given curve class
$\beta\in H^2(X,\mathbb{Z})$, and turn on a world volume
flux through this curve. This gives two complex constraints, however since the linear system is a
$\mathbb{P}^1$, there are generically no solutions. As a result, the
coefficients of the potential polar terms in $h_{P_1,1}$ and
$h_{P_1,2}$ vanish. This fixes $h_{P_1,\mu}$ to be the weight -3/2
modular form, 
\be
\label{hP1mu}
h_{P_1,\mu}(\tau)=\frac{2\,f_{P_1,\mu}(\tau)}{\eta(\tau)^{24}},
\ee
with
\be
\begin{split}
f_{P_1,\mu}(\tau)&=\frac{11}{18}
E_4(\tau)\,E_6(\tau)\,\Theta_\mu(\tau) \\
&\quad -6\, E_4(\tau)^2\,D_{1/2}\Theta_\mu(\tau)\\
&\quad +16\,E_6(\tau)\,D_{5/2}D_{1/2}\Theta_\mu(\tau),
\end{split} 
\ee
with $\Theta_\mu(\tau)=\sum_{n\in 4\mathbb{Z}+\mu} q^{n^2/8}$, $D_k$ the
modular derivative acting on weight $k$ modular forms, $D_k=\frac{1}{2\pi i}
\frac{d}{d\tau}-\frac{k}{12}E_2(\tau)$, and $E_\ell$ the Eisenstein series
of weight $\ell$. The $f_{P_1,\mu}$ also appeared in the context
of Gromov-Witten theory \cite{Klemm:2004km, 2007arXiv0705.1653M}.
Finally since $P_1^3=0$, $h_{nP_1,\mu}$ is also expected to be a weakly
holomorphic modular form (rather than a mock modular form), and given
in terms of Hecke transformations of (\ref{hP1mu})
\cite{Vafa:1994tf, Bouchard:2016lfg}.

Returning to our original problem of $\Phi_\bfmu$ and $\Psi_\bfmu$, note that we have for the products of quadratic forms $D_1D_3^{-1}=\left(
\begin{array}{cc}
 0 & 0 \\
 1 & 0 \\
\end{array}
\right)$,
$D_1D_3^{-1}D_1=\begin{pmatrix}
0 & 0\\
0 & 0
\end{pmatrix}$, $D_2D_3^{-1}D_1=D_1D_3^{-1}D_2=D_1$
which produces the quadratic form (\ref{underD}),
\begin{eqnarray}
\underline D &=& \begin{pmatrix*}[c]
0 & 0 & 0 & 0\\
0 & 4 & 0 & 4\\ 
0 & 0 & 0 & 8\\
0 & 4 & 8 & 4
\end{pmatrix*}.
\end{eqnarray}
Excluding the null direction the determinant of $\underline{D}=-256$.
The quadratic form on the lattice of electric charge lattice is given by the pseudo-inverse of $\underline{D}$ as follows:
\begin{eqnarray}\label{dinv2}
\underline{D}^{-1}=\left(
\begin{array}{cccc}
0 & 0 & 0 & 0 \\
0 & \frac{1}{4} & -\frac{1}{8} & 0 \\
0 & -\frac{1}{8} & 0 & \frac{1}{8} \\
0 & 0 & \frac{1}{8} & 0 \\
\end{array}
\right).
\end{eqnarray}
Electric charge vectors $Q_j\in \Lambda_j^*$ have the following form
\[Q_1=\begin{pmatrix} 0 \\ q_{12} \end{pmatrix},\quad
Q_2=\begin{pmatrix} q_{21} \\ q_{22} \end{pmatrix}.\] There are 64 conjugacy classes of the form
\begin{eqnarray}
q_{12}&=&4k_{12}+\mu_{12},\\ \nn
q_{21}&=&4k_{22}+\mu_{21},\quad q_{22}=4k_{21}+2k_{22}+\mu_{22},
\end{eqnarray}
where $\mu_{ij}\in\{0,1,2,3\}$, $k_{ij}\in\mathbb{Z}$. These are
$N_q=64$ (\ref{SimpNgNq}) conjugacy classes for fixed $\mu$. 

Using equations in (\ref{bfqq}), we get
\begin{eqnarray}
\bfQ_1 &=& \begin{pmatrix}
0\\
4k_{12}+4k_{22}+\mu_{12}-\mu_{31}
\end{pmatrix},\\ \nn
\bfQ_2 &=& \begin{pmatrix}
8k_{22}+\mu_{21}-\mu_{31}\\
8k_{21}+4k_{22}+4k_{12}+\mu_{22}-\mu_{32}
\end{pmatrix}.
\end{eqnarray}
When $\mu_1+\mu_2+\mu_3=0$, we have,
\begin{eqnarray}\nn
\bfQ_1 &=& \begin{pmatrix}
0\\
4k_{12}+4k_{22}+\mu_{12}+\mu_{21}
\end{pmatrix}, \quad \bfmu_1=\begin{pmatrix}
0\\ \nn
\mu_{12}+\mu_{21}
\end{pmatrix},\\
\bfQ_2 &=& \begin{pmatrix}
8k_{22}+2\mu_{21}\\
8k_{21}+4k_{22}+4k_{12}+2\mu_{22}+\mu_{12}
\end{pmatrix}, \quad \bfmu_2=\begin{pmatrix}
2\mu_{21}\\ \nn
2\mu_{22}+\mu_{12}
\end{pmatrix}.
\end{eqnarray}
\subsection*{List of $\Phi_{\bfmu}$}
We list $\Phi_\bfmu$ for a various choices of $\bfmu$. For this choice
of charges, we find that some of the inequalities for convergence are
sometimes saturated. This however do not give rise to divergences in the $q$ series.
As in explained in Section \ref{BoStaLat}, $\bfmu$ is fully determined by specifying $(\mu_1,\mu_2,\mu_3)$ and $\mu=\mu_1+\mu_2+\mu_3$. The overall sign of the $q$-expansion is given by
$(-1)^{a+b+c}=(-1)^{\mu_{12}}$.
The contributions $A_\ell$ are for this example 
\be
A_1=A_2=A_3=\frac{2}{\pi}\,\arctan(-3/4),
\ee
which is an irrational number.

For $\mu_1=\mu_2=0$, we have:
\begin{eqnarray}\nn
\Phi_{\bfmu} &=& 4q^6+4q^{24}+12q^{30}+6q^{40}+4q^{54}+12q^{64}+12q^{70}+12q^{78}+12q^{88}+4q^{96}\\ \nn
&&+\frac{(A_1+A_2+A_3+1)}{2}+(A_1+A_2+A_3+3)(q^8+q^{32}+q^{72})+\dots
\end{eqnarray}
For $\mu_1=(0,0),\;\mu_2=(1,0)$, we have:
\begin{eqnarray}\nn
\Phi_{\bfmu} &=& 2q^{33/8}\left(1+q^4+q^7+q^{16}+2q^{21}+q^{22}+q^{24}+q^{30}+3q^{31}+q^{34}+q^{41}+q^{42}+q^{44} \right.\\ \nn
&+& \left. 2q^{53}+q^{54}+q^{56}+2q^{58}+q^{60}+2q^{66}+4q^{67}+q^{72}+2q^{74}+q^{76}+2q^{77}+2q^{81} \right.\\ \nn
&+& \left. q^{82}+q^{84}+q^{91}+q^{92}+2q^{93}\right)+ \frac{(A_3+1)}{2}\; q^{1/8}(1+q^6+q^{10}+q^{28}+q^{36}+q^{66}+q^{78})
\end{eqnarray}
For $\mu_1=(0,0),\;\mu_2=(2,0)$, we have:
\begin{eqnarray}\nn
\Phi_{\bfmu} &=& 2\left(\frac{1}{2}q^{5/2}+q^{21/2}+q^{29/2}+q^{33/2}+q^{41/2}+\frac{3}{2}q^{45/2}+q^{61/2}+q^{65/2}+q^{77/2}+2q^{81/2}+ q^{85/2}\right.\\ \nn
&+&\left. q^{89/2}+3q^{101/2}+2q^{105/2}+q^{109/2}+q^{117/2}+q^{121/2}+\frac{1}{2}q^{125/2}+2q^{129/2}+q^{133/2}+q^{145/2}\right.\\ \nn
&+& \left. q^{149/2}+q^{153/2}+2q^{157/2}+q^{161/2}+q^{165/2}+2q^{173/2}+q^{181/2}+4q^{185/2}+2q^{189/2}\right)\\ \nn
&&+q^{1/2}\;\frac{(A_3+1)}{2}(1+q^4+q^{12}+q^{24}+q^{40})+\dots
\end{eqnarray}
For $\mu_1=(0,0),\;\mu_2=(3,0)$, we have:
\begin{eqnarray} \nn
\Phi_{\bfmu} &=& 2q^{105/8}\left(2+q^5+q^6+q^{13}+q^{14}+2q^{24}+q^{30}+2q^{31}+2q^{33}+q^{34}+q^{38}+q^{40}+q^{43}+2q^{45} \right.\\ \nn
&+& \left. q^{46}+q^{52}+3q^{55}+2q^{60}+q^{70}+2q^{71}+2q^{73}+q^{74}+q^{76}+2q^{78}+2q^{82}+q^{84}+q^{86}\right)\\ \nn
&&+\frac{(A_3+1)}{2} q^{9/8}(1+q^2+q^{14}+q^{20}+q^{44}+q^{54}+q^{90})+\dots
\end{eqnarray}
For $\mu_1=(0,0),\;\mu_2=(0,1)$, we have:
\begin{eqnarray}\nn
\Phi_{\bfmu} &=& 2\left(q^4+q^8+q^{12}+q^{20}+q^{24}+2q^{26}+q^{28}+2q^{34}+2q^{36}+q^{40}+2q^{46}+q^{48}+q^{56}+2q^{58}\right.\\ \nn
&+& \left. 2q^{60}+2q^{64}+4q^{70}+2q^{72}+2q^{76}+q^{80}+2q^{82}+2q^{84}+2q^{86}+q^{88}+2q^{96}+2q^{98}+q^{100}\right)+\dots\\ \nn
\end{eqnarray}
For $\mu_1=(0,0),\;\mu_2=(0,2)$, we have:
\begin{eqnarray}\nn
\Phi_{\bfmu} &=& 2\left(q^{10}+2q^{16}+2q^{22}+2q^{32}+2q^{38}+2q^{42}+2q^{48}+4q^{52}+2q^{58}+2q^{62}
+2q^{66}+4q^{76}\right.\\ \nn
&+& \left. 2q^{80}+2q^{82}+3q^{90}+4q^{94}+2q^{96}\right)
+q^2\; (A_3+1)(1+q^{16}+q^{48}+q^{96})+\dots
\end{eqnarray}
For $\mu_1=(0,1),\;\mu_2=(1,0)$, we have:
\begin{eqnarray}\nn
\Phi_{\bfmu} &=& -2q^{19/4}\left( 1+q^2+q^6+q^{17}+2q^{21}+q^{23}+q^{24}+q^{26}+q^{27}+q^{29}+q^{33}+q^{34}+q^{38}+q^{39}\right.\\ \nn
&+& \left. q^{46}+q^{52}+q^{54}+q^{56}+q^{57}+q^{58}+q^{61}+2q^{62}+2q^{64}+q^{66}+2q^{68}+q^{69}+q^{71}+q^{72} \right.\\ \nn
&+& \left. q^{75} +q^{76}+q^{77}+q^{78}+q^{81}+q^{82}+q^{86}+q^{87}+q^{88}+q^{89}+q^{90}+q^{95}\right)+\dots
\end{eqnarray}
For $\mu_1=(0,1),\;\mu_2=(2,0)$, we have:
\begin{eqnarray}\nn
\Phi_{\bfmu} &=& -2q^{71/8}\left(1+q^5+q^9+q^{12}+q^{16}+q^{21}+q^{25}+q^{26}+q^{28}+q^{30}+q^{35}+q^{36}+q^{39}+q^{41}+q^{43}\right.\\ \nn
&+& \left. q^{45}+ q^{47}+q^{48}+q^{50}+2q^{54}+q^{58}+q^{60}+q^{63}+q^{65}+q^{67}+q^{68}+q^{69}+q^{71}+q^{72}+q^{75} \right.\\ \nn
&+& \left. q^{76}+q^{78}+ q^{80}+q^{81}+q^{82}+q^{86}+q^{88}+q^{89}+q^{90}\right)+\dots \\ \nn
\end{eqnarray} 
For $\mu_1=(0,0),\;\mu_2=(1,1)$, we have:
\begin{eqnarray}\nn
\Phi_{\bfmu} &=& q^{45/8}\left(3+2q^4+q^{10}+4q^{18}+2q^{22}+2q^{23}+4q^{25}+2q^{26}+2q^{30}+4q^{35}
+2q^{36}+2q^{43}+4q^{48} \right.\\ \nn
&+& \left. 2q^{56}+4q^{57} +4q^{59}+4q^{60}+2q^{62}+4q^{66}+2q^{68}+q^{70}+4q^{71}+2q^{72}+4q^{73}+2q^{74}\right.\\ \nn
&+& \left. 4q^{82}+8q^{85}+ 4q^{90}+2q^{92}\right)+\dots\\ \nn
\end{eqnarray}
For $\mu_1=(0,0),\;\mu_2=(1,3)$, we have:
\begin{eqnarray}\nn
\Phi_{\bfmu} &=& q^{5/8}\left(1+2q^2+2q^{10}+2q^{13}+2q^{14}+2q^{16}+2q^{20}+4q^{21}+q^{30}+2q^{32}+4q^{39}+2q^{40}+4q^{42} \right. \\ \nn
&+& \left. 2q^{44}+3q^{50}+8q^{51}+2q^{54}+4q^{56}+4q^{63}+2q^{64}+2q^{66}+2q^{73}
+4q^{77}+2q^{78}\right.\\ \nn
&+& \left. 2q^{80}+4q^{84}+2q^{86}+2q^{87}+2q^{90}+4q^{92}+8q^{93}\right)+\dots\\ \nn
\end{eqnarray}
For $\mu_1=(0,0),\;\mu_2=(2,1)$, we have:
\begin{eqnarray}\nn
\Phi_{\bfmu} &=& 4q^{15/2}\left(1+q^{12}+q^{20}+q^{24}+2q^{28}+q^{40}+2q^{48}+q^{52}+q^{60}+3q^{64}+2q^{72}+2q^{76}\right.\\ \nn
&& \left. +q^{80}+2q^{92}\right)+\dots
\end{eqnarray}
For $\mu_1=(0,0),\;\mu_2=(2,3)$, we have:
\begin{eqnarray}\nn
\Phi_{\bfmu} &=& 4q^{3/2}\left(1+q^{12}+3q^{16}+q^{36}+3q^{44}+3q^{48}+3q^{56}+q^{72}+3q^{84}+3q^{92}+3q^{96}\right)+\dots
\end{eqnarray}
For $\mu_1=(0,1),\;\mu_2=(0,0)$, we have:
\begin{eqnarray}\nn
\Phi_{\bfmu} &=& -2q^{39/8}\left(1+q^2+q^3+q^{17}+q^{21}+2q^{22}+q^{24}+q^{26}+q^{27}+2q^{28}+2q^{33}+q^{39}+q^{46}+q^{52}\right.\\ \nn
&+& \left. 2q^{55}+q^{57}+2q^{60}+q^{61}+2q^{63}+q^{64}+q^{66}+q^{67}+q^{68}+4q^{70}+2q^{76}+q^{78}+2q^{79}+2q^{81}\right.\\ \nn
&+& \left. q^{87}+2q^{89}+q^{95}\right)+\dots \\ \nn
\end{eqnarray}
For $\mu_1=(0,1),\;\mu_2=(1,1)$, we have:
\begin{eqnarray}\nn
\Phi_{\bfmu} &=& -2q^{45/4}\left(1+q^3+q^6+q^{10}+q^{14}+q^{18}+q^{23}+2q^{28}+q^{31}+q^{33}+q^{34}+q^{36}+2q^{39} +q^{42}\right.\\ \nn
&+& \left. q^{45}+q^{46}+q^{49}+q^{51}+q^{53}+q^{55}+q^{58}+q^{59}+q^{62}+q^{65}+q^{67}+q^{70}+2q^{72}+q^{75}+q^{77}\right. \\  \nn
&+& \left. q^{78}+q^{80}+ q^{81}+q^{82}+q^{83}+2q^{85}+2q^{88}\right)\dots
\end{eqnarray}
For $\mu_1=(0,1),\;\mu_2=(1,2)$, we have:
\begin{eqnarray}\nn
\Phi_{\bfmu} &=& -2q^{47/4}\left(1+q^3+q^5+q^{9}+q^{13}+q^{17}+q^{23}+q^{28}+q^{29}+q^{31}+q^{32}+q^{35}+q^{36}+q^{37}+q^{38} \right.\\ \nn
&+&\left. q^{42}+2q^{44}+q^{49}+q^{50}+q^{52}+q^{53}+2q^{58}+q^{60}+q^{65}+q^{68}+q^{70}+q^{71}+q^{74}+q^{76} \right. \\ \nn
&+& \left. q^{77}+q^{78}+q^{79}+q^{80}+q^{83}+q^{84}+3q^{85}+q^{87}\right)+\dots
\end{eqnarray}
For $\mu_1=(0,1),\;\mu_2=(1,3)$, we have:
\begin{eqnarray}\nn
\Phi_{\bfmu} &=& -2q^{13/4}\left(1+q^6+q^{10}+q^{15}+q^{19}+q^{20}+q^{21}+q^{27}+q^{30}+q^{33}+q^{34}+q^{35}+q^{39}+q^{42} + q^{45}\right. \\ \nn
&+& \left. q^{46}+q^{50}+q^{51}+q^{52}+q^{54}+q^{56}+q^{58}+q^{60}+q^{63}+q^{64}+q^{65}+q^{70}+q^{71}+q^{72}+q^{74} \right. \\ \nn
&+&\left. q^{75}+ q^{76}+q^{78}+q^{80}+q^{81}+q^{84}+q^{85}+q^{86}+q^{87}+q^{93}+q^{94}+q^{95}+q^{96}\right)+\dots
\end{eqnarray}
For $\mu_1=(0,1),\;\mu_2=(2,2)$, we have:
\begin{eqnarray}\nn
\Phi_{\bfmu} &=& -2q^{15/8}\left(1+q^{10}+q^{13}+q^{15}+q^{16}+2q^{18}+q^{27}+q^{33}+q^{38}+2q^{40}+q^{42}+q^{45}+3q^{47} \right. \\ \nn
&+& \left. q^{48}+2q^{52}+2q^{53}+2q^{58}+q^{60}+q^{68}+2q^{69}+q^{75}+2q^{79}+q^{81}+3q^{86}+2q^{88}+q^{90}\right.\\ \nn
&+& \left. q^{93}+q^{94}+q^{95}+3q^{96}+2q^{97}\right)+\dots
\end{eqnarray}
For $\mu_1=(0,2),\;\mu_2=(0,1)$, we have:
\begin{eqnarray}\nn
\Phi_{\bfmu} &=& 2q^{11/2}\left(1+q^4+q^{10}+2q^{18}+q^{22}+q^{24}+2q^{26}+q^{30}+q^{34}+2q^{36}+q^{42}+q^{44} \right.\\ \nn
&+& \left. 2q^{48}+2q^{56}+2q^{58}+3q^{60}+q^{64}+2q^{68}+q^{70}+3q^{72}+2q^{74}+q^{80}+3q^{84}+2q^{86}\right. \\ \nn
&+& \left. 3q^{90}+q^{94}\right)+\dots \\ \nn
\end{eqnarray}
For $\mu_1=(0,2),\;\mu_2=(1,0)$, we have:
\begin{eqnarray}\nn
\Phi_{\bfmu} &=& 2q^{41/8}\left(1+2q^5+q^{10}+2q^{11}+4q^{18}+q^{21}+q^{25}+2q^{26}+2q^{27}+2q^{32}+2q^{35}+2q^{36} \right.\\ \nn
&+& \left. 2q^{37}+q^{41}+2q^{44}+3q^{45}+4q^{48}+2q^{55}+2q^{56}+2q^{57}+q^{58}+2q^{59}+2q^{60}+2q^{61}\right. \\ \nn
&+& \left. q^{66}+2q^{68}+3q^{70}+2q^{71}+2q^{73}+2q^{74}+4q^{75}+2q^{81}+2q^{83}+4q^{86}+4q^{87}+q^{88}\right.\\ \nn
&+& \left. 4q^{90}+2q^{93}\right)+\dots
\end{eqnarray}
For $\mu_1=(0,2),\;\mu_2=(1,1)$, we have:
\begin{eqnarray}\nn
\Phi_{\bfmu} &=& q^{29/8}\left(1+q^4+3q^{9}+q^{15}+2q^{16}+q^{19}+q^{22}+2q^{23}+2q^{24}+q^{29}+q^{30}+4q^{33}+2q^{38} \right.\\ \nn
&+& \left. q^{39}+q^{40}+2q^{42}+2q^{43}+2q^{44}+q^{49}+2q^{51}+q^{52}+3q^{54}+2q^{55}+2q^{56}+q^{60}+2q^{62}\right. \\ \nn
&+& \left. 3q^{65}+4q^{66}+4q^{69}+q^{70}+q^{72}+2q^{78}+3q^{79}+2q^{80}+2q^{82}+4q^{83}+2q^{84}+q^{85}+q^{87}\right.\\ \nn
&+& \left. 2q^{91}+2q^{93}+q^{94}+4q^{95}+2q^{96}\right)+\dots
\end{eqnarray}
For $\mu_1=(0,2),\;\mu_2=(2,0)$, we have:
\begin{eqnarray}\nn
\Phi_{\bfmu} &=& 4\left(q^{13}+q^{19}+q^{27}+q^{37}+q^{43}+q^{45}+q^{47}+q^{53}+q^{57}+q^{59}+q^{67}+q^{69}+q^{73}+q^{83}\right. \\ \nn
&+& \left. q^{85}+q^{87}+q^{89}+q^{93}+q^{97}+q^{99}\right)+\dots
\end{eqnarray}

\subsection*{List of $\Psi_{\bfmu}$}
Similarly as the $\Phi_{\bfmu}$ $q$-series we list the $q$-series of $\Psi_\bfmu$ for a various choices of $\bfmu$. In this example $(-1)^{a+b+c}=(-1)^{\mu_{12}}$.
For $\mu_1=\mu_2=(0,0)$, we have:
\begin{eqnarray}\nn
\Psi_{\bfmu} &=& 16 (q^6 + 4 q^{24} + 3 q^{30} + 9 q^{54} + 12 q^{64} + 3 q^{70} + 
 3 q^{78} + 16 q^{96})+\dots
\end{eqnarray}
For $\mu_1=(0,0)\;\mu_2=(1,0)$, we have:
\begin{eqnarray}\nn
\Psi_{\bfmu} &=& 2q^{33/8}\left(1+4q^4+9q^{16}+2q^{21}+4q^{22}+16q^{24}+4q^{30}+8q^{31}+q^{42}+25q^{44}+18q^{53}+16q^{54} \right.\\ \nn
&+& \left. 36q^{56}+2q^{58}+4q^{60}+17q^{66}+40q^{67}+4q^{72}+8q^{74}+8q^{81}+9q^{82}
+49q^{84}+2q^{93}\right)+\dots
\end{eqnarray}
For $\mu_1=(0,0)\;\mu_2=(2,0)$, we have:
\begin{eqnarray}\nn
\Psi_{\bfmu} &=& 8q^{21/2}\left(1+q^6+q^{12}+4q^{22}+q^{28}+2q^{30}+4q^{32}+2q^{40}+5q^{42}+q^{48}+2q^{54}+9q^{56}+\right. \\ \nn \nn
&+&\left. 4q^{66}+8q^{68}+9q^{70}+q^{72}+2q^{76}+10q^{82}+13q^{84}\right)+\dots
\end{eqnarray}
For $\mu_1=(0,0)\;\mu_2=(3,0)$, we have:
\begin{eqnarray}\nn
\Psi_{\bfmu} &=& 2q^{105/8}\left(5+4q^6+q^{14}+25q^{24}+4q^{30}+2q^{31}+8q^{33}+16q^{34}
+4q^{40}+8q^{45}+9q^{46}+2q^{55}\right. \\ \nn 
&+& \left. 61q^{60}+16q^{70}+18q^{71}+32q^{73}+36q^{74}+4q^{76}+2q^{78}+8q^{82}+16q^{84}+q^{86}\right)+\dots
\end{eqnarray}
For $\mu_1=(0,0)\;\mu_2=(0,1)$, we have:
\begin{eqnarray}\nn
\Psi_{\bfmu} &=& 2q^4\left(1+q^4+q^8+9q^{16}+q^{20}+2q^{22}+9q^{24}+2q^{30}+10q^{32}+q^{36}+2q^{42}+25q^{44}\right.\\ \nn
&+& \left. 9q^{52}+18q^{54}+26q^{56}+2q^{60}+20q^{66}+34q^{68}+2q^{72}+9q^{76}
+2q{78}+2q^{80}+18q^{82}\right. \\ \nn
&+& \left. 49q^{84}+2q^{92}+2q^{94}+25q^{96}\right)+\dots
\end{eqnarray}
For $\mu_1=(0,0)\;\mu_2=(0,2)$, we have:
\begin{eqnarray}\nn
\Psi_{\bfmu} &=& 16\left(q^{16}+q^{32}+4q^{42}+q^{48}+2q^{52}+4q^{66}+2q^{76}+9q^{80}+4q^{90}+8q^{94}+q^{96}\right)+\dots
\end{eqnarray}
For $\mu_1=(0,0)\;\mu_2=(1,1)$, we have:
\begin{eqnarray}\nn
\Psi_{\bfmu} &=& 2q^{45/8}\left(1+q^4+13q^{18}+q^{22}+2q^{25}+9q^{26}+q^{30}+2q^{32}
+4q^{36}+41q^{48}+9q^{56}+8q^{57}\right. \\ \nn
&+& \left. 18q^{59}+ 26q^{60}+2q^{66}+9q^{68}+2q^{71}+q^{72}+18q^{73}+16q^{74}+2q^{82}+10q^{85}+85q^{90}\right)+\dots
\end{eqnarray}
For $\mu_1=(0,0)\;\mu_2=(1,3)$, we have:
\begin{eqnarray}\nn
\Psi_{\bfmu} &=& 2q^{21/8}\left(1+4q^8+q^{12}+9q^{14}+q^{18}+2q^{19}+16q^{30}+8q^{37}+9q^{38}
+25q^{40}+q^{42}+9q^{48}\right. \\ \nn
&+& \left. 20q^{49}+q^{52}+2q^{54}+2q^{61}+4q^{62}+36q^{64}+32q^{75}+25q^{76}+49q^{78}+8q^{82}+9q^{84}+q^{88}\right.\\ \nn
&&\left. +29q^{90}+ 68q^{91}\right)\dots
\end{eqnarray}
For $\mu_1=(0,0)\;\mu_2=(2,1)$, we have:
\begin{eqnarray}\nn
\Psi_{\bfmu} &=& 4q^{15/2}\left(1+q^{12}+9q^{20}+q^{24}+2q^{28}+9q^{40}+2q^{48}+25q^{52}+9q^{60}+19q^{64}+2q^{72}+2q^{76}\right.\\ \nn
&+&\left. 25q^{80}+18q^{92}\right)
+\dots
\end{eqnarray}
For $\mu_1=(0,0)\;\mu_2=(2,3)$, we have:
\begin{eqnarray}\nn
\Psi_{\bfmu} &=& 4q^{3/2}\left(1+9q^{12}+3q^{16}+25q^{36}+27q^{44}+3q^{48}+3q^{56}+49q^{72}+75q^{84}
+27q^{92}+3q^{96}\right)
+\dots
\end{eqnarray}
For $\mu_1=(0,2)\;\mu_2=(2,0)$, we have:
\begin{eqnarray}\nn
\Psi_{\bfmu} &=& 4\left(q^{13}+q^{19}+q^{27}+9q^{37}+q^{43}+q^{45}+9q^{47}+q^{53}+q^{57}+9q^{59}+q^{67}+q^{69}\right.\\ \nn
&+& \left. 25q^{73}+9q^{83}+9q^{85}+25q^{87}+q^{89}+q^{93}+9q^{97}+q^{99}\right)+\dots
\end{eqnarray}
For $\mu_1=(0,1)\;\mu_2=(1,0)$, we have:
\begin{eqnarray}\nn
\Psi_{\bfmu} &=& -\frac{q^{19/4}}{2}\left(1+q^2+q^6+25q^{17}+26q^{21}+q^{23}+9q^{24}+q^{26}+q^{27}+25q^{29}+q^{33}+9q^{34}+q^{38}\right.\\ \nn
&+& \left. 9q^{39}+81q^{46}+81q^{52}+25q^{54}+25q^{56}+49q^{57}+q^{58}+25q^{61}+26q^{62}+90q^{64}+q^{66}+2q^{68}\right. \\ \nn
&+& \left. 49q^{71}+ 25q^{72}+q^{75}+q^{76}+25q^{77}+49q^{78}+9q^{81}+9q^{82}+q^{86}+169q^{87}+9q^{88}+q^{89}\right. \\ \nn
&& \left. +9q^{90} +169q^{95}\right)+\dots
\end{eqnarray}
For $\mu_1=(0,1)\;\mu_2=(2,0)$, we have:
\begin{eqnarray}\nn
\Psi_{\bfmu} &=& -\frac{q^{71/8}}{2}\left(1+q^5+9q^9+q^{12}+9q^{16}+25q^{21}+9q^{25}+q^{26}+q^{28}+25q^{30}+q^{35}+49q^{36}\right. \\ \nn
&+& \left. q^{39}+25q^{41}+9q^{43}+9q^{45}+49q^{47}+q^{48}+q^{50}+90q^{54}+9q^{58}+49q^{60}+25q^{63}\right. \\ \nn
&+& \left. 25q^{65}+81q^{67}+q^{68}+9q^{69}+9q^{71}+q^{72}+121q^{75}+25q^{76}+q^{78}+25q^{80}\right.\\ \nn
&+& \left. q^{81}+81q^{82}+49q^{86}+49q^{88}+q^{89}+121q^{90}\right)+\dots
\end{eqnarray}
For $\mu_1=(0,1)\;\mu_2=(0,0)$, we have:
\begin{eqnarray}\nn
\Psi_{\bfmu} &=& -\frac{q^{39/8}}{2}\left(1+9q^2+q^3+25q^{17}+49q^{21}+2q^{22}+25q^{24}+9q^{26}+q^{27}+18q^{28}+2q^{33}+9q^{39}\right. \\ \nn
&+& \left. 81q^{46} 121q^{52}+50q^{55}+81q^{57}+2q^{60}+49q^{61}+98q^{63}+25q^{64}+9q^{66}+q^{67} +q^{68}\right. \\ \nn
&+& \left. 68q^{70}+18q^{76}+49q^{78}+2q^{79}+2q^{81}+169q^{87}+18q^{89}+225q^{95}\right)+\dots
\end{eqnarray}
For $\mu_1=(0,1)\;\mu_2=(1,1)$, we have:
\begin{eqnarray}\nn
\Psi_{\bfmu} &=&  -\frac{q^{45/4}}{2}\left(1+q^3+q^6+9q^{10}+9q^{14}+9q^{18}+25q^{23}+26q^{28}+q^{31}+25q^{33}+q^{34}+q^{36}\right. \\ \nn
&+& \left. 50q^{39}+q^{42}+49q^{45}+9q^{46}+9q^{49}+49q^{51}+9q^{53}+9q^{55}+81q^{58}+9q^{59}+9q^{62}+81
q^{65} \right. \\ \nn
&+& \left. 25q^{67}+25q^{70}+82q^{72}+25q^{75}+25q^{77}+q^{78}+121q^{80}+q^{81}+25q^{82}+q^{83}+26q^{85}+122q^{88}\right)\\ \nn
&&+\dots
\end{eqnarray}
For $\mu_1=(0,1)\;\mu_2=(1,2)$, we have:
\begin{eqnarray}\nn
\Psi_{\bfmu} &=&  -\frac{q^{47/4}}{2}\left(9+9q^3+q^5+9q^9+q^{13}+q^{17}+49q^{23}+49q^{28}+9q^{29}+9q^{31}+25q^{32}+9q^{35}\right. \\ \nn
&+& \left. 9q^{36}+q^{37}+49q^{38}+q^{42}+34q^{44}+9q^{49}+25q^{50}+q^{52}+q^{53}+122q^{58}+q^{60}+121q^{65}+49q^{68} \right. \\ \nn
&+& \left. 49q^{70}+81q^{71}+9q^{74}+49q^{76}+49q^{77}+9q^{78}+121q^{79}+25q^{80}
+9q^{83}+9q^{84}+25q^{85}+81q^{87}\right)\\ \nn
&&+\dots
\end{eqnarray}
For $\mu_1=(0,1)\;\mu_2=(1,3)$, we have:
\begin{eqnarray}\nn
\Psi_{\bfmu} &=&  -\frac{q^{13/4}}{2}\left(1+9q^6+q^{10}+25q^{15}+q^{19}+q^{20}+9q^{21}+49q^{27}+q^{30}+9q^{33}+9q^{34}+25q^{35}\right. \\ \nn
&+& \left. q^{39}+81q^{42}+q^{45}+9q^{46}+25q^{50}+25q^{51}+49q^{52}+q^{54}+q^{56}+9q^{58}+121q^{60}\right. \\ \nn
&+& \left. q^{63}+9q^{64}+25q^{65}+49q^{70}+49q^{71}+81q^{72}+q^{74}+q^{75}+9q^{76}+9q^{78}+25q^{80}\right. \\ \nn
&+& \left. 169q^{81}+q^{84}+9q^{85}+25q^{86}+49q^{87}+81q^{93}+81q^{94}+121q^{95}+q^{96}\right)+\dots
\end{eqnarray}
For $\mu_1=(0,1)\;\mu_2=(2,2)$, we have:
\begin{eqnarray}\nn
\Psi_{\bfmu} &=&  -\frac{q^{15/8}}{2}\left(1+9q^{10}+25q^{13}+9q^{15}+q^{16}+2q^{18}+q^{27}+49q^{33}+81q^{38}+18q^{40}+49q^{42}+25q^{45}\right. \\ \nn
&+& \left. 59q^{47}+q^{48}+2q^{52}+18q^{53}+2q^{58}+25q^{60}+121q^{68}+2q^{69}+169q^{75}
+98q^{79}+121q^{81}\right. \\ \nn
&+& \left. 99q^{86}+162q^{88}+49q^{90}+25q^{93}+9q^{94}+9q^{95}+99q^{96}+50q^{97}\right)+\dots
\end{eqnarray}
For $\mu_1=(0,2)\;\mu_2=(0,1)$, we have:
\begin{eqnarray}\nn
\Psi_{\bfmu} &=&  16q^{47/2}\left(1+q^8+q^{18}+4q^{30}+q^{38}+q^{40}+4q^{42}+q^{50}+q^{54}+4q^{56}+q^{66}+q^{68}+9q^{72}\right)+\dots
\end{eqnarray}
For $\mu_1=(0,2)\;\mu_2=(1,0)$, we have:
\begin{eqnarray}\nn
\Psi_{\bfmu} &=& 2q^{81/8}\left(1+q^6+5q^{13}+q^{21}+9q^{22}+q^{27}+q^{30}+4q^{31}+9q^{32}+q^{39}+q^{40}+25q^{43}+4q^{50}\right.\\ \nn
&+& \left. q^{51}+q^{52}+4q^{54}+9q^{55}+25q^{56}+q^{63}+9q^{65}+q^{66}+9q^{68}+16q^{69}+26q^{70}+q^{76}\right.\\ \nn
&+& \left. 4q^{78}+10q^{81}+13q^{82}+61q^{85}+q^{88}\right)+\dots
\end{eqnarray}
For $\mu_1=(0,2)\;\mu_2=(1,1)$, we have:
\begin{eqnarray}\nn
\Psi_{\bfmu} &=& 2q^{101/8}\left(1+4q^7+q^{14}+4q^{15}+13q^{24}+q^{29}+q^{33}+4q^{34}+16q^{35}+4q^{42}+4q^{45}+9q^{46}\right.\\ \nn
&+& \left. 16q^{47}+q^{53}+4q^{56}+5q^{57}+41q^{60}+9q^{69}+4q^{70}+4q^{71}+9q^{73}+17q^{74}\right.\\ \nn
&+& \left. 36q^{75}+q^{82}+4q^{84}+17q^{86}+4q^{87}\right)+\dots
\end{eqnarray}

\vspace{1cm} \noindent {\bf Acknowledgement:} We thank Sergey
Alexandrov and Boris Pioline for useful discussions and comments on the draft. The authors are supported by Laureate Award 15175 ``Modularity in Quantum Field Theory and Gravity" of the Irish Research Council.

\bibliographystyle{JHEP}   

 \providecommand{\href}[2]{#2}\begingroup\raggedright\endgroup

\end{document}